\documentclass[prc,unsortedaddress,superscriptaddress,twocolumn,amsmath,amsfonts,amstex,amssymb,showpacs,floatfix]{revtex4}
\usepackage[dvips]{graphicx}
\usepackage{epsfig}   
\begin{document}

\title{Theory of deuteron stripping. From surface integrals to generalized $R$-matrix approach.}

\author{A.\,M.~Mukhamedzhanov}
\affiliation{Cyclotron Institute, Texas A\&M University,
College Station, TX 77843}

\date{\today}
                               
\begin{abstract}
There are two main reasons for absence of the practical theory of stripping to resonance states which could be used by experimental groups: numerical problem of the convergence of the DWBA matrix element when the full transition operator is included and it is unclear what spectroscopic information can be extracted from the analysis of transfer reactions populating the resonance states. The purpose of this paper is to address both questions.  The theory of the deuteron stripping is developed, which is based on the post continuum discretized coupled channels (CDCC) formalism going beyond of the DWBA and surface integral formulation of the reaction theory  [A. S. Kadyrov et al., Ann. Phys.  {\bf 324}, 1516 (2009)].  First, the formalism is developed for the DWBA and then extended to the CDCC formalism, which is ultimate goal of this work. The CDCC wave function takes into account not only the initial elastic $d+ A$ channel but also its coupling to the deuteron breakup channel $p + n + A$ missing in the DWBA. Stripping to both bound states and resonances are included. The convergence problem for stripping to resonance states is solved in the post CDCC formalism. The reaction amplitude is parametrized in terms of the reduced width amplitudes (ANCs), inverse level matrix, boundary condition and channel radius, that is the same parameters which are used in the conventional $R$-matrix method. For stripping to resonance states many-level, one and two-channel cases are considered. The theory provides a consistent tool to analyze both binary resonant reactions and deuteron stripping in terms of the same parameters.
 
\end{abstract}

\maketitle

\section{Introduction}
Production of unstable nuclei close to proton and neutron drip lines has become possible in recent years, making deuteron stripping reactions $(d,p)$ and $(d,n)$ on these nuclei (in inverse kinematics) not only more and more feasible as beam intensity increasing but also a unique tool to study unstable nuclei and astrophysical $(n,\,\gamma)$, $\,(p,\gamma)$ and $(p,\,\alpha)$ processes.  
The deuteron stripping reactions populating resonance states of final nuclei are important and most challenging part of reactions on unstatble nuclei. If for nucleon transfer reactions populating bound states for about fifty years experimentalists used the standard distorted waves Born approximation (DWBA), an adequate theory for transfer reactions to resonance states yet to be developed.
By standard DWBA I mean the approach in which the one-step transfer matrix element is evaluated with incoming and outgoing distorted waves calculated by fitting the deuteron and proton elastic scattering with local optical potentials. The transition operator contains finite range effects as well as the full complex remnant term. The main idea of the DWBA is that the transition matrix element is so small that one can use the first order perturbation theory. 
Since the nuclear potential is quite large by itself ($\sim 100$ MeV), the smallness of the transition operator can be fulfilled only if the reaction is peripheral enough, so that the non-diagonal matrix element, representing the transfer reaction amplitude, becomes small. However, since the resonance wave function is large in the nuclear interior and different channels are coupled in the nuclear interior, the character of the stripping to resonances can be quite different from the stripping to bound states. Nowadays the standard DWBA is gradually being replaced by more advanced approaches like continuum dicretized coupled channels (CDCC) \cite{rawitscher,kamimura,austern},  adiabatic distorted wave (ADWA) \cite{johnson},  coupled reaction channels  (CRC) and the coupled channels in Born approximation (CCBA) available in FRESCO code \cite{thompson}. There are two main reasons for absence of the practical theory of stripping to resonance states which could be used by experimental groups. First one is the numerical problem of the convergence of the DWBA matrix element when the full transition operator is included. However, it is only a technical problem. The second pure scientific unsolved problem is what spectroscopic information can be extracted from the analysis of transfer reactions populating the resonance states.  Besides, since the standard DWBA is deficient to more advanced methods like CDCC or ADWA, a new approach should go beyond of the DWBA. 

Majority of theoretical works devoted to the development of the theory of single-nucleon stripping into unbound states of the residual nucleus have been published in 1970-s \cite{huby,levin,fortune,vincent,bang,bunakov70,bunakov70a,bunakov71,huby70,cole,cole71,lipperheide,noble,dol73,zeldovich,berggren}. Great interest in these reactions at that time stemmed primarily from the fact that they
allow one to extract reliable information on the properties of nuclear resonant states
by means of the combined analysis of the data on stripping and elastic resonant
scattering of nucleons from the target nucleus \cite{fortune,bunakov71,cole,cole71}.
In most of the cited works the theory of stripping into resonant states was developed
within the standard DWBA by analogy with usual stripping
to bound states. In this case the expression for the reaction amplitude obtained instead
of the bound-state wave function for the captured nucleon (form factor) contained
a continuum wave function which leads to slow convergence of the radial integrals
or even to their divergence depending on the choice of this wave function. In
Refs. \cite{huby,vincent,bunakov70} the form factor was taken to be a scattering wave function, which described
the resonant scattering of the nucleon from the target nucleus. This wave function was
calculated using a single-particle potential whose parameters were adjusted to give a
resonance with the corresponding properties. The Gamov decaying-state wave function
and the Weinberg wave function which are regular at the origin and purely outgoing
at infinity were used in Refs. \cite{bang} and \cite{huby70}, respectively.
Various methods were suggested to calculate radial integrals practically with the
above-mentioned form factors: (i) the introduction of the convergence factor $exp
(-\alpha\,r)$ into the integrand \cite{huby}; the integral obtained was calculated for various $\alpha > 0$ and then its values were extrapolated numerically to the limit of $\alpha = 0$; (ii) the method of contour integration in the complex $r$-plane (complex scaling) \cite{vincent}; (iii) the method based on the correct account of the boundary conditions in the three-body scattering problem \cite{bunakov70};
(iv) the Zeldowich-Berggren method \cite{zeldovich} of the regularization of integrals containing the Gamov function in which the convergence factor $exp (-\alpha\,r^{2})$ was introduced \cite{bang};
(v) the pseudo-bound-states method \cite{huby70}. The methods (ii) and (iii) were most convenient for numerical calculations. 
Although the above methods allow one to avoid formal difficulties, nevertheless all the methods are rather complicated because of cumbersome numerical calculations and carry on the shortcomings of the standard DWBA for stripping to bound states. 

Even if  we put aside the technical problem of convergence of the matrix element for stripping to resonance states, there is more important question remains: the spectroscopic information which can be extracted from analysis of deuteron stripping reactions (and other transfer reactions) into resonant states. This is really a crucial question because the answer 
determines the reason why we measure nuclear reactions.
For more than 50 years transfer reactions to bound states, and deuteron stripping in particular, have been used to determine the spectroscopic factors, which measure the weight of the single-particle state in the overlap function of the initial and final nuclei. That is why there was always a temptation to develop a theory of stripping into resonant states which is fully similar to stripping to bound states.  For example, in \cite{bunakov71} it was assumed that the spectroscopic factor could be extracted from deuteron stripping into resonance states. In this case the spectroscopic factor is the ratio of the observable and single-particle resonance widths.
However, the spectroscopic factor is not observable and depends on the single-particle potential used to calculate the single-particle width. In \cite{muk2010} it has been shown that spectroscopic factors are not invariant under finite-range unitary transformations and, hence, in exact approach nuclear reactions cannot be a tool to determine spectroscopic factors. In \cite{muk2010} it was called separation of nuclear reactions and spectroscopic factors. However, there is a model-independent information, which can be extracted from deuteron stripping reactions. I mean the asymptotic normalization coefficients (ANCs), which are the amplitudes of the tails of the overlap functions \cite{blokhintsev} and are invariant under finite range unitary transformations. The most model-independent definition of the ANC is that it determines the residue of the elastic scattering $S$ matrix in the pole corresponding to bound, virtual or resonance states. For the resonance state the ANC and partial resonance widths are related \cite{muk77,muktr99}:
\begin{equation}
[C_{A\,j\,l}^{F}]^{2}= (-1)^{l}e^{2\,\phi_{j\,l}(k_{xA(0)j\,l})}\,\frac{\mu_{xA}}{k_{xA(0)j\,l}}\,\Gamma_{xA\,j\,l}.
\label{ANCGamma1}
\end{equation}
Here $l$ and $j$ are the orbital and total angular momentum of particle $x$ in the resonance state $F= (A\,x)$, $\mu_{xA}$ is the reduced mass of $x$ and $A$, $k_{xA(0)j\,l}$ is the real part of the resonance relative momentum of $x$ and $A$, $\phi_{j\,l}(k_{xA})$ is the non-resonant scattering phase shift, $C_{A\,j\,l}^{F}$ and $\Gamma_{xA\,j\,l}$ are the ANC and partial resonance width in the channel $x+ A$ with the quantum numbers $l$ and $j$. 
Eq. (\ref{ANCGamma1}) stands for narrow resonance, i.e. for $k_{xA(I)j\,l} << k_{xA(0)j\,l}$,
where $k_{xA(I)j\,l}$ is the imaginary part of the resonance momentum $k_{xA(R)j\,l}= k_{xA(0)j\,l} - i\,k_{xA(I)j\,l}$, which determines the location of the resonance pole in the momentum plane. 
Due to relation (\ref{ANCGamma1}), the resonance width is also invariant under finite-range unitary transformations and can be determined from the experiment. 

Nowadays, it is quite well understood that the ANCs can be determined from peripheral transfer reactions,
see \cite{goncharov82,goncharov,gulyamov,gagliardi,muk2003,mohr} and references therein. However, the ANC method has been applied only for transfer reactions populating bound states. 
It is well known that from binary resonance scattering and reactions using the conventional $R$-matrix approach one can determine the resonance partial widths, which, as we have underscored are related to the ANCs. $R$-matrix method is one of the most popular tools among the experimental groups worldwide because the approach is comparatively simple even for many-body, many-channel cases and deals with the formal partial resonance widths determined from the fit to the experimental data. These formal widths can be easily related with the observable partial widths. Using the $R$-matrix approach one can fit simultaneously data for all available channels. It allows one to control the consistency of the obtained physical parameters. The question is whether the theory of stripping to resonance states can be formulated in terms of the same parameters which are used in the $R$-matrix analysis of the binary resonance reactions.  

It is the purpose of this paper to address a theory of the deuteron stripping, which will solve all the above mentioned problems for the deuteron stripping into resonant states. The delivered theory is based on the post CDCC formalism going beyond of the DWBA and surface integral formulation of the reaction theory. The CDCC wave function takes into account not only the initial elastic $d+ A$ channel but also its coupling to the deuteron breakup channel $p + n + A$ missing in the DWBA. The convergence problem is also resolved in this formalism. The reaction amplitude is parametrized in terms of the reduced width amplitudes (ANCs), inverse level matrix, boundary condition and channel radius, that is the same parameters which are used in the $R$-matrix method. Thus the theory provides a consistent tool to analyze both binary resonant reactions and deuteron stripping in terms of the same parameters.

The theory is based on the surface-integral formulation of nuclear reactions and valid for stripping to both bound and resonance states. First, just for demonstration of the formalism, the transformation of the DWBA amplitude for stripping to the bound state is presented. The reaction matrix element is split into two parts: internal (over the relative coordinate between the transferred nucleon and target) and external. The idea of such separation is based on the fact that in the post formalism the main contribution to the stripping amplitude comes from the nuclear exterior while the prior form amplitude is dominated by the internal region. It will be shown that the dominant external post (internal prior) amplitude using the Green's theorem can be written as the dominant surface integral encircling the internal volume plus small addition from the prior external (post internal) part. Thus, both post and prior forms lead to the same reaction amplitude given by the sum of small internal post form, small external prior form and the dominant surface integral. The contribution of the post internal part can be minimized by a proper choice of the final-state optical potential, and the other two amplitudes are parameterized in terms of the reduced widths amplitudes (ANCs).  After that the theory is extended to the CDCC formalism.
Then the theory is applied for stripping to resonance states. First it is developed for the standard DWBA and then the post CDCC formalism based on the surface integrals is developed. One of the most important results of this paper is that the post CDCC form for stripping into resonant states can be written as the sum of the small internal (over the coordinate $r_{nA}$) post form and the dominant surface part. The absence of the diverging (or poor converging) external part solves the problem of convergence of the matrix element for stripping to resonance state.

In the developed approach the information about the resonance subprocess is contained in the scattering wave function of the fragments formed by resonance decay. This wave function is written in a standard $R$-matrix form using its separation into the internal and external parts. It allows us to generalize the $R$-matrix method for binary reactions to stripping reactions. Since the deuteron stripping into resonant states is $2 \to 3$ particles reaction, the excitation of the resonance occurs in the subsystem, while the third particle causes the distortion. The extracted partial resonance widths can be used for calculation of the $(n,\gamma)$ processes. If the cross section for $(n,\,\gamma)$ resonant capture is available, the simultaneous fit to the deuteron stripping and $(n,\,\gamma)$ resonance capture can be done. The method can be also applied for analysis of the Trojan Horse reactions \cite{lacognata}. Concrete calculations and the application of the theory for deuteron stripping and Trojan Horse reactions will be presented in the following up papers.
In what follows we use the system of units in which $\hbar=c=1$. We also neglect the spins of the particles 
if not specified otherwise. 

\section{Surface integral formulation for deuteron stripping to bound state.}
Before the theory of the deuteron stripping to resonant states will be outlined I will present a surface integral formulation of the theory for stripping populating bound states. First, just for demonstration, I consider the DWBA and then extend it by including the CDCC wave functions. As it has been explained in Introduction, the transfer reaction matrix element will be split into two parts in the subspace determining the relative motion of the transferred nucleon and target: internal and external parts. After that replacing the potentials in the transition operators by the kinetic energy operators and using the Green's theorem the matrix element in terms of the surface integral will be obtained. 

\subsection{Stripping to bound state. Post form of  DWBA.} 
\label{postDWBA}
In this section we consider the post form DWBA amplitude, which we split into the internal and external part in the subspace over the relative coordinate between the transferred $n$ and $A$. Due to the choice of the transition operator in the post form, the internal part turns out to be small. The external part, which is parameterized in terms of the ANC, will be transformed into the dominant surface integral encircling the internal volume and small external prior DWBA amplitude. 

We start consideration from the exact reaction amplitude for the deuteron stripping
to bound states 
\begin{equation}
d + A \to p + F,
\label{deutstrbndst1}
\end{equation}
where $F=(A\,n)$ is the bound state. 
The post form of the exact reaction amplitude 
\begin{equation}
 M^{(post)}({\rm {\bf k}}_{pF},\,{\rm {\bf k}}_{dA})= < \Phi_{f}^{(-)}\big|\,\Delta V_{pF} \big|\Psi_{i}^{(+)}>,
\label{postreactampl1}
\end{equation}
where $\Psi_{i}^{(+)}$ is the exact scattering wave function in the initial state with the two-body incident wave $d + A$,
$\,\Phi_{f}^{(-)}=\chi_{pF}^{(-)}\,\varphi_{F }^{*}$  is the channel function in the exit state $p+ F$,
 $\,\varphi_{i}$ is the bound-state wave function of nucleus $i$, $\,\chi_{ij}^{(+)} \equiv \chi_{{\rm {\bf k}}_{ij}}^{(+)}({\rm {\bf r}}_{ij})$ is the distorted wave describing the relative motion of particles $i$ and $j$ with the relative momentum ${\rm {\bf k}}_{ij}$; $\,\,\Delta\,V_{pF} = V_{pA} + V_{pn} - U_{pF}$ is the transition operator in the post form, $V_{ij}$ is the microscopic interaction potential between nuclei $i$ and $j$, $\,U_{ij}$ is the optical potential between nuclei $i$ and $j$; $\,{\rm {\bf r}}_{ij}$ is the radius-vector connecting the center of mass of particles $i$ and $j$. I remind that the exact wave function $\Psi_{i}^{(+)}$ is fully antisymmetrized but the channel wave function $\Phi_{f}^{(-)}$ is not antisymmetrized with respect to exchange of the exiting proton and nucleons in $F$. However, the internal wave function of $F$ $\,\varphi_{F}$ in $\Phi_{f}^{(-)}$ is fully antisymmetrized. The reason why we can drop the antisymmetrization in the channel wave function is the presence of the fully antisymmetrized exact wave function in the initial state and fully symmetric transition operator what can be seen below when the transition operator is expressed in terms of the kinetic energy operators.  

To obtain the post form of the DWBA from Eq. (\ref{postreactampl1}) we replace $\Psi_{i}^{(+)}$ by the channel wave function $\Phi_{i}^{(+)}= \varphi_{d}\,\varphi_{A}\,\chi_{dA}^{(+)}$ in the initial $d + A$ state: 
\begin{equation}
 {\tilde M}^{(post)}({\rm {\bf k}}_{pF},\,{\rm {\bf k}}_{dA})= <\Phi_{f}^{(-)}|\,\Delta\,V_{pF}|\Phi_{i}^{(+)}>.
\label{postdw1}
\end{equation}
Then we use approximation 
\begin{equation}
\varphi_{F} \approx I_{A}^{F}\,\varphi_{A}, 
\label{overlapfunct1}
\end{equation}
where 
$\,I_{A}^{F}({\rm {\bf r}}_{nA})$  is the overlap function of the bound state wave functions of nuclei $F$ and $A$:
\begin{equation}
I_{A}^{F}({\rm {\bf r}}_{nA})= \big< \varphi_{A}|\,\varphi_{F} \big>.
\label{overlapfunction1}
\end{equation}
Note that the integration in Eq. (\ref{overlapfunction1}) is taken over all the internal coordinates of nucleus $A$. 
Then the transition operator in Eq. (\ref{postdw1}) takes the form 
$<\varphi_{A}|\Delta\,V_{pF}|\varphi_{A}>\,=\, <\varphi_{A}|V_{pA}|\varphi_{A}> 
+ V_{pn} - U_{pF}$. Potential $<\varphi_{A}|V_{pA}|\varphi_{A}> $ is replaced by the optical potential $U_{pA}$ and we obtain a standard post form of the DWBA amplitude:
\begin{equation}
 M^{DW(post)}({\rm {\bf k}}_{pF},\,{\rm {\bf k}}_{dA})= <\Phi_{f}^{(-)}|\,\Delta{\overline V}_{pF}|\Phi_{i}^{(+)}>,
\label{standpostdw1}
\end{equation}
where $\Delta\,{\overline V}_{pF}= U_{pA} + V_{pn} - U_{pF}$. 
Now we will transform this volume integral into the surface one. First, we adopt ${\rm {\bf r}}_{nA}$ and ${\rm {\bf r}}_{pF}$ as Jacobian variables and split the configuration space over  ${\rm {\bf r}}_{nA}$ into the internal and external regions, while the integral over the second Jacobian variable, ${\rm {\bf r}}_{pF}$, is taken over all the coordinate space. Splitting the reaction amplitude into internal and external amplitudes we get
\begin{align}
&M^{DW(post)}({\rm {\bf k}}_{pF},\,{\rm {\bf k}}_{dA}) = M_{int}^{DW(post)}({\rm {\bf k}}_{pF},\,{\rm {\bf k}}_{dA}) \nonumber\\
&+ M_{ext}^{DW(post)}({\rm {\bf k}}_{pF},\,{\rm {\bf k}}_{dA}), 
\label{intextdwpost1}
\end{align}
where the internal amplitude $M_{int}^{DW(post)}$ is given by
\begin{align}
&M_{int}^{DW(post)}({\rm {\bf k}}_{pF},\,{\rm {\bf k}}_{dA})                      \nonumber\\
&= <\chi_{pF}^{(-)}\,I_{A}^{F}\,|\,\Delta\,{\overline V}_{pF}| \varphi_{d}\,\chi_{dA}^{(+)}>\Big |_{r_{nA} \leq R_{nA}}. 
\label{intdwpost1}
\end{align}
Correspondingly, the external amplitude is given by
\begin{align}
&M_{ext}^{DW(post)}({\rm {\bf k}}_{pF},\,{\rm {\bf k}}_{dA})                                         \nonumber\\
&= <\chi_{pF}^{(-)}\,I_{A}^{F}\,|\,\Delta\,{\overline V}_{pF}|\varphi_{d}\,\chi_{dA}^{(+)}>\Big |_{r_{nA} > R_{nA}}.
\label{extdwpost1}
\end{align}
Here, $R_{nA}$ is the channel radius similar to the one introduced in the $R$-matrix approach, which separates the internal and external regions.

The splitting of the amplitude into the internal a
nd external parts in the subspace over the Jacobian variable ${\rm {\bf r}}_{nA}$ is natural and evident. The overlap function $I_{A}^{F}({\rm {\bf r}}_{nA})$ is the only object in the reaction amplitude which provides spectroscopic and structure information. In the external region the overlap function has a standard radial shape given by the spherical Hankel function (for neutrons) with the amplitude called the ANC (see below). To determine the behavior of the overlap function in the nuclear interior, which bring one of the main uncertainties in the analysis of the deuteron stripping, microscopic calculations are required \cite{brida}. In a standard approach the internal part of the overlap function is approximated by the single-particle bound state wave function calculated in the adopted mean field. The proportionality coefficient is the square root of the spectroscopic factor.  Due to the structure of the transition operator the external matrix element $M_{ext}^{DW(post)}$ in the post form is dominant compared to a small contribution coming from the internal part $M_{int}^{DW(post)}$. This simple observation stems from the following. 

In the internal matrix element, $r_{nA} \leq R_{nA}$, due absorption of the protons inside nucleus $F$, effective $r_{pn} \sim r_{pA}\approx r_{pF}> R_{F}$, where $R_{F}$ is the radius of nucleus $F$. For the protons outside of $F$ and neutrons inside or on the surface of $A$ each nuclear interaction in the operator $\Delta\,{\overline V}_{pF} =\,U_{pA} + V_{pn} - U_{pF}$ is small. 
Potential $U_{pF}$ is arbitrary and often $U_{pF}$ is chosen to compensate for $U_{pA}$  so that the transition operator reduces to $V_{pn}$. Since the DWBA is the first order perturbation theory, the minimization of the whole transition operator $\Delta\,{\overline V}_{pF}$ provides smaller higher order terms and, hence, better serves the theory. This choice is more preferable in the formalism presented here and we adopt $U_{pF}$, which minimizes $\Delta\,{\overline V}_{pF} = U_{pA} + V_{pn} - U_{pF}$ at $r_{nA} \leq R_{nA}$ making the contribution from the internal matrix element small compared to the external one.

In the external matrix element ($r_{nA} > R_{nA}$), which is dominant, the overlap function  $I_{A}^{F}$ can be replaced by its asymptotic tail. Although $M_{ext}^{DW(post)}$ can be easily calculated for stripping to the bound state, here we transform this matrix element into an alternative form, which has clear advantage in case of stripping to resonance states discussed below where convergence becomes a main impediment. 

Now we proceed to the transformation of the volume integral defining the external matrix element 
in terms of the dominant surface integral encircling the sphere at $r_{nA}=R_{nA}$ and a small, due to the structure of the transition operator in the prior form (see Eq. ({\ref{DeltaVdAext1})), external volume integral in the prior form.  
Note that the transformation is exact within the DWBA formalism.

To transform the external volume integral to the surface one, we rewrite the transition operator as 
\begin{align}
\Delta\,{\overline V}_{pF} = U_{pA} + V_{pn} -  U_{pF}=  [V_{pn} + U_{dA}] - [U_{pF}]  \nonumber\\
+ (U_{pA} - U_{dA}).
\label{deltavpfext1}
\end{align}
The bracketed operators are the right-hand-side operators in the Schr\"odinger equations for the initial and final channel wave functions in the external region:
\begin{equation}
(E- T)\,\varphi_{d}\,\chi_{dA}^{(+)} = (V_{pn} + U_{dA})\,\varphi_{d}\,\chi_{dA}^{(+)}
\label{shreqphi1}
\end{equation}
and 
\begin{equation}
(E - T)\,I_{A}^{F}\,\chi_{pF}^{(-)*}=  U_{pF}\,I_{A}^{F}\,\chi_{pF}^{(-)*}.
\label{shreqpphif1}
\end{equation}
To derive Eq. (\ref{shreqpphif1}) we took into account that at $r_{nA} > R_{nA}$ $\,\,I_{A}^{F}$ satisfies the asymptotic Schr\"odinger equation $(\varepsilon_{n\,A} - T_{nA})\,I_{A}^{F}=0$,  where $\varepsilon_{ij}$ is the binding energy of the bound state 
$(i\,j)$ and $T_{i\,j}$ is the kinetic energy operator of the relative motion of $i$ and $j$.   
These equations imply the following connection between the external post form DWBA amplitude and the matrix element 
$M_{S}^{DW}$ containing the surface integral:
\begin{align}
M_{ext}^{DW(post)}({\rm {\bf k}}_{pF},\,{\rm {\bf k}}_{dA}) = M_{S}^{DW}({\rm {\bf k}}_{pF},\,{\rm {\bf k}}_{dA}) \nonumber\\
+ M_{ext}^{DW(prior)}({\rm {\bf k}}_{pF},\,{\rm {\bf k}}_{dA}),
\label{dwpostexttildeprior1}
\end{align}
where
\begin{align}
&M_{ext}^{DW(prior)}({\rm {\bf k}}_{pF},\,{\rm {\bf k}}_{dA})         \nonumber\\
&=  <\chi_{pF}^{(-)}\,I_{A}^{F}|\,\Delta\,{\overline V}_{dA} |\varphi_{d}\,\chi_{dA}^{(+)}>\Big |_{r_{nA} > R_{nA}}
\label{dwpriorext1}
\end{align}
and
\begin{align}
&M_{S}^{DW}({\rm {\bf k}}_{pF},\,{\rm {\bf k}}_{dA})                \nonumber\\
&= <\chi_{pF}^{(-)}\,I_{A}^{F}|\,{\overleftarrow T} - {\overrightarrow T}|\varphi_{d}\,\chi_{dA}^{(+)}>\Big |_{r_{nA} > R_{nA}}.
\label{SDW1}
\end{align}
Here, the transition operator in the prior form $\Delta\,{\overline V}_{dA}$ in the external region, where the nuclear $n-A$ interaction disappears, takes the form
\begin{align}
\Delta\,{\overline V}_{dA} = U_{pA} - U_{dA}.
\label{DeltaVdAext1}
\end{align}

The overlap function is given by
\begin{align}
I_{A}^{F}({\rm {\bf r}}_{nA}) &= \sum\limits_{j_{nA}\,m_{j_{nA}\,m_{l_{nA}}}}\,<J_{A}\, M_{A}\,\, j_{nA}\, m_{j_{nA}}| J_{F}\,M_{F}> \nonumber\\
&\times <J_{n}\,M_{n}\,\,l_{nA}\,m_{l_{nA}}|j_{nA}\,m_{j_{nA}}> \nonumber\\
&\times Y_{l_{nA}\,m_{l_{nA}}}({\rm {\bf {\hat r}}}_{nA})\,I_{A\,j_{nA}\,l_{nA}}(r_{nA}).
\label{overlapfunction2}
\end{align}
Here, $<j_{1}\,m_{1}\,\,j_{2}\,m_{2}|j_{3}\,m_{3}>$ is the Clebsch-Gordan coefficient, $\,l_{nA}$ ($m_{l_{nA}}$) is the orbital angular momentum (its projection) of the relative motion of $n$ and $A$, $j_{nA}$ ($m_{j_{nA}}$) is the total angular momentum (its projection) of $n$ in the bound state $F=(nA)$, $J_{i}\,(M_{i})$ is the spin (its projection) of nucleus $i$; $\,I_{A\,\,l_{nA}\,j_{nA}}^{F}(r_{nA})$ is the radial overlap function, which is a real function \cite{blokhintsev}, $ Y_{l\,m}({\rm {\bf {\hat r}}})$ is the spherical harmonics and ${\rm {\bf {\hat r}}}= {\rm {\bf r}}/r$ is the unit vector. We assume that only one value of  $\,l_{nA}\,$ contributes to expansion (\ref{overlapfunction2}). If the channel radius is taken larger than the range of the nuclear interaction, the radial overlap function can be replaced by its asymptotic term, 
\begin{align}
I_{A\,\,j_{nA}\,l_{nA}}^{F}(R_{nA}) \stackrel{r_{nA} > R_{nA}}{\approx} C^{F}_{A\,\,j_{nA}\, l_{nA}}\,i^{l_{nA}+1} \nonumber\\
\times \,\kappa_{nA}\,h_{l_{nA}}^{(1)}(i\,\kappa_{nA}\,r_{nA}),
\label{overlapasympt1}
\end{align}
where $h_{l_{nA}}^{(1)}(i\,\kappa_{nA}\,r_{nA})$ is the spherical Hankel function of the first order, 
$C^{F}_{A\,\,j_{nA}\, l_{nA}}$ is the ANC of the overlap function, 
$\kappa_{nA}= \sqrt{2\,\mu_{nA}\,\varepsilon_{nA}}\,$  is the bound state wave number.

It is also useful to introduce the reduced-width amplitude used in the $R$-matrix approach, which can be expressed in terms of the ANC \cite{muktr99}:
\begin{align}
&\gamma_{nA\,j_{nA}\,l_{nA}} =  \,\sqrt{ \frac{R_{nA} }{ 2\,\mu _{nA} }}\,I_{A\,\,j_{nA}\, l_{nA}}^{F}(R_{nA})    \nonumber\\
&= \,\sqrt{ \frac{R_{nA} }{ 2\,\mu _{nA} }}\,i^{l_{nA} + 1}\,\kappa _{nA}\,C_{A\,j_{nA}\,l_{nA} }^F\,h_{l_{nA}}^{(1)}(i\,\kappa _{nA}\,R_{nA}).
\label{gammaredwamplnA1}
\end{align}
Correspondingly, the reduced width is
\begin{align}
&\gamma_{nA\,j_{nA}\,l_{nA}}^2 =  \,\frac{{R_{nA}}}{{2\,{\mu _{nA}}}}\,[I_{A\,\,j_{nA}\, l_{nA}}^{F}(R_{nA})]^{2}       \nonumber\\
&= \,\frac{{R_{nA}}}{{2\,{\mu _{nA}}}}{( - 1)^{{l_{nA}} + 1}}\kappa _{nA}^2\,{[C_{A{\kern 1pt} {\kern 1pt} {j_{nA}}\,{l_{nA}}\,}^F\,h_{{l_{nA}}}^{(1)}(i{\kappa _{nA}}{R_{nA}})]^2}.
\label{gammanA1}
\end{align}
It is worth mentioning that, due to the presence of the channel radius $R_{nA}$, the reduced width, in contrast to the ANC, is model-dependent. 
The dependence on the channel radius becomes crucial with increase of the binding energy. We are going to use also the boundary condition, which is the logarithmic derivative of the overlap function at $r_{nA}= R_{nA}$:
\begin{align}
{B_{nA}} = \frac{1}{{h_{{l_{nA}}}^{(1)}(i{\kappa _{nA}}{R_{nA}})}}\,\frac{{d[{r_{nA}}h_{{l_{nA}}}^{(1)}(i{\kappa _{nA}}{r_{nA}})]}}{{dr}}{\Big |_{{r_{nA}} = {R_{nA}}}}.
\label{boundarycond1}
\end{align}

Due to Eq. (\ref{overlapasympt1}), the amplitude $M_{ext}^{DW(prior)}$ can be parametrized in terms of the ANC. We note that this amplitude is also small. In the external region, $r_{nA} > R_{nA}$, the nuclear $n-A$ interaction can be neglected. Besides in this region the overlap function exponentially fades away. Also, if the proton absorption is strong in the internal region of $A$, the dominant contribution comes from $r_{pA} > R_{A}$, where $R_{A}$ is the radius of nucleus $A$. If the adopted radius channel $R_{nA}$ is larger than the $n-A$ nuclear interaction radius we can neglect $n-A$ nuclear interaction in the external region.
In this region each nuclear potential $U_{pA}^{N}$ and $U_{dA}^{N}$ and their difference $U_{pA} - U_{dA}$ are small. The Coulomb part $U_{pA}^{C} - U_{dA}^{C} \approx Z_{A}\,e^{2}\,R_{d}/(2\,R_{A}^{2})$, where $R_{d}$ is the deuteron size and $Z_{A}\,e$ is the charge of nucleus $A$, is also too small compared to the nuclear potential. Thus the dominant contribution to the  post DWBA amplitude $M_{ext}^{DW(post)}$, Eq. (\ref{dwpostexttildeprior1}), and, hence, to the total post form DWBA amplitude $M^{DW(post)}$ comes from the surface integral $M_{S}^{DW}$. Here and in what follows all the amplitudes with the transition operator ${\overleftarrow T} - {\overrightarrow T}$ are assigned the subscript $S$, which is abbreviation of "surface", because the volume matrix elements of these amplitudes can be transformed into the surface ones in the subspace over variable ${\rm {\bf r}}_{nA}$
while over the second Jacobian variable ${\rm {\bf r}}_{pF}$ we always keep the volume integral. 

Now we express  
$M_{S}^{DW}$ in terms of the surface integral over variable ${\rm {\bf r}}_{nA}$ and the same technique will be used throughout the paper.  The kinetic energy operator can be written as $T = T_{pF} + T_{nA}$. $T_{pF}$ is a Hermitian operator in the subspace spanned by the bra and ket states in Eq. (\ref{SDW1}). It can be proved if we take into account that at $r_{pF} \to \infty$ the integrand in this equation vanishes exponentially due to the presence of the bound state wave function $\varphi_{d}({\rm {\bf r}}_{pn})$ and the overlap function $I_{A}^{F}({\rm {\bf r}}_{nA})$. Hence, integrating by parts twice the integral over ${\rm {\bf r}}_{pF}$ we obtain
\begin{align}
& <\chi_{pF}^{(-)}\,I_{A}^{F}|\,{\overleftarrow T}_{pF} - {\overrightarrow T}_{pF}|\,\varphi_{d}\,\chi_{dA}^{(+)}>\Big |_{r_{nA} > R_{nA}}  \nonumber\\
&=   <\chi_{pF}^{(-)}\,I_{A}^{F}|\,{\overrightarrow T}_{pF} - {\overrightarrow T}_{pF}|\,\varphi_{d}\,\chi_{dA}^{(+)}>\Big |_{r_{nA} > R_{nA}} = 0.
\label{SDW11}
\end{align}
Then $M_{S}^{DW}$ reduces to 
\begin{align}
&M_{S}^{DW}({\rm {\bf k}}_{pF},\,{\rm {\bf k}}_{dA})       \nonumber\\
&= <\chi_{pF}^{(-)}\,I_{A}^{F}|\,{\overleftarrow T}_{nA} - {\overrightarrow T}_{nA}\,|\,\varphi_{d}\,\chi_{dA}^{(+)}>\Big |_{r_{nA} > R_{nA}}.
\label{SDWnA1}
\end{align}

We apply now Green's theorem to transform the volume integral into the surface one, which encircles the inner volume over the coordinate ${\rm {\bf r}}$: 
\begin{align}
&\int\limits_{r \leq R}\,{\rm d}\,{\rm {\bf r}}\,f({\rm {\bf r}})\left [ {\overleftarrow T} - {\overrightarrow T} \right ]\,g({\rm {\bf r}}) \nonumber\\
&= - \frac{1}{2\,\mu}\, \,\oint\limits_{r = R} {\rm d}{\rm {\bf S}}\, 
\left[ g({\rm {\bf r}})\,{\rm {\bf \nabla}}_{{\rm {\bf r}}}\, f({\rm {\bf r}})\,  
-  f({\rm {\bf r}})\,{\rm {\bf \nabla}}_{{\rm {\bf r}}}\,g({\rm {\bf r}}) \right ] \nonumber\\
&= - \frac{1}{2\,\mu}\, \,R^{2}\,\int {\rm d}{\Omega}_{{\rm {\bf r}}}\, 
\left[ g({\rm {\bf r}})\,\frac{\partial f({\rm {\bf r}})}{{\partial r}}\,  
-  f({\rm {\bf r}})\,\frac{\partial g({\rm {\bf r}})}{{\partial r}} \right ]_{r=R}. 
\label{greentheorem1}
\end{align}
Here, ${\rm d}\,{\rm {\bf S}}= R^{2}\,{\rm d}\,{\Omega}\,{\rm {\bf {\hat r}}}$, where $\Omega$ is the solid angle. 
Note that the unit vector ${\rm {\bf {\hat r}}}$ is the normal vector to the sphere directed outside of the restricted by the surface volume. 
The integration in Eq. (\ref{SDWnA1}) over ${\rm {\bf r}}_{nA}$ is taken over the external volume restricted by two spherical surfaces: the inner surface with the radius $R_{nA}$ and the external surface with the radius $R_{nA}^{'} \to \infty$, that is 
\begin{align}
M_{S}^{DW}({\rm {\bf k}}_{pF},\,{\rm {\bf k}}_{dA}) = -M_{S_{R_{nA}}}^{DW}({\rm {\bf k}}_{pF},\,{\rm {\bf k}}_{dA}) \nonumber\\
+ M_{S_{\infty}}^{DW}({\rm {\bf k}}_{pF},\,{\rm {\bf k}}_{dA}).
\label{SSDWnA1}
\end{align}

The first term in this equation is the surface integral encircling the inner surface of the external volume at $r_{nA} = R_{nA}$ while the second term is the surface integral taken at $r_{nA}= R_{nA}^{'} \to \infty$. A negative sign in front of the first term
appears because the normal to the surface is directed inward to the center of the volume, i.e. opposite to the normal to the external surface (at infinitely large radius). The second term vanishes because of the presence of the overlap function $I_{A}^{F}$, which decreases exponentially at $r_{nA} \to \infty$. 
Then for $M_{S}^{DW}$ we get 
\begin{widetext}
\begin{align}
&M_{S}^{DW}({\rm {\bf k}}_{pF},\,{\rm {\bf k}}_{dA}) = -M_{S_{R_{nA}}}^{DW}({\rm {\bf k}}_{pF},\,{\rm {\bf k}}_{dA}) \nonumber\\
&= \frac{1}{2\,\mu_{nA}}\,R_{nA}^{2}\,\int\,{\rm d}\,{\rm {\bf r}}_{pF}\,\chi_{- {\rm {\bf k}}_{pF} }^{(+)}({\rm {\bf r}}_{pF})\,\int\,{\rm d}\,\Omega_{{\rm {\bf r}}_{nA}}\,
\Big[\varphi_{d}(r_{pn})\,\chi_{ {\rm {\bf k}}_{dA} }^{(+)}({\rm {\bf r}}_{dA})\,\frac{ \partial\,[I_{A}^{F}({\rm {\bf r}}_{nA})]^{*}}{ \partial\,r_{nA} }  
-\,[I_{A}^{F}({\rm {\bf r}}_{nA})]^{*}\,\frac{ \partial\,\varphi_{d}(r_{pn})\,\chi_{{\rm {\bf k}}_{dA}}^{(+)})({\rm {\bf r}}_{dA})}{\partial
\,r_{nA}} \Big]\Big |_{r_{nA}= R_{nA}}. 
\label{SdwdtnA12}
\end{align} 
\end{widetext}
Here we took into account that $\chi_{{\rm {\bf k}}}^{(-)*}({\rm{\bf r}})= \chi_{-{\rm{\bf k}}}^{(+)}({\rm {\bf r}})$.
Invoking Eqs. (\ref{overlapfunction2}) and (\ref{overlapasympt1}) we can rewrite $M_{S}^{DW}$ in the form explicitly showing parametrization in terms of the reduced width amplitude (ANC) and boundray condition, the quantities used in the $R$-matrix approach:
\begin{widetext}
\begin{align}
&M_{S}^{DW}({\rm {\bf k}}_{pF},\,{\rm {\bf k}}_{dA}) 
= \frac{1}{2\,\mu_{nA}}\,i^{l_{nA}+\,1}\,\kappa_{nA}\,R_{nA}\,h_{l_{nA}}^{(1)}(i\,\kappa_{nA}\,R_{nA})\,\sum\limits_{j_{nA}\,m_{j_{nA}}\,m_{l_{nA}}\,M_{n}}\,<J_{A}\, M_{A}\,\, j_{nA}\, m_{j_{nA}}| J_{F}\,M_{F}>  \nonumber\\
& \times \,<J_{n}\,M_{n}\,l_{nA}\,m_{l_{nA}}|j_{nA}\,m_{j_{nA}}>\, 
<J_{p}\,M_{p}\,\,J_{n}\,M_{n}| J_{d}\,M_{d}>\,C^{F}_{A\,\,j_{nA}\,l_{nA}}\, \nonumber\\
&\times \,\int\,{\rm d}\,{\rm {\bf r}}_{pF}\,\chi_{- {\rm {\bf k}}_{pF} }^{(+)}({\rm {\bf r}}_{pF})\,\int\,{\rm d}\,\Omega_{{\rm {\bf r}}_{nA}}\,Y_{l_{nA}\,m_{l_{nA}}}^{*}({\rm {\bf {\hat r}}}_{nA})\,\left[\varphi_{d}(r_{pn})\,\chi_{ {\rm {\bf k}}_{dA} }^{(+)}({\rm {\bf r}}_{dA})\,(B_{nA} -\,1) 
- \,R_{nA}\,\,
\frac{ \partial\,\varphi_{d}(r_{pn})\,\chi_{{\rm {\bf k}}_{dA}}^{(+)})({\rm {\bf r}}_{dA})}{\partial
\,r_{nA}} \right]\Big |_{r_{nA}= R_{nA}}
\label{Sgammadw1}  \\
&=\,\sqrt{\frac{R_{nA}}{2\,\mu_{nA}}}\sum\limits_{j_{nA}\,m_{j_{nA}}\,m_{l_{nA}}\,M_{n}}\,<J_{A}\, M_{A}\,\, j_{nA}\, m_{j_{nA}}| J_{F}\,M_{F}> \,<J_{n}\,M_{n}\,\,l_{nA}\,m_{l_{nA}}|j_{nA}\,m_{j_{nA}}> \nonumber\\
&\times\,<J_{p}\,M_{p}\,\,J_{n}\,M_{n}| J_{d}\,M_{d}>\,\gamma_{nA\,j_{nA}\,l_{nA}}\,\int\,{\rm d}\,{\rm {\bf r}}_{pF}\,\chi _{{-{\rm {\bf k}}_{pF}}}^{(+)}({\rm {\bf r}}_{pF})\,\int\,{\rm d}\,\Omega_{{\rm {\bf r}}_{nA}}\,Y_{l_{nA}\,m_{l_{nA}}}^{*}({\rm {\bf {\hat r}}}_{nA})  \nonumber\\
& \times\,\Big[\varphi_{d}(r_{pn})\,\chi_{ {\rm {\bf k}}_{dA} }^{(+)}({\rm {\bf r}}_{dA})\,(B_{nA} - \,1) - \,R_{nA}\,
\frac{ \partial\,\varphi_{d}(r_{pn})\,\chi_{{\rm {\bf k}}_{dA}}^{(+)})({\rm {\bf r}}_{dA})}{\partial
\,r_{nA}}\Big]\Big |_{r_{nA}= R_{nA}}.
\label{Sgammadw2}
\end{align} 
\end{widetext}

Finally, the total post form DWBA amplitude is given by
\begin{align}
M^{DW(post)}({\rm {\bf k}}_{pF},\,{\rm {\bf k}}_{dA}) = M_{int}^{DW(post)}({\rm {\bf k}}_{pF},\,{\rm {\bf k}}_{dA}) \nonumber\\
+ M_{ext}^{DW(prior)}({\rm {\bf k}}_{pF},\,{\rm {\bf k}}_{dA}) +  M_{S}^{DW}({\rm {\bf k}}_{pF},\,{\rm {\bf k}}_{dA}). 
\label{finaldwpost1}
\end{align}
Taking into account that $M_{S}^{DW} = M_{ext}^{DW(post)} - M_{ext}^{DW(prior)}$ we can rewrite Eq. (\ref{finaldwpost1}) in a different form:
\begin{align}
&M^{DW(post)}({\rm {\bf k}}_{pF},\,{\rm {\bf k}}_{dA}) = M_{int}^{DW(post)}({\rm {\bf k}}_{pF},\,{\rm {\bf k}}_{dA}) \nonumber\\
&+ M_{ext}^{DW(prior)}({\rm {\bf k}}_{pF},\,{\rm {\bf k}}_{dA})                            \nonumber\\
&+  \Big[M_{ext}^{DW(post)}({\rm {\bf k}}_{pF},\,{\rm {\bf k}}_{dA}) - M_{ext}^{DW(prior)}({\rm {\bf k}}_{pF},\,{\rm {\bf k}}_{dA})\Big]. 
\label{finaldwpost2}
\end{align}

Thus, the main result of this section is that the post form of the DWBA amplitude can be written 
as the sum of the peripheral parts, $M_{ext}^{DW(prior)} +  M_{S}^{DW}$,  and small internal term $M_{int}^{DW(post)}$.
The peripheral part itself consists of the dominant surface amplitude $M_{S}^{DW}$ and small external prior form
$M_{ext}^{DW(prior)}$. The peripheral part is parametrized in terms of the ANC (reduced width amplitude), channel radius $R_{nA}$ and the logarithmic boundary condition, that is in terms of the parameters used in the $R$-matrix fitting.  
The model dependence of these two peripheral amplitudes is caused by the ambiguity of the optical potentials and channel radius $R_{nA}$. The strongest model dependence comes from $M_{int}^{DW(post)}$, because, in addition to the ambiguity of the optical potentials, to calculate it one needs to know the behavior of the overlap function in the internal region.
For peripheral reactions contribution of $M_{int}^{DW(post)}$ can be neglected.

\subsection{Prior form of  DWBA. Stripping to bound state.}
\label{priorDWBA}
In subsection A the post form of the DWBA amplitude has been considered. However, all the results hold also for the prior form 
\begin{align}
&M^{DW(prior)}({\rm {\bf k}}_{pF},\,{\rm {\bf k}}_{dA}) = \, <\chi_{pF}^{(-)}\,I_{A}^{F}|\,\Delta\,{\overline V}_{dA}|\,\varphi_{d}\,\chi_{dA}^{(+)}> \nonumber\\
& = M_{int}^{DW(prior)}({\rm {\bf k}}_{pF},\,{\rm {\bf k}}_{dA}) + M_{ext}^{DW(prior)}({\rm {\bf k}}_{pF},\,{\rm {\bf k}}_{dA})
\label{dwprior1}
\end{align}
where
\begin{align}
&M_{int}^{DW(prior)}({\rm {\bf k}}_{pF},\,{\rm {\bf k}}_{dA})                    \nonumber\\
&= <\chi_{pF}^{(-)}\,I_{A}^{F}\,|\,\Delta\,{\overline V}_{dA}| 
\,\varphi_{d}\,\chi_{dA}^{(+)}>\Big |_{r_{nA} \leq R_{nA}}. 
\label{intdwprior1}
\end{align}
and
\begin{align}
&M_{ext}^{DW(prior)}({\rm {\bf k}}_{pF},\,{\rm {\bf k}}_{dA})                    \nonumber\\
&= <\chi_{pF}^{(-)}\,I_{A}^{F}\,|\,\Delta\,{\overline V}_{dA}|\,\varphi_{d}\,\chi_{dA}^{(+)}>\Big |_{r_{nA} > R_{nA}}.
\label{extdwprior1}
\end{align}
with the transition operator 
\begin{align}
\Delta\,{\overline V}_{dA} = U_{pA} + {\overline V}_{nA} - U_{dA}.
\label{DeltaVdA1}
\end{align}
The $n-A$ interaction potential ${\overline V}_{nA} = <\varphi_{A}|V_{nA}|\varphi_{A}> $ is the mean field real potential supporting the bound state $(n\,A)$. 
The splitting of the amplitude into the internal and external terms in the subspace over the coordinate ${\rm {\bf r}}_{nA}$ helps us to further transform the prior DWBA amplitude. Due to the structure of the transition operator the external matrix element $M_{ext}^{DW(prior)}$ in the prior form is small (see the discussion in 
subsection \ref{postDWBA}) and the main contribution in the prior form comes from the internal part $M_{int}^{DW(prior)}$. Since the internal part is given by the volume integral, its calculation requires the knowledge of the overlap function in the internal region. The model dependence of the overlap function in the nuclear interior ($r_{nA} \leq R_{nA}$) brings one of the main problems and main uncertainty in the calculation of the internal matrix element. However, using the surface integral we can redistribute the internal contribution in terms of dominant the surface term (over variable ${\rm {\bf r}}_{nA}$) plus  small internal part written in terms of the volume integral in the post form.
 With reasonable choice of the channel radius $R_{nA}$ the contribution from the internal volume integral in the post form can be significantly decreased compared the surface matrix element. The latter can be expressed in terms of the $R$-matrix parameters - the observable reduced width amplitude (ANC), boundary condition and channel radius.   
To transform $M_{int}^{DW(prior)}$ into the surface integral in the subspace over variable ${\rm {\bf r}}_{nA}$ we rewrite the transition operator in the internal region as
\begin{align}
&\Delta\,V_{dA} = U_{pA} + {\overline V}_{nA} - U_{dA}                 \nonumber\\
& = [{\overline V}_{nA} + U_{pF}] + (U_{pA} + V_{pn} - U_{pF}) - [V_{pn} + U_{dA}]. 
\label{DeltaVdAint1}
\end{align}
The bracketed transition operators are the potential operators in the Schr\"odinger equations for the initial and final channel wave functions. Hence, for the internal prior form of the DWBA we obtain 
\begin{align}
&M_{int}^{DW(prior)}({\rm {\bf k}}_{pF},\,{\rm {\bf k}}_{dA})             \nonumber\\
&= M_{int}^{DW(post)}({\rm {\bf k}}_{pF},\,{\rm {\bf k}}_{dA})
+  M_{S}^{DW(prior)}({\rm {\bf k}}_{pF},\,{\rm {\bf k}}_{dA}),
\label{dwpriorint1}
\end{align}
where 
\begin{align}
&M_{S}^{DW}({\rm {\bf k}}_{pF},\,{\rm {\bf k}}_{dA}) = - <\chi_{pF}^{(-)}\,I_{A}^{F}|\,{\overleftarrow T} - {\overrightarrow T}|\,\varphi_{d}\,\chi_{dA}^{(+)}>    \nonumber\\
&= - <\chi_{pF}^{(-)}\,I_{A}^{F}|\,{\overleftarrow T}_{nA} - {\overrightarrow T}_{nA}|\,\varphi_{d}\,\chi_{dA}^{(+)}> \nonumber\\
&= - M_{S_{R_{nA}}}^{DW}({\rm {\bf k}}_{pF},\,{\rm {\bf k}}_{dA}).
\label{Sintdwprior1}
\end{align}
Note that here $M_{S_{R_{nA}}}^{DW}$ is the surface integral encircling the border of the internal volume at $r_{nA}=R_{nA}$ with the normal directed outward. 
Thus we have demonstrated, what should be expected from the very beginning, that $M^{DW(prior)} = M^{DW(post)}$.
Hence all the equations obtained in the previous subsection \ref{postDWBA} are also valid in the prior formalism.

It is worth mentioning that in the post formalism, in contrast to the prior one, we have obtained two surface integrals (in the subspace over ${\rm {\bf r}}_{nA}$) with the radii $r_{nA}= R_{nA}$ and $r_{nA}= R_{nA}^{'} \to \infty$ and then proved that the second integral is zero. From the equality of the post and prior DWBA amplitudes we could conclude that the surface matrix element over infinitely large sphere $r_{nA}= R_{nA}^{'} \to \infty$, which appears only in the post formalism, vanishes. 

There is another interesting point to discuss which explains the advantage of the above outlined formulation of the stripping. As we have discussed, due to different structure of the transition operators in the post and prior forms, 
the main contribution to the post (prior) form comes from the external (internal) part (in the subspace over variable ${\rm {\bf r}}_{nA}$). Since both forms give identical amplitudes, that is, describe the same reaction mechanism and the same physics,
such redistribution of the main contribution is possible only if the main contribution to each form comes from the border between external and internal parts. In the post (prior) form this border attributed to the external (internal) form  and can be expressed in term of the surface integral.
Let us rewrite equality $M^{DW(prior)} = M^{DW(post)}$ in the following form:
\begin{align}
&M_{int}^{DW(prior)}({\rm {\bf k}}_{pF},\,{\rm {\bf k}}_{dA}) + M_{ext}^{DW(prior)}({\rm {\bf k}}_{pF},\,{\rm {\bf k}}_{dA})                                           \nonumber\\
&= M_{int}^{DW(post)}({\rm {\bf k}}_{pF},\,{\rm {\bf k}}_{dA}) + M_{ext}^{DW(post)}({\rm {\bf k}}_{pF},\,{\rm {\bf k}}_{dA}).
\label{postpriorDWintext1}
\end{align}
In this form the dominant terms are $M_{int}^{DW(prior)}$ and $M_{ext}^{DW(post)}$ while the rest two terms, $ M_{ext}^{DW(prior)}$  and $M_{int}^{DW(post)}$ are smaller. From Eq. (\ref{postpriorDWintext1}) we get
\begin{align}
&M_{ext}^{DW(post)}({\rm {\bf k}}_{pF},\,{\rm {\bf k}}_{dA})  -  M_{ext}^{DW(prior)}({\rm {\bf k}}_{pF},\,{\rm {\bf k}}_{dA})               \nonumber\\
&=M_{int}^{DW(prior)}({\rm {\bf k}}_{pF},\,{\rm {\bf k}}_{dA}) -  M_{int}^{DW(post)}({\rm {\bf k}}_{pF},\,{\rm {\bf k}}_{dA})                                            \nonumber\\                                                               
&= M_{S}^{DW}({\rm {\bf k}}_{pF},\,{\rm {\bf k}}_{dA})= - M_{S_{R_{nA}}}^{DW}({\rm {\bf k}}_{pF},\,{\rm {\bf k}}_{dA}).
\label{postpriorDWintextS1}
\end{align}
Thus the difference between the post and prior external amplitudes (or the prior and post internal ones) is the surface integral in the subspace over ${\rm {\bf r}}_{nA}$. 

There is one more point left to discuss. When deriving the post form of the DWBA amplitude from Eq. (\ref{postdw1}) we used 
approximation $\varphi_{F} \approx I_{A}^{F}\,\varphi_{A}$ neglecting the contribution from the channels $n + A_{n}$, $n>0$, where $A_{n}$ is the excited state of $A$. However, I will show now that the surface integral formulation doesn't require this
approximation. To this end let us split $ {\tilde M}^{(post)}$ into the internal and external parts in the subspace over variable ${\rm {\bf r}}_{nA}$. In the internal part we use a standard DWBA approximation $\varphi_{F} \approx I_{A}^{F}\,\varphi_{A}$ to arrive to the standard internal post DWBA amplitude. In the external part we rewrite the transition operator as 
\begin{align}
&\Delta\,V_{pF} = V_{pA} + V_{pn} - U_{pF}                       \nonumber\\
&=  - \big[V_{A} + U_{pF} \big]  +  \big[V_{pn} + V_{A} + U_{dA} \big]  + \big( V_{pA} - U_{dA} \big).
\label{DeltatildVpF1}
\end{align}
The bracketed operators are the right-hand side operators of the Schr\"odinger equations
\begin{align}
\big( E - T \big)\,\Phi_{i}^{(+)}  = \big( V_{pn} + V_{A} + U_{dA} \big)\,\Phi_{i}^{(+)}
\label{ShreqPhi11}
\end{align}
and
\begin{align}
\big( E - T \big)\,\Phi_{f}^{(-)*} = \big( V_{A} + U_{pF} \big)\,\Phi_{f}^{(-)*}.
\label{ShreqPhif11}
\end{align}

Hence, the external part of ${\tilde M}^{(post)}$ reduces to 
\begin{align}
&{\tilde M}_{ext}^{(post)}({\rm {\bf k}}_{pF},\,{\rm {\bf k}}_{dA})                          \nonumber\\
&= {\tilde M}_{S(ext)}({\rm {\bf k}}_{pF},\,{\rm {\bf k}}_{dA}) + {\tilde M}_{ext}^{(prior)}({\rm {\bf k}}_{pF},\,{\rm {\bf k}}_{dA}),
\label{tildepostSextpr1}
\end{align}
where 
\begin{align}
&{\tilde M}_{ext}^{(prior)}({\rm {\bf k}}_{pF},\,{\rm {\bf k}}_{dA})                                             \nonumber\\
&= <\Phi_{f}^{(-)}|\,V_{pA} - U_{dA}\,|\Phi_{i}^{(+)}>\Big|_{r_{nA} > R_{nA}}
\label{tildeMextprior1}
\end{align}
and
\begin{align}
{\tilde M}_{S(ext)}({\rm {\bf k}}_{pF},\,{\rm {\bf k}}_{dA})= <\Phi_{f}^{(-)}|\,{\overleftarrow T} - {\overrightarrow T}\,|\Phi_{i}^{(+)}>\Big|_{r_{nA} > R_{nA}}.
\label{tildeMS1}
\end{align}
In the matrix element ${\tilde M}_{ext}^{(prior)}$ we can use a standard DWBA approximation $\varphi_{F} \approx I_{A}^{F}\,\varphi_{A}$ which leads to the standard external prior DWBA amplitude. The matrix element ${\tilde M}_{S(ext)}$ can be rewritten as
\begin{align}
&{\tilde M}_{S(ext)}({\rm {\bf k}}_{pF},\,{\rm {\bf k}}_{dA})              \nonumber\\
&= <\Phi_{f}^{(-)}|\,{\overleftarrow T}_{nA} - {\overrightarrow T}_{nA}\,|\Phi_{i}^{(+)}>\Big|_{r_{nA} > R_{nA}} \nonumber\\
&= <\chi_{pF}^{(-)}\,\varphi_{F}|\,{\overleftarrow T}_{nA} - {\overrightarrow T}_{nA}\,|\varphi_{d}\,\varphi_{A}\,\chi_{dA}^{(+)}    >\Big|_{r_{nA} > R_{nA}}                                                                                                    \nonumber\\
&= <\chi_{pF}^{(-)}\,I_{A}^{F}|\,{\overleftarrow T}_{nA} - {\overrightarrow T}_{nA}\,|\varphi_{d}\,\chi_{dA}^{(+)}    >\Big|_{r_{nA} > R_{nA}}                                   \nonumber\\
&= -  M_{ S_{R_{nA}}}^{DW}({\rm {\bf k}}_{pF},\,{\rm {\bf k}}_{dA}),
\label{tildeMS2}
\end{align} 
We took into account that  $<\Phi_{f}^{(-)}|{\overleftarrow T}_{pF} - {\overrightarrow T}_{pF} + {\overleftarrow T}_{A} - {\overrightarrow T}_{A}|\Phi_{i}^{(+)}>\,=\,0$, where $T_{A}$ is the internal motion kinetic energy operator of nucleus $A$, and 
$T_{nA}\,\varphi_{A}\,=\, \varphi_{A}\,T_{nA}\,$. 
Thus ${\tilde M}_{S(ext)}$ can be transformed to the surface integral over variable ${\rm {\bf r}}_{nA}$ encircling the inner volume with the radius $r_{nA}=R_{nA}$ without invoking approximation $\varphi_{F} \approx I_{A}^{F}\,\varphi_{A}$.
It means that, when deriving the post form of the DWBA amplitude, the approximation $\varphi_{F} \approx I_{A}^{F}\,\varphi_{A}$
is required only to obtain two small terms,  $M_{int}^{DW(post)}$ and $M_{ext}^{DW(prior)}$, but not the dominant surface term $- M_{ S_{R_{nA}}}^{DW}$. In this sense the surface integral formalism is an improvement of the DWBA.

\subsection{Deuteron stripping to bound states. Post CDCC formalism}
\label{postCDCC}
In the previous sections we succeeded to parametrize the DWBA amplitude in terms of the ANC
except for a small term, $M_{int}^{DW(post)}$. The most serious shortcoming of the DWBA is that it neglects the coupling to open reaction and breakup channels. This coupling can be taken into account if an exact wave function in the initial or final states is used. However, the exact wave functions are not yet available (if they would be available in the whole configuration space, we don't need to calculate the matrix element because the asymptotic terms of the exact wave functions provide the reaction amplitudes in all the open channels). Here we use the CDCC formalism, which takes into account the elastic $d + A$ and the deuteron breakup channel $p + n +A$ in the initial state.

In this subsection the surface integral formulation  of  the reaction theory will be applied to the post form of the CDCC amplitude for deuteron stripping to bound states. It will allow us to parametrize the stripping amplitude in the CDCC approach in terms of the $R$-matrix parameters - the reduced width amplitude, boundary condition and the channel radius. 
To obtain the CDCC wave function describing the initial state of the stripping reaction, first the exact initial scattering wave function $\Psi_{i}^{(+)}$ is replaced by the three-body wave function $\Psi_{i}^{3B(+)}$, which takes into account the coupling of the initial channel $d + A$ and the deuteron breakup channel $p + n + A$
 \cite{rawitscher,kamimura,austern} and satisfies the Schr\"odinger equation
(in the three-body $p+n+A$ model space)
\begin{equation}
(E - T - U_{pA} - U_{nA} - V_{pn})\,\Psi_{i}^{3B(+)}=0
\label{SchreqthreebodywfT1}
\end{equation}
with the outgoing waves in the elastic channel $d+ A$ and the breakup channel $p + n + A$. 
A general solution of this equation with the $d+ A$ incident wave has outgoing waves in the elastic, breakup and two rearrangement channels, $n +(p\,A)$ and $p + (n\,A)$. To damp rearrangement channels in the asymptotic behavior of the wave function $\Psi_{i}^{3B(+)}$ the optical potentials $U_{pA}$ and $U_{nA}$ with strong imaginary terms can be used \cite{kawai}. 
$\Psi_{i}^{3B(+)}$  is given by 
\begin{align}
\Psi_{i}^{3B(+)}({\rm {\bf r}}_{dA},\,{\rm {\bf r}}_{pn}) = \varphi_{d}(r_{pn})\,\chi_{{\rm {\bf k}}_{dA}}^{(+)}({\rm {\bf r}}_{dA})
\nonumber\\
+ \int{\rm d}\,{\rm {\bf p}}_{pn}\,
\psi_{{\rm {\bf p}}_{pn}}^{(+)}({\rm {\bf r}}_{pn})\,\chi_{{\rm {\bf P}}^{(+)}({\rm {\bf p}}_{pn})
}({\rm {\bf r}}_{dA}).
\label{3Bwfunct1}
\end{align} 
Here, $\varphi_{d}(r_{pn})$ is the deuteron bound state wave function, $\psi_{{\rm {\bf p}}_{pn}}^{(+)}({\rm {\bf r}}_{pn})$ the $p-n$ scattering wave function with the relative momentum ${\rm {\bf p}}_{pn}$, $\,\,\chi_{{\rm {\bf k}}_{dA}}^{(+)}({\rm {\bf r}}_{dA})$ and $\,\chi_{{\rm {\bf P}}({\rm {\bf p}}_{pn})
}^{(+)}({\rm {\bf r}}_{dA})$ are the expansion coefficients, $\,E_{dA} - \varepsilon_{pn}=  P^{2}/(2\,\mu_{dA}) + p_{pn}^{2}/(2\,\mu_{pn})$.  

In practical application the wave function $\Psi_{i}^{3B(+)}$ is replaced by the CDCC  wave function, which is a solution of the projected Schr\"odinger equation 
\begin{align}
(E - T - U_{pA}^{(P_{pn})} - U_{nA}^{(P_{pn})} - V_{pn})\,\Psi_{i}^{CDCC(+)}=0.
\label{PSchreqCDCC11}
\end{align}
Here, $U_{iA}^{(P_{pn})}= {\hat P}_{pn}\,U_{pA}\,{\hat P}_{pn}$, 
and 
\begin{align}
&{\hat P}_{pn}= \sum\limits_{l_{pn}=0}^{l_{pn}^{max}}\,\sum\limits_{\,m_{l_{pn}}=-l_{pn}}^{
l_{pn}}\,\int{\rm d}\,\Omega_{{\rm {\bf r}}_{pn}}\,
Y_{l_{pn}\,m_{l_{pn}}}({\rm {\bf {\hat r}}}_{pn})              \nonumber\\
&\times Y_{l_{pn}\,m_{l_{pn}}}^{*}({\rm {\bf {\hat r}}}_{pn}^{'})
\label{projectorPlpn1}
\end{align}
is the projection operator, which truncates the number of the spherical harmonics $Y_{l_{pn}\,m_{l_{pn}}}({\rm {\bf {\hat r}}}_{pn})$ in the coordinate ${\rm {\bf r}}_{pn}$. Application of this operator to the three-body wave function
suppresses the rearrangement channels in the asymptotic wave function. The CDCC wave function    
is taken in the form
\begin{align}
\Psi_{i}^{CDCC(+)}({\rm {\bf r}}_{pn},\,{\rm {\bf r}}_{dA}) = {\hat P}_{pn}\,\sum\limits_{n=0}^{n_{max}}\,
\psi_{pn}^{(n)}({\rm {\bf r}}_{pn}^{'})\,\chi_{i}^{(n)(+)}({\rm {\bf r}}_{dA}),
\label{cdcc1}
\end{align}
where $\psi_{pn}^{(0)}({\rm {\bf r}}_{pn})= \varphi_{d}({\rm {\bf r}}_{pn})$ is the deuteron bound state wave function, $\psi_{pn}^{(n)}({\rm {\bf r}}_{pn})$, $\,n \geq 1$, is the $n$-th discretized continuum state of the $p-n$ pair obtained by averaging continuous breakup states in the $n$-th bin,
$\chi_{i}^{(n)(+)}({\rm {\bf r}}_{dA})$ are the functions, which describe the relative motion of the center-of-mass of the $p-n$ pair in the $n$-th state and $A$.  Note that $\chi_{i}^{(0)(+)}({\rm {\bf r}}_{dA})$ asymptotically behaves as the incident Coulomb distorted $d-A$ plane wave plus outgoing scattered wave, while $\chi_{i}^{(n)(+)}({\rm {\bf r}}_{dA})$ for $n>0$ asymptotically do not contain any plane wave having only the outgoing scattered wave.  

To derive the post form of the CDCC amplitude from the exact one, first we replace the initial exact scattering wave function 
$\Psi_{i}^{(+)}$ by $\varphi_{A}\,\Psi_{i}^{3B(+)}$. Note that $\Psi_{i}^{3B(+)}$ is the three-body model ($p+n+A$) wave function which treats nucleus $A$ as a constituent particle leaving its internal degrees of freedom intact. That is why the wave function $\Psi_{i}^{(+)}$ is approximated by the product of the bound state wave function $\varphi_{A}$ and $\Psi_{i}^{3B(+)}$.
Correspondingly, the transition operator $\Delta V_{pF}= V_{pA} +V_{pn} - U_{pF}$ is replaced by $\Delta\,{\overline V}_{pF}= U_{pA} + V_{pn} - U_{pF}$. This replacement of the microscopic potential $V_{pA}$ in the exact post form amplitude by $U_{pA}$ 
is evident because the $p-A$ interaction potential in the transition operator should be the same as the one 
in the Schr\"odinger equation for the initial scattering wave function $\Psi_{i}^{3B(+)}$. Potential $V_{pn}$ remains the same
when we approximate the initial exact scattering wave function ny the three-body one.  The final state optical potential $U_{pF}$ is arbitrary and we discuss the optimal choice of this potential later on.  
These approximations lead to the expression for the post form stripping amplitude in the three-body model in the initial state:
\begin{align}
&M^{3B(post)}({\rm {\bf k}}_{pF},\,{\rm {\bf k}}_{dA}) \nonumber\\
&= <\chi_{pF}^{(-)}\,\varphi_{F}\,|\,\Delta\,{\overline V}_{pF}|\,\varphi_{A}\,\Psi_{i}^{3B(+)}>  \nonumber\\
&=  <\chi_{pF}^{(-)}\,I_{A}^{F}\,|\,\Delta\,{\overline V}_{pF}|\,\Psi_{i}^{3B(+)}>      
\label{post3B1}`	
\end{align}
Thus, even if we treat the $d+A$ collision in the initial channel in the three-body approach, the final state 
contains the overlap function, which is essentually many-body object. Eq. (\ref{post3B1}) is impractical to use 
because it requires the knowledge of the three-body wave function $\Psi_{i}^{3B(+)}$, Eq. (\ref{3Bwfunct1}), which contains unknown expansion coefficients $\chi_{{\rm {\bf k}}_{dA}}({\rm {\bf r}}_{dA})$ and $\chi_{{\rm {\bf P}}({\rm {\bf p}}_{pn})}({\rm {\bf r}}_{dA})$. In practical applications the $\Psi_{i}^{3B(+)}$ is approximated by the CDCC wave function $\Psi_{i}^{CDCC(+)}$, which requires the knowledge of the finite number of the expansion coefficients. They can be found from the coupled equations. Correspondingly, the transition operator $\Delta {\overline V}_{pF}= U_{pA} +V_{pn} - U_{pF}$ in Eq. (\ref{post3B1}) is replaced by $\Delta {\overline V}_{pF}^{P_{pn}}= U_{pA}^{P_{pn}} + V_{pn} - U_{pF}$. Note that only the potential $U_{pA}({\rm {\bf r}}_{pA})$, where ${\rm {\bf r}}_{pA}=  {\rm {\bf r}}_{dA} + 1/2\,{\rm {\bf r}}_{pn}$ is affected by the projector ${\hat P}_{pn}$. Then the expression for the post form of the CDCC amplitude takes the form:    
\begin{align}
&M^{CDCC(post)}({\rm {\bf k}}_{pF},\,{\rm {\bf k}}_{dA}) \nonumber\\
&= <\chi_{pF}^{(-)}\,I_{A}^{F}\,|\,\Delta\,{\overline V}_{pF}^{P_{pn}}|\,\Psi_{i}^{CDCC(+)}>.     
\label{postPCDCC1}`	
\end{align}
Now we split $M^{CDCC}$ into the internal and external parts in the subspace ${\rm {\bf r}}_{nA}$: 
\begin{align}
M^{CDCC(post)}({\rm {\bf k}}_{pF},\,{\rm {\bf k}}_{dA}) = M_{int}^{CDCC(post)}({\rm {\bf k}}_{pF},\,{\rm {\bf k}}_{dA}) \nonumber\\
+ M_{ext}^{CDCC(post)}({\rm {\bf k}}_{pF},\,{\rm {\bf k}}_{dA}). 
\label{postCDCC2}
\end{align}
The internal amplitude $M_{int}^{CDCC(post)}$ is given by
\begin{align}
&M_{int}^{CDCC(post)}({\rm {\bf k}}_{pF},\,{\rm {\bf k}}_{dA})   \nonumber\\                
&= <\chi_{pF}^{(-)}\,I_{A}^{F}|\,\Delta\,{\overline V}_{pF}^{P_{pn}}|\Psi_{i}^{CDCC(+)}>\Big |_{r_{nA} \leq R_{nA}}. 
\label{intCDCCpost11}
\end{align}
Correspondingly, the external amplitude is 
\begin{align}
&M_{ext}^{CDCC(post)}({\rm {\bf k}}_{pF},\,{\rm {\bf k}}_{dA})                     \nonumber\\
&= <\chi_{pF}^{(-)}\,I_{A}^{F}|\,\Delta\,{\overline V}_{pF}^{P_{pn}}|\Psi_{i}^{CDCC(+)}>\Big |_{r_{nA} > R_{nA}}.
\label{extCDCCpost1}
\end{align}
I remind that the integral over the second Jacobian variable, ${\rm {\bf r}}_{pF}$, is taken over all the coordinate space.
Similarly to the DWBA case, the internal part is small if the channel radius $R_{nA}$ is not too large. Due to the strong absorption of the proton inside $A$, which is controlled by the imaginary part of the optical potential $U_{pA}^{P_{pn}}$, the effective distances are $r_{pA} > R_{A}$. Besides, in the internal region, $r_{nA} \leq R_{nA}$, and large $r_{pA}$, where $\,r_{pA} \sim r_{pn}= |{\rm {\bf r}}_{pA} - {\rm {\bf r}}_{nA}|$, $U_{pA}^{P_{pn}} + V_{pn}$ can be well approximated by a properly chosen optical potential $U_{pF}$ minimizing $\Delta\,{\overline V}_{pF}^{P_{pn}}$ and the internal matrix element. The next step is to transform the external matrix element to the surface one. To this end
we rewrite the transition operator in the form
\begin{align}
\Delta\,{\overline V}_{pF}^{P_{pn}} = U_{pA}^{P_{pn}} + V_{pn}  - U_{pF} = [ - U_{pF}] + [U_{pA}^{P_{pn}} + V_{pn}].
\label{extdltvcdcc1} 
\end{align}  
The bracketed operators in (\ref{extdltvcdcc1}) are the right-hand-side potential operators in the Schr\"odinger equations
in the external region $r_{nA} > R_{nA}$, where the nuclear $n-A$ interaction vanishes:
\begin{equation}
(E - T)\,\Psi_{i}^{CDCC(+)}= (U_{pA}^{P_{pn}} + V_{pn})\,\Psi_{i}^{CDCC(+)}
\label{shreqcdccpsi1}
\end{equation} 
and 
\begin{equation}
(E - T)\,\chi_{pF}^{(-)*}\,I_{A}^{F\,*}=U_{pF}\,\chi_{pF}^{(-)*}\,I_{A}^{F\,*}.
\label{shreqphif11}
\end{equation}
Note that the second equation follows from  
\begin{align}
(-\varepsilon_{nA} - T_{nA})\,I_{A}^{F}\,=\, <\varphi_{A}|V_{nA}|\,\varphi_{F}>.
\label{eqiaf1}
\end{align}

In the external region, $r_{nA} > R_{nA}$, the source term on the right-hand-side disappears and Eq. (\ref{shreqphif11})
becomes evident.
Taking into account Eqs (\ref{shreqcdccpsi1}) and (\ref{shreqphif11}) we get  
\begin{align}
M_{ext}^{CDCC(post)}({\rm {\bf k}}_{pF},\,{\rm {\bf k}}_{dA}) \equiv
M_{S}^{CDCC(post)}({\rm {\bf k}}_{pF},\,{\rm {\bf k}}_{dA})              \nonumber\\
= <\chi_{pF}^{(-)}\,I_{A}^{F}|{\overleftarrow T} - {\overrightarrow T}|\,\Psi_{i}^{CDCC(+)}>\Big |_{r_{nA} > R_{nA}},
\label{extCDCCpostT1}
\end{align}
where $T = T_{pF} + T_{nA}$. Here, as in the previous section, for the surface integral we use the subscript "S".
 Since the CDCC wave function doesn't propagate into the final state (its asymptotic terms have only elastic and breakup terms) the operator $T_{pF}$ is Hermitian, i.e. 
\begin{align}
<\chi_{pF}^{(-)}\,I_{A}^{F}|{\overleftarrow T}_{pF} - {\overrightarrow T}_{pF}|\Psi_{i}^{CDCC(+)}>\Big |_{r_{nA} > R_{nA}} \nonumber\\
= <\chi_{pF}^{(-)}\,I_{A}^{F}|{\overrightarrow T}_{pF} - {\overrightarrow T}_{pF}|\Psi_{i}^{CDCC(+)}>\Big |_{r_{nA} > R_{nA}}=0.
\label{leftrightT1}
\end{align}
It can be also shown explicitly taking into account that the volume integral over ${\rm {\bf r}}_{pF}$ can be transformed into the surface integral over the sphere with the radius $r_{pF} = R_{pF} \to \infty$. Since the overlap function decays exponentially at $r_{nA} \to \infty$, the integration over $r_{nA}$ is limited. Hence, at $r_{pF} \to \infty$ using Eqs (\ref{rpFrnArpn1}) we get that $r_{dA} \sim r_{pF} \to \infty$ and $r_{pn} \sim r_{pF} \to \infty$. The first term of the CDCC wave function decays exponentially at $r_{pF} \to \infty$ because of the presence of the deuteron bound state wave function. The terms with $n \geq 1$ decay as $1/r_{pF}^{3}$ \cite{sawada}.  The distorted wave $\chi_{pF}^{(-)*}({\rm{\bf p}}_{pF})$ decays as $1/r_{pF}$, see Eq. (\ref{asymplwave1}). Hence the surface integral vanishes at $R_{pF} \to \infty$ as $R_{pF}^{2}/R_{pF}^{4} \to 0$.  

Then $M_{S}^{CDCC(post)}$ takes the form
\begin{align}
&M_{S}^{CDCC(post)}({\rm {\bf k}}_{pF},\,{\rm {\bf k}}_{dA})                        \nonumber\\
&= <\chi_{pF}^{(-)}\,I_{A}^{F}|{\overleftarrow T}_{nA} - {\overrightarrow T}_{nA}|\,\Psi_{i}^{CDCC(+)}>\Big |_{r_{nA} > R_{nA}} \nonumber\\
&= - M_{S_{R_{nA}}}^{CDCC(post)}({\rm {\bf k}}_{pF},\,{\rm {\bf k}}_{dA})  + M_{S_{\infty}}^{CDCC(post)}({\rm {\bf k}}_{pF},\,{\rm {\bf k}}_{dA}). 
\label{extCDCCpostTnA1}
\end{align}
Thus, the volume integral at $r_{nA} > R_{nA}$ in the matrix element $M_{S}^{CDCC(post)}$ can be written as the sum of two surface integrals encircling the external volume, the sphere with the radius $r_{nA}= R_{nA}$ and the sphere with $r_{nA} = R_{nA}^{'} \to \infty$. Note that the integral over ${\rm {\bf r}}_{pF}$ is taken over all the coordinate space. Evidently that the integral over the infinitely large sphere vanishes
because the overlap function $I_{A}^{F}$ exponentially decreases. Hence, 
\begin{align}
&M_{S}^{CDCC(post)}({\rm {\bf k}}_{pF},\,{\rm {\bf k}}_{dA})= - M_{S_{R_{nA}}}^{CDCC(post)}({\rm {\bf k}}_{pF},\,{\rm {\bf k}}_{dA}) . 
\label{extCDCCpostTnA1}
\end{align}
The negative sign in front of the inner surface integral appears because the normal vector to the inner surface is directed to the center, i.e. opposite to the direction of the normal to the external surface at $r_{nA}=R_{nA}^{'} \to \infty$. 
Now we can use equations from subsection A replacing the initial channel wave function by the CDCC one. 
For $M_{S}^{CDCC(post)}$  we get
\begin{widetext}
\begin{align}
M_{S}^{CDCC(post)}({\rm {\bf k}}_{pF},\,{\rm {\bf k}}_{dA})= - M_{S_{R_{nA}}}^{CDCC(post)}({\rm {\bf k}}_{pF},\,{\rm {\bf k}}_{dA}) \qquad\qquad\qquad 
\nonumber\\
= \,\frac{R_{nA}^{2}}{2\,\mu_{nA}}\,\int\,{\rm d}\,{\rm {\bf r}}_{pF}\,\chi_{-{\rm {\bf k}}_{pF}}^{(+)}({\rm {\bf r}}_{pF})\,\int{\rm d}\,\Omega_{{\rm {\bf r}}_{nA}}\,{\rm {\bf {\hat r}}}_{nA}\, \Big[ [I_{A}^{F}({\rm {\bf r}}_{nA})]^{*}({\overleftarrow {\rm {\bf \nabla }}}_{{\rm {\bf r}}_{nA}} - {\overrightarrow {\rm {\bf \nabla }}}_{{\rm {\bf r}}_{nA}})\Psi_{i}^{CDCC(+)}({\rm {\bf r}}_{pF},\,{\rm {\bf r}}_{nA})  \,\Big]\,\Big |_{r_{nA} > R_{nA}}  \nonumber\\
= \frac{R_{nA}^{2}}{2\,\mu_{nA}}\,\int\,{\rm d}\,{\rm {\bf r}}_{pF}\,\chi_{-{\rm {\bf k}}_{pF}}^{(+)}({\rm {\bf r}}_{pF})\,\int{\rm d}\,\Omega_{{\rm {\bf r}}_{nA}}\,
\Big[\Psi_{i}^{CDCC(+)}({\rm {\bf r}}_{pF},\,{\rm {\bf r}}_{nA})\,\frac{\partial\,[I_{A}^{F}({\rm {\bf r}}_{nA})]^{*}}{{\partial {r_{nA}}}}\, - \,[I_{A}^{F}({\rm {\bf r}}_{nA})]^{*}\,\frac{\partial\,\Psi_{i}^{CDCC(+)}({\rm {\bf r}}_{pF},\,{\rm {\bf r}}_{nA})}{{\partial {r_{nA}}}}\Big]\,\Big |_{r_{nA} = R_{nA}}.
\label{SCDCCpostTnA2}
\end{align}
\end{widetext} 
Natural Jacobian variables for $\Psi_{i}^{CDCC(+)}$ are ${\rm {\bf r}}_{dA}$ and ${\rm {\bf r}}_{pn}$, but here  we use another set of Jacobian variables, ${\rm {\bf r}}_{pF}$ and ${\rm {\bf r}}_{nA}$.
Taking into account Eq. (\ref{overlapfunction2}) and (\ref{overlapasympt1}) we get
\begin{widetext}
\begin{align}
&M_{S}^{CDCC(post)}({\rm {\bf k}}_{pF},\,{\rm {\bf k}}_{dA})= - M_{S_{R_{nA}}}^{CDCC(post)}({\rm {\bf k}}_{pF},\,{\rm {\bf k}}_{dA})         \nonumber\\
&= \,\sqrt{\frac{R_{nA}}{2\,\mu_{nA}}}\,\sum\limits_{j_{nA}\,m_{j_{nA}}\,m_{l_{nA}}\,M_{n}}\,<J_{A}\, M_{A}\,\, j_{nA}\, m_{j_{nA}}| J_{F}\,M_{F}>
 \nonumber\\
&\times \,<J_{n}\,M_{n}\,\,l_{nA}\,m_{l_{nA}}|j_{nA}\,m_{j_{nA}}>\,\gamma_{nA\,j_{nA}\,l_{nA}}\int\,{\rm d}\,{\rm {\bf r}}_{pF}\,\chi _{{-{\rm {\bf k}}_{pF}}}^{(+)}({\rm {\bf r}}_{pF})\,\int\,{\rm d}\,\Omega_{{\rm {\bf r}}_{nA}}\,Y_{l_{nA}\,m_{l_{nA}}}^{*}({\rm {\bf {\hat r}}}_{nA}) \nonumber\\
& \times \,\left[\Psi_{i}^{CDCC(+)}({\rm {\bf r}}_{pF},\,{\rm {\bf r}}_{nA})\,(B_{nA} - \,1) - \,R_{nA}\,
\frac{ \partial\,\Psi_{i}^{CDCC(+)}({\rm {\bf r}}_{pF},\,{\rm {\bf r}}_{nA})}{\partial
\,r_{nA}}\right]\Big |_{r_{nA}= R_{nA}}.
\label{extpostCDCCA1}
\end{align} 
\end{widetext}
Note that the CDCC wave function itself also depends on quantum numbers of $p-n$ and $d-A$ subsystems, which we don't specify here. It will be done in the following up paper where concrete calculations will be presented.

Thus we have obtained a remarkable result: the post form of the CDCC amplitude, in contrast to the DWBA one, is given by the sum of only two terms:
\begin{align}
M^{CDCC(post)}({\rm {\bf k}}_{pF},\,{\rm {\bf k}}_{dA}) = M_{int}^{CDCC(post)}({\rm {\bf k}}_{pF},\,{\rm {\bf k}}_{dA}) \nonumber\\
- M_{S_{R_{nA}}}^{CDCC(post)}({\rm {\bf k}}_{pF},\,{\rm {\bf k}}_{dA}),
\label{postCDCC22}
\end{align}
where the first term, which is the internal post form  of the CDCC amplitude, can be minimized by a proper choice of $U_{pF}$ and the channel radius $R_{nA}$, while the second term, which is dominant, represents the surface integral with the radius $R_{nA}$, which encircles the internal volume in the subspace over the coordinate ${\rm {\bf r}}_{nA}$. If the channel radius is larger than the $n-A$ nuclear interaction radius the second term is parametrized in terms of the reduced width amplitude (ANC of the projection of  the bound state wave function of $F$ on the two-body state $n+ A$) and the boundary condition at $r_{nA}= R_{nA}$. If $ M_{int}^{CDCC(post)}$ is small enough, 
\begin{align}
M^{CDCC(post)}({\rm {\bf k}}_{pF},\,{\rm {\bf k}}_{dA}) \approx - M_{S_{R_{nA}}}^{CDCC(post)}({\rm {\bf k}}_{pF},\,{\rm {\bf k}}_{dA}).
\label{cdccpostms11}
\end{align}
Thus we succeeded to parametrize the post form of the CDCC amplitude in terms of the $R$-matrix parameters. 
Eq. (\ref{postCDCC22}) and parametrization of the surface term of the post CDCC amplitude in terms of the $R$-matrix parameters, Eq. (\ref{extpostCDCCA1}),  are one of the main results of this paper.   

Although it is assumed that $M_{int}^{CDCC(post)}$ can be minimized so that the second term in Eq. (\ref{postCDCC22}) becomes dominant, I would like to present a different form for $M_{int}^{CDCC(post)}({\rm {\bf k}}_{pF},\,{\rm {\bf k}}_{dA})$, which leads to a different form for the whole amplitude $M^{CDCC(post)}({\rm {\bf k}}_{pF},\,{\rm {\bf k}}_{dA})$. To this end, let us rewrite the transition operator $\Delta\,{\overline V}_{pF}^{P_{pn}}$ in $M_{int}^{CDCC(post)}({\rm {\bf k}}_{pF},\,{\rm {\bf k}}_{dA})$ as 
\begin{align}
&\Delta\,{\overline V}_{pF}^{P_{pn}}= U_{pA}^{P_{pn}} + V_{pn} - U_{pF}                              \nonumber\\
&= [U_{pA}^{P_{pn}} + U_{nA}^{P_{pn}} + V_{pn}] - [{\overline V}_{nA} + U_{pF}] + {\overline V}_{nA} - U_{nA}^{P_{pn}}.
\label{deltavpfint1}
\end{align}  
Here, ${\overline V}_{nA}$ is the mean field potential supporting the bound state $(n\,A)$ while $U_{nA}^{P_{pn}}$ is the projected optical potential describing the $n-A$ interaction in the initial state of the reaction and entering the Schr\"odinger equation for the projected CDCC wave function in the initial state. 
The bracketed potential operators are the right-hand-side operators of the Schr\"odinger equations in the internal region, $r_{nA} \leq R_{nA}$,  
\begin{equation}
(E - T)\,\Psi_{i}^{CDCC(+)}= (U_{pA}^{P_{pn}} + U_{nA}^{P_{pn}} + V_{pn})\,\Psi_{i}^{CDCC(+)}
\label{intshreqcdcc1}
\end{equation} 
and 
\begin{equation}
(E - T)\,\chi_{pF}^{(-)*}\,I_{A}^{F\,*}= ({\overline V}_{nA} + U_{pF})\,\chi_{pF}^{(-)*}\,I_{A}^{F\,*}.
\label{intshreqphif1}
\end{equation}
Replacing the bracketed potential operators $[U_{pA}^{P_{pn}} + U_{nA}^{P_{pn}} + V_{pn}]$ and $[{\overline V}_{nA} + U_{pF}]\,$ by $\,E - {\overrightarrow T}$ and $\,E - {\overleftarrow T}$, correspondingly, we get for $M_{int}^{CDCC(post)}$ a new form:
\begin{align}
&M_{int}^{CDCC(post)}({\rm {\bf k}}_{pF},\,{\rm {\bf k}}_{dA}) = M_{S_{R_{nA}}}^{CDCC(post)}({\rm {\bf k}}_{pF},\,{\rm {\bf k}}_{dA})  \nonumber\\
& + M_{aux}^{CDCC(post)}({\rm {\bf k}}_{pF},\,{\rm {\bf k}}_{dA}), 
\label{intpostCDCCnew1}
\end{align}
\begin{align}
&M_{aux}^{CDCC(post)}({\rm {\bf k}}_{pF},\,{\rm {
\bf k}}_{dA})                                                    \nonumber\\       
&= <\chi_{pF}^{(-)}\,I_{A}^{F}|\,\Delta\,{\overline V}_{nA}^{P_{pn}}|\,\Psi_{i}^{CDCC(+)}>\Big |_{r_{nA} \leq R_{nA}},
\label{auxCDCCpost1}
\end{align}
\begin{align}
\Delta\,{\overline V}_{nA}^{P_{pn}} = {\overline V}_{nA} - U_{nA}^{P_{pn}}.
\label{dltoverlvnAPpn1}
\end{align}
Then the total post form of the CDCC amplitude can be written as 
\begin{align}
&M^{CDCC(post)}({\rm {\bf k}}_{pF},\,{\rm {\bf k}}_{dA})= M_{int}^{CDCC(post)}({\rm {\bf k}}_{pF},\,{\rm {\bf k}}_{dA}) \nonumber\\
& - M_{S_{R_{nA}}}^{CDCC(post)}({\rm {\bf k}}_{pF},\,{\rm {\bf k}}_{dA})      \nonumber\\
&=  M_{S_{R_{nA}}}^{CDCC(post)}({\rm {\bf k}}_{pF},\,{\rm {\bf k}}_{dA})  - M_{S_{R_{nA}}}^{CDCC(post)}({\rm {\bf k}}_{pF},\,{\rm {\bf k}}_{dA})                              \nonumber\\
&+ M_{aux}^{CDCC(post)}({\rm {\bf k}}_{pF},\,{\rm {\bf k}}_{dA})             
=  M_{aux}^{CDCC(post)}({\rm {\bf k}}_{pF},\,{\rm {\bf k}}_{dA})     \nonumber\\
&= <\chi_{pF}^{(-)}\,I_{A}^{F}|\,{\overline V}_{nA} - U_{nA}^{P_{pn}}|\Psi_{i}^{CDCC(+)}>\Big |_{r_{nA} \leq R_{nA} }.
\label{finalcdccpost1}
\end{align}
Thus, we obtained another important result. The CDCC amplitude in the post form is equal to the inner volume integral over variable ${\rm {\bf r}}_{nA}$ with the transition operator  ${\overline V}_{nA} - U_{nA}^{P_{pn}}$.  This transition operator is the difference between the bound state potential ${\overline V}_{nA}$ supporting the final bound state $(n\,A)$ and the projected optical potential describing the $n-A$ interaction in the initial state. It is worth mentioning that Eqs (\ref{postCDCC22}) and (\ref{finalcdccpost1}) are exact within the CDCC approach. If $M_{int}^{CDCC(post)}$ is small enough, then   
\begin{align}
M_{aux}^{CDCC(post)}({\rm {\bf k}}_{pF},\,{\rm {\bf k}}_{dA}) \approx - \,M_{S_{R_{nA}}}^{CDCC(post)}({\rm {\bf k}}_{pF},\,{\rm {\bf k}}_{dA}).    
\label{auxmsrna1}
\end{align}
However, I prefer Eq. (\ref{postCDCC22}) rather than (\ref{finalcdccpost1}). To calculate $M_{aux}^{CDCC(post)}$ 
one needs to know the overlap function in the internal region, where the overlap function is model-dependent and requires microscopic calculations. In contrast, in Eq. (\ref{postCDCC22}) the dominant part is the surface integral, which is parametrized in terms of the reduced width amplitude (ANC). The model dependence of the 
surface part is related with the ambiguity of the optical potentials and the value of the cut-off orbital angular momentum in the $p-n$ subsystem in the CDCC approach. Comparison with experiment allows one to extract the reduced width amplitude. The model-dependent internal part in Eq. (\ref{postCDCC22}) is small.  
Eqs (\ref{postCDCC22}) and (\ref{finalcdccpost1}) is prelude to the theory of the stripping to resonance, where the convergence problem of the external part is one of the main issue. As we have demonstrated in the post CDCC formalism the external part doesn't appear at all. It resolves the convergence problem related with the external part.

\subsection{Deuteron stripping to bound states. Prior CDCC formalism}

A priori, the amplitudes in the post and prior forms of the CDCC formalism are not equal. That is why the obtained equations using the surface integrals are expected to be different in both formalisms. 
The prior form of the CDCC stripping amplitude is
\begin{align}
&M^{CDCC(prior)}({\rm {\bf k}}_{pF},\,{\rm {\bf k}}_{dA}) \nonumber\\
&= <\Psi_{f}^{CDCC(-)}\,|\,\Delta\,{\overline V}_{dA}^{P_{nA}}|\varphi_{d}\,\chi_{dA}^{(+)}>, 
\label{prCDCCbst1}	
\end{align}
where
\begin{align}
\Delta\,{\overline V}_{dA}^{P_{nA}}= U_{pA}^{P_{nA}} + V_{nA} - U_{dA}.
\label{deltaVdAPnA1}
\end{align}
The projected CDCC wave function in the final state is a solution of the three-body Schr\"odinger equation
\begin{align}
(E - T - U_{pA}^{P_{nA}} - V_{nA} - V_{pn}^{P_{nA}})\,\Psi_{f}^{CDCC(-)*}=0.
\label{prSchreqPCDCC1}
\end{align}
Here,
\begin{align}
&{\hat P}_{nA}= \sum\limits_{l_{nA}=0}^{l_{nA}^{max}}\,\sum\limits_{\,m_{l_{nA}}=-l_{nA}}^{
l_{nA}}\,\int{\rm d}\,\Omega_{{\rm {\bf r}}_{nA}}\,
Y_{l_{nA}\,m_{l_{nA}}}({\rm {\bf {\hat r}}}_{nA})            \nonumber\\
& \times Y_{l_{nA}\,m_{l_{nA}}}^{*}({\rm {\bf {\hat r}}}_{nA}^{'}).
\label{projectorprPlnA1}
\end{align}
is the projection operator, which truncates the number of the spherical harmonics $\,Y_{l_{nA}\,m_{l_{nA}}}({\rm {\bf {\hat r}}}_{nA})\,$ in the coordinate $\,{\rm {\bf r}}_{nA}$. 

Now, as usually, we split the amplitude $M^{CDCC(prior)}$ into the internal and external parts in the subspace over variable ${\rm {\bf r}}_{nA}$:
\begin{align}
&M^{CDCC(prior)}({\rm {\bf k}}_{pF},\,{\rm {\bf k}}_{dA}) \nonumber\\
&= M_{int}^{CDCC(prior)}({\rm {\bf k}}_{pF},\,{\rm {\bf k}}_{dA}) + M_{ext}^{CDCC(prior)}({\rm {\bf k}}_{pF},\,{\rm {\bf k}}_{dA}),  
\label{prCDCCintext1}`	
\end{align}
where 
\begin{align}
&M^{CDCC(prior)}({\rm {\bf k}}_{pF},\,{\rm {\bf k}}_{dA}) \nonumber\\
&= <\Psi_{f}^{CDCC(-)}\,|\,U_{pA}^{P_{nA}} + V_{nA} - U_{dA}|\varphi_{d}\,\chi_{dA}^{(+)}>\Big|_{r_{nA} \leq R_{nA}} 
\label{prCDCCbstint1}	
\end{align}
and
\begin{align}
&M_{ext}^{CDCC(prior)}({\rm {\bf k}}_{pF},\,{\rm {\bf k}}_{dA}) \nonumber\\
&= <\Psi_{f}^{CDCC(-)}\,|\,U_{pA}^{P_{nA}} - U_{dA}|\varphi_{d}\,\chi_{dA}^{(+)}>\Big|_{r_{nA} > R_{nA}}. 
\label{prCDCCbstext1}	
\end{align}

The external part of the prior amplitude (see discussion in subsection \ref{priorDWBA}), due to the structure 
of the transition operator,  is small and the dominant contribution comes from the internal amplitude. We will rewrite this 
amplitude singling out the surface integral over variable ${\rm {\bf r}}_{nA}$. To do it we rewrite the transition operator  
\begin{align}
\Delta\,{\overline V}_{dA}^{P_{nA}}= U_{pA}^{P_{nA}} + V_{nA} - U_{dA}     \nonumber\\
=[U_{pA}^{P_{nA}} + V_{nA} + V_{pn}^{P_{nA}}] - [V_{pn} + U_{dA}].             
&+ (V_{pn} -  V_{pn}^{P_{nA}})
\label{deltaVdAPnAtr1}
\end{align}
The bracketed operators are the right-hand-side operators of the Schr\"odinger equations 
\begin{align}
(E - T)\,\Psi_{f}^{CDCC(-)*} = (U_{pA}^{P_{nA}} + V_{nA} + V_{pn}^{P_{nA}})\,\Psi_{f}^{CDCC(-)*}
\label{psifcdcc1}
\end{align}
and
\begin{align}
(E- T)\,\varphi_{d}\,\chi_{dA}^{(+)}= (V_{pn} + U_{dA})\,\varphi_{d}\,\chi_{dA}^{(+)}.
\label{varphidchida1}
\end{align}

Taking into account these equations we can rewrite $M_{int}^{CDCC(prior)}({\rm {\bf k}}_{pF},\,{\rm {\bf k}}_{dA})$ in the form: 
\begin{align}
&M_{int}^{CDCC(prior)}({\rm {\bf k}}_{pF},\,{\rm {\bf k}}_{dA})= M_{S}^{CDCC(prior)}({\rm {\bf k}}_{pF},\,{\rm {\bf k}}_{dA})  \nonumber\\
&+ M_{aux}^{CDCC(prior)}({\rm {\bf k}}_{pF},\,{\rm {\bf k}}_{dA}),  
\label{totprCDCC11}	
\end{align}
where
\begin{align}
&M_{aux}^{CDCC(prior)}({\rm {\bf k}}_{pF},\,{\rm {\bf k}}_{dA})              \nonumber\\
&= <\Psi_{f}^{CDCC(-)}\,|\,V_{pn} -  V_{pn}^{P_{nA}}\,|\,\varphi_{d}\,\chi_{dA}^{(+)}> \Big|_{r_{nA} \leq R_{nA}}
\label{auxprior1}
\end{align}
and
\begin{align}
&M_{S}^{CDCC(prior)}({\rm {\bf k}}_{pF},\,{\rm {\bf k}}_{dA})  \nonumber\\
&= - <\Psi_{f}^{CDCC(-)}\,|\,{\overleftarrow T} - {\overrightarrow T}\,|\,\varphi_{d}\,\chi_{dA}^{(+)}>\Big|_{r_{nA} \leq R_{nA}}.
\label{SprCDCC12}	
\end{align}
Here, the kinetic energy operator $T= T_{pF} + T_{nA}$.  In $M_{S}^{CDCC(prior)}$ the volume integral over ${\rm {\bf r}}_{pF}$ can be transformed into the surface one taken over the sphere with the infinitely large radius: $r_{pF}= R_{pF} \to \infty$. For $r_{nA} \leq R_{nA}$, due to the presence of the deuteron bound state wave function, the integrand goes to zero exponentially, that is this surface integral vanishes. Hence, only the surface integral encircling the inner volume with the radius $r_{nA}=R_{nA}$:
\begin{align}
&M_{S}^{CDCC(prior)}({\rm {\bf k}}_{pF},\,{\rm {\bf k}}_{dA})  \nonumber\\
&= - <\Psi_{f}^{CDCC(-)}\,|\,{\overleftarrow T}_{nA} - {\overrightarrow T}_{nA}\,|\,\varphi_{d}\,\chi_{dA}^{(+)}>\Big|_{r_{nA} \leq R_{nA}}     \nonumber\\
&= -\,M_{S_{R_{nA}}}^{CDCC(post)}({\rm {\bf k}}_{pF},\,{\rm {\bf k}}_{dA}).
\label{SRnASprCDCCint11}	
\end{align}
$M_{S_{R_{nA}}}^{CDCC(post)}$ is given by Eq. (\ref{extpostCDCCA1}).
$M_{aux}^{CDCC(prior)}({\rm {\bf k}}_{pF},\,{\rm {\bf k}}_{dA})$ is an auxiliary internal part, which is small because at 
$r_{nA} \leq R_{nA}$ and $r_{pF} > R_{F}$ due to the proton absorption in the nuclear interior, $p-n$ nuclear interaction is significantly depleted, and so the difference $V_{pn} -  V_{pn}^{P_{nA}}$. 
Then 
\begin{align}
&M^{CDCC(prior)}({\rm {\bf k}}_{pF},\,{\rm {\bf k}}_{dA}) \nonumber\\
&= M_{aux}^{CDCC(prior)}({\rm {\bf k}}_{pF},\,{\rm {\bf k}}_{dA}) -  M_{S_{R_{nA}}}^{CDCC(post)}({\rm {\bf k}}_{pF},\,{\rm {\bf k}}_{dA})                                                                  \nonumber\\
&+  M_{ext}^{CDCC(prior)}({\rm {\bf k}}_{pF},\,{\rm {\bf k}}_{dA}),
\label{prCDCCbstfinal1}	
\end{align}

Thus the total prior form CDCC amplitude consists of three terms, small auxiliary internal part, small external prior form and the dominant surface term. We can see that post and prior CDCC formalisms are not equivalent. In the approach used in the paper the configuration space over variable ${\rm {\bf r}}_{nA}$ was split into the internal and external parts. As it has been discussed in Introduction, such a splitting is natural because the main object of interest in the analysis of deuteron stripping is the overlap function $I_{A}^{F}$ of the bound states wave functions of the target $A$ and final nucleus $F$. Its external part 
($r_{nA} > R_{nA}$) is parametrized in terms of the observable ANC while the internal part is model-dependent. 

In the post formalism the external part is dominant. Invoking the post CDCC formalism  allows us to rewrite the external CDCC matrix element in the form of the surface integral over variable ${\rm {\bf r}}_{nA}$, which can be parametrized in terms of the parameters used in the $R$-matrix method for binary reactions, while the model-dependent internal part gives small contribution. Thus the volume part of the matrix element over variable ${\rm {\bf r}}_{nA}$ is transformed to the surface integral. For transfer to bound states such a transformation doesn't bring any significant advantages because the volume matrix element converges. However, for stripping to resonance states (see subsection \ref{postCDCCres}) this transformation provides a decisive benefit because it solves the convergence problem of the matrix element. Here, the transformation of the post CDCC matrix element has been presented mostly for demonstration but the results will be used below in subsection \ref{postCDCCres} for stripping to resonance states. 

The prior CDCC formalism would be preferable if we split the matrix element into the internal and external parts over variable ${\rm {\bf r}}_{p\,n}$ to separate the internal and peripheral parts of the deuteron bound state wave function. But this wave function is well known and is not an object of study.  That is why below, when considering the stripping to resonance states, we use only the post CDCC formalism.   

\section{Deuteron stripping into resonance states}
Now we proceed to the main goal of this paper, formulation of the deuteron stripping into resonance states
using the surface integrals what will lead us to the generalized $R$-matrix approach for the stripping into resonance states.
Let us consider the deuteron stripping  
\begin{equation}
d + A \to p + b + B.
\label{dpnstripping}
\end{equation}
We assume that the resonance formed in the system $F=A+n$ can decay into channel $B + b$, which can be different from the entry channel $A + n$. 
We start from the post form and transform it to the surface integral following the method applied for the stripping to bound states. Now the application of the $R$-matrix approach looks natural.  Although we consider the deuteron stripping leading to a specific final channel $d + A \to p + b +B$, there can be a few open channels coupled to the channel $n + A$, which is formed after neutron is transferred to the target $A$. As in the previous sections, follow the $R$-matrix approach, we split the integration region over ${\rm {\bf r}}_{nA}$ into two regions: internal and external. Internal region is determined as the one where all open channels are coupled with each other, so that the transition from one channel to another can occur only in the internal region. The external region is the one where all the channels are decoupled. We obtain new forms for the DWBA and then for the post form of the CDCC amplitude. 
For the DWBA both post and prior approach will lead to the same final expression. In the standard approach the post form of the DWBA amplitude is mainly contributed by the external part in the subspace ${\rm {\bf r}}_{nA}$, where the convergence question of the DWBA matrix element, which contains the integration over ${\rm {\bf r}}_{pF}$ and ${\rm {\bf r}}_{nA}$, becomes a main issue. In the prior form the main contribution to the DWBA matrix element mainly comes from the internal region in the subspace ${\rm {\bf r}}_{nA}$, where a strong coupling between different open channels becomes an issue. In a new approach formulated below the DWBA amplitude (in the post and prior forms) is written as the sum of three amplitudes: small internal post and external prior forms, and the dominant surface integral in the subspace over ${\rm {\bf r}}_{nA}$. This surface term is parametrized in terms of the reduced width amplitudes, resonance energies and boundary condition, that is the quantities used in a standard $R$-matrix approach.  
In the post CDCC approach the amplitude is given by the sum of the small internal post form and the dominant surface term, that is, in contract to the DWBA, no external prior form appears in the CDCC method. This resolves the issue of the convergence for stripping into resonant states. 

\subsection{Stripping to resonance states. Post form of DWBA.}
\label{postDWBAres} 
The post form of the DWBA amplitude can be obtained by generalizing the corresponding equation for the deuteron stripping to the bound state. As a starting point, we use Eq. (\ref{standpostdw1}) in which, to get the amplitude for the deuteron stripping to resonance states, we should replace the overlap function $I_{A}^{F}$ by the exact scattering wave function $\Psi_{bB}^{(-)}$ with the incident wave in the channel $b + B$:
\begin{align}
&M^{DW(post)}(P,\,{\rm {\bf k}}_{dA})  \nonumber\\
&= <\chi_{pF}^{(-)}\,\Psi _{bB}^{(int)(-)}|
 \,|\,\Delta\,{\overline V}_{pF}|\varphi_{d}\,\varphi_{A}\,\chi_{dA}^{(+)}>,
\label{dwpostres1}
\end{align}
where $\Delta {\overline V}_{pF}= U_{pA} +V_{pn} - U_{pF}$ and  
\begin{align}
\Psi_{bB}^{(-)} \equiv \Psi_{ {\rm {\bf k}}_{bB} }^{(-)} = \Psi_{-{\rm {\bf k}}_{bB}}^{(+)*}.
\label{psibb(-)psibB(+)}
\end{align}

 Since we consider the stripping to the resonance state, which decays into two fragments $b$ and $B$, there are three particles, $p,\,b$ and $B$, in the final state. Hence, the kinematics of the final state of the reaction depends on two Jacobian momenta, for which we adopt the relative momentum of two fragments $b$ and $B$ and by the momentum corresponding to the relative motion of the exiting proton and the center of mass of the system $b + B$.  Thus the deuteron stripping reaction amplitude depends on the 
momentum $P=\{{\rm {\bf k}}_{pF},\,{\rm {\bf k}}_{bB}\}$, which is the six-dimensional momentum conjugated to the Jacobian coordinates of the system $p + b + B$  $\,Y=\{{\rm {\bf r}}_{pF},\,{\rm {\bf r}}_{bB}\}$. 

Then repeating the steps used in derivation of the expression for the post form of the DWBA amplitude for deuteron stripping to the bound state we get
\begin{align}
M^{DW(post)}(P,\,{\rm {\bf k}}_{dA}) 
= M_{int}^{DW(post)}(P,\,{\rm {\bf k}}_{dA})    \nonumber\\ 
+ M_{ext}^{DW(prior)}(P,\,{\rm {\bf k}}_{dA}) +  M_{S}^{DW}(P,\,{\rm {\bf k}}_{dA}). 
\label{finaldwpostres1}
\end{align}
Here, internal post amplitude $M_{int}^{DW(post)}(P,\,{\rm {\bf k}}_{dA})$ and external prior amplitude  $M_{ext}^{DW(prior)}(P,\,{\rm {\bf k}}_{dA})$ are given by
\begin{align}
&M_{int}^{DW(post)}(P,\,{\rm {\bf k}}_{dA})  \nonumber\\
&= <\chi_{pF}^{(-)}\,\Upsilon_{nA}^{(int)(-)}
 \,|\,\Delta\,{\overline V}_{pF}| 
\varphi_{d}\,\chi_{dA}^{(+)}>\Big |_{r_{nA} \leq R_{nA}}
\label{intdwpostres1}
\end{align}
and
\begin{align}
&M_{ext}^{DW(prior)}(P,\,{\rm {\bf k}}_{dA})       \nonumber\\
& =  <\chi_{pF}^{(-)}\,\Upsilon_{nA}^{(ext)(-)}|\,\Delta\,{\overline V}_{pF} |\varphi_{d}\,\chi_{dA}^{(+)}>\Big |_{r_{nA} > R_{nA}}.
\label{extdwpriorres1}
\end{align}
Here, $\Upsilon_{nA}^{(int)(-)}({\rm {\bf r}}_{nA})= <\varphi_{A}|\Psi _{bB}^{(int)(-)}>$ and 
$\Upsilon_{nA}^{(ext)(-)}({\rm {\bf r}}_{nA})= < \varphi_{A}|\Psi _{bB}^{(ext)(-)}>$.

The last term of Eq. (\ref{finaldwpostres1}), which will be transformed to the surface integral, is
\begin{align}
&  M_{S}^{DW}(P,\,{\rm {\bf k}}_{dA})      \nonumber\\
&= <\chi_{pF}^{(-)}\,\Upsilon_{nA}^{(ext)(-)}|\,{\overleftarrow T} - {\overrightarrow T}\,|\,\varphi_{d}\,\chi_{dA}^{(+)}>\Big |_{r_{nA} > R_{nA}}.
\label{SDWnAres1}
\end{align}

Let us discuss the advantage of this new form of the DWBA amplitude for the deuteron stripping to resonance state(s).  
Since the internal part $M_{int}^{DW(post)}$ is given by the volume integral, its calculation requires the knowledge of $\Psi _{bB}^{(int)(-)}$ in the internal region. The model dependence of this function in the nuclear interior ($r_{nA} \leq R_{nA}$), where different coupled channels do contribute, brings one of the main problems and main uncertainty in the calculation of the internal matrix element. However, as it has been discussed in subsection \ref{postDWBA}, this matrix element gives a small contribution to the  total post form amplitude $M^{DW(post)}$ due to the structure of the transition operator $\Delta\,{\overline V}_{pF}$ and constrain $r_{nA} \leq R_{nA}$. These arguments are also valid when considering the stripping into resonance states. A proper choice of the optical potential $U_{pF}$ and the channel radius $R_{nA}$ may significantly reduce the contribution from the internal post form DWBA amplitude. 
Due to the structure of the transition operator $\Delta\,{\overline V}_{dA}$, which has been also discussed in subsection \ref{postDWBA}, the external matrix element $M_{ext}^{DW(prior)}$ in the prior form is   also small and in some cases with reasonable choice of the channel radius $R_{nA}$ even can be neglected. Note that in order to keep small $M_{int}^{DW(post)}$ the channel radius $R_{nA}$ cannot be too large  
and  in order to keep small $M_{ext}^{DW(prior)}$ cannot be too small. Thus with optimal choice of the channel radius the dominant part is the surface part $M_{S}^{DW}$, which contains only one volume integral over ${\rm {\bf r}}_{pF}$.  Eq. (\ref{finaldwpostres1}), which presents a new form of the DWBA amplitude for stripping to resonance states, is quite important for analysis of the stripping to resonance In this sense the usage of the external prior amplitude $M_{ext}^{DW(prior)}$ has clear benefit because it is small and better converges the external post form. Also small is the internal amplitude $M_{int}^{DW(post)}$. The main contribution to $M^{DW(post)}$ comes from the surface term $M_{S}^{DW}$. 
Using the $R$-matrix representation of the scattering wave function $\Psi_{bB}^{(-)*}$ we are able to express the total DWBA amplitude in terms of the reduced width amplitudes, level matrix, boundary condition and the channel radius, that is parameters used in a standard $R$-matrix method to analyze binary resonant reactions $n + A \to b + B$. Since the reaction under consideration is the deuteron stripping, the presence of the deuteron in the initial state and exiting proton causes the distortions. That is why the reaction amplitude, in addition to the $R$-matrix parameters describing the binary subprocess, contains additional factors - distorted waves in the initial and the final state.
That is why we can call the obtained expression for the DWBA amplitude a generalized $R$-matrix for deuteron stripping to resonance states. 

Now we proceed to the derivation of the expressions for each amplitude in the right-hand-side of Eq. (\ref{finaldwpostres1})
and the total post form DWBA amplitude.
Since the stripping into resonance states can lead to rearrangement, the exit channel $b + B$ may differ from the entry channel
$n + A$. To proceed further we now use the equations for $\Psi_{bB}^{(+)}$ obtained in Appendix \ref{psibB1}. 
Taking into account Eqs. (\ref{psibb(-)psibB(+)}) and (\ref{PsibBresint1})  
we get
\begin{widetext}
\begin{align}
&M_{int}^{DW(post)}(P,\,{\rm {\bf k}}_{dA}) = \frac{2\,\pi}{k_{bB}}\,\sqrt{\frac{k_{bB}}{\mu_{bB}}} \,\sum\limits_{J_{F}\,M_{F},\,l\,m_{l}\,M_{n}}\,\,<s\,m_{s}\,\,l\,m_{l}|J_{F}\,M_{F}>\,<J_{n}\,M_{n}\,\,J_{p}\,M_{p}|J_{d}\,M_{d}>                 \nonumber\\
&\,e^{-i\,\delta_{bB\,l}^{hs}}\,i^{l}\,Y_{l\,m_{l}}^{*}(-{\rm {\bf {\hat k}}}_{bB})\,\sum\limits_{\nu,\tau = 1}^N [\Gamma_{\nu\,bB\,s\,l\,J_{F}}(E_{bB})]^{1/2}\,[{\bf A}^{ - 1} ]_{\nu \tau}\,<\chi_{pF}^{(-)}\,\Xi_{\tau\,nA}^{J_{F}\,M_{F}}|\Delta\,{\overline V}_{pF}| \varphi_{d}\,\chi_{dA}^{(+)}>\Big |_{r_{nA} \leq R_{nA}}.
\label{intdwpostres11}
\end{align}
\end{widetext}
In this equation we assume that the channel spin $s$ and its projection $m_{s}$ in the exit channel $c=b+B$ are fixed \footnote[1]{Note that when considering the wave function in Appendix \ref{psibB1} the channel $c=b+ B$ was the entry channel. But when
considering the deuteron stripping reaction amplitude, this channel is the exit channel while the channel $c'=n + A$ becomes the entry channel in the resonant subreaction $n + A \to b + B$.}. $J_{F}$ is the resonance spin ($M_{F}$ its projection)
in the subsystem $F=n+ A= b + B$ and $l$ is the $b+ B$ orbital angular momentum in the resonance state.  The sum over 
$J_{F}$ and $l$ assumes that a few resonances with different spins may contribute to the reaction. 
The subscript $c$ used in Appendix \ref{psibB1} for the channel $b + B$  is replaced here by $bB$.
Also $\Xi_{\tau\,nA}^{J_{F}\,M_{F}}=<\varphi_{A}|X_{\tau}^{J_{F}\,M_{F}}>$ is projection of $X_{\tau}^{J_{F}\,M_{F}}$ introduced in Appendix \ref{psibB1} on the bound state $\varphi_{A}$. 
The bound-state like wave function $X_{\tau}^{J_{F}\,M_{F}}$ describes the system  $F= n + A= b + B$ in the internal region. A priori, it can be calculated using, for example, the shell model approach \cite{mahaux}. In Appendix \ref{psibB1} $X_{\tau}^{J_{F}\,M_{F}}$ is written as a nonorthogonal sum of coupled channels, see Eq. (\ref{XtauJM1}). If we neglect the contribution from the channel $c$, then $\Xi_{\tau\, nA}^{J_{F}M_{F}}$ can be approximated by the internal part of the overlap function, see Eq. (\ref{overlapfunction2}). Taking into account this equation (rewritten in $LS$-coupling scheme) we get   
\begin{widetext}
\begin{align}
&M_{int}^{DW(post)}(P,\,{\rm {\bf k}}_{dA}) = \frac{2\,\pi}{k_{bB}}\,\sqrt{\frac{k_{bB}}{\mu_{bB}}} \,
\sum\limits_{J_{F}\,M_{F}\,s'\,l\,l'\,m_{s'}\,m_{l}\,m_{l'}\,M_{n}}\,i^{l}\,\,<s\,m_{s}\,\,l\,m_{l}|J_{F}\,M_{F}>\,< s'\,m_{s'}\,l'\,m_{l'}\,|J_{F}\,M_{F} >  \nonumber\\
& \times\,<J_{n}\,M_{n}\,\,J_{A}\,M_{A}|s'\,m_{s'}>\,<J_{n}\,M_{n}\,\,J_{p}\,M_{p}|J_{d}\,M_{d}>\,e^{-i\,\delta_{bB\,l}^{hs}}\,Y_{l\,m_{l}}^{*}(-{\rm {\bf {\hat k}}}_{bB})     \nonumber\\
&\times  
 \sum\limits_{\nu,\tau = 1}^N [\Gamma_{\nu\,bB\,s\,l\,J_{F}}(E_{bB})]^{1/2}\,[{\bf A}^{ - 1} ]_{\nu \tau} 
\,<\chi_{pF}^{(-)}\,Y_{l'\,m_{l'}}^{*}({\rm {\bf {\hat r}}}_{nA})\,I_{A\,\,s'\,l'\,J_{F}}^{F}(r_{nA})\,|\Delta\,{\overline V}_{pF}| \varphi_{d}\,\chi_{dA}^{(+)}>\Big |_{r_{nA} \leq R_{nA}}.
\label{intdwpostres12}
\end{align}
\end{widetext}
Here we added the sum over the channel spin $s'$ (its projection $m_{s'}$ in the entry channel $c'=n+A$ of the resonant
subreaction $n + A \to F \to b + B$ and over the $n + A$ orbital angular momentum $l'$. The sum over $M_{n}$ and $s'$ appears because the transferred neutron is intermediate (virtual). 
It is important that with a proper choice of the optical potential $U_{pF}$ the matrix element $M_{int}^{DW(post)}$ can be minimized so that its model dependence wouldn't have impact on the total matrix element $M^{DW(post)}$. 

To obtain the expression for $M_{ext}^{DW(prior)}$ we use for the external part $\Psi _{bB}^{(ext)(-)}$, which can be obtained from
Eq. (\ref{psirc11}), assuming that the resonance contribution to this wave function is dominant. 
In the sum over $J_{F}$ in Eq. (\ref{psirc11}) we keep only those total angular momenta  at which resonances contributing to the reaction occur. 
Let us consider two possible cases.
\\
(i) The exit channel $c=b+B$ in the resonant sub-process $n + A \to b + B$ is different from channel $c'=n+ A$. In this case 
the external resonant wave function is given by Eq. (\ref{psiccprext1}) and its projection on the bound state $\xi_{c'}=\varphi_{A}$  is determined by Eq. (\ref{Upsiloncprext1}).  Then $M_{ext}^{DW(prior)}$ reduces to
\begin{widetext}
\begin{align}
&M_{ext}^{DW(prior)}(P,\,{\rm {\bf k}}_{dA}) =  -\,i\,\frac{2\,\pi}{k_{bB}}\,\sqrt {\frac{v_{bB}}{v_{nA}}}\,         \sum\limits_{J_{F}\,M_{F}\,s'\,l\,l'\,m_{s'}\,m_{l}\,m_{l'}\,M_{n}}\,i^{l}\,< l\,m_{l}\,\,s\,{m_s}|J_{F}\,M_{F} >\,< l'\,m_{l'}\,\,s'\,m_{s'}|J_{F}\,M_{F} >                     \nonumber\\ 
&\times  <J_{n}\,M_{n}\,\,J_{A}\,M_{A}|s'\,m_{s'}><J_{n}\,M_{n}\,\,J_{p}\,M_{p}|J_{d}\,M_{d}>\,Y_{l\,m_{l}}^{*}(-{\rm {\bf {\hat k}}}_{bB})\,S_{bB\,s\,l;nA\,s'\,l'}^{J_{F}}     \nonumber\\
&\times <\chi_{pF}^{(-)}\,\frac{O_{l'}^{*}(k_{nA},\,r_{nA})}{r_{nA}}\,Y_{l'\,m_{l'}}^{*}({\rm {\bf {\hat r}}}_{nA})|\,\Delta\,{\overline V}_{dA} |\varphi_{d}\,\chi_{dA}^{(+)}>\Big |_{r_{nA} > R_{nA}}.
\label{extdwpriorres2}
\end{align}
\end{widetext}
Here, ${\overline V}_{dA}$ is given by Eq. (\ref{DeltaVdA1}). In the external region ${\overline V}_{nA}=0$
and ${\overline V}_{dA}= U_{pA} - U_{dA}$. Also has been added the sum over the orbital angular momentum $l$ and its projection $m_{l}$ ($l'$ and $m_{l'}$) in the exit (entry) channel $c=b + B$ ($c'=n + A$) of the resonant subreaction $n + A \to b + B$, the sum over the channel spin $s'$ and its projection $m_{s'}$ in the entry channel $c'=n + A$ of the resonance subprocess $n + A \to b + B$ and the sum over $M_{n}$ because the neutron is transferred particle.
The projections of the spins of the incident deuteron $M_{d}$, the exiting proton $M_{p}$, the channel spin $s$ and its projection $m_{s}$ of the exiting particles $b$ and $B$ are fixed.
We also use the symmetry of the $S$ matrix:  $S_{c'\,s'\,l';c\,s\,l}^{J_{F}}=S_{c\,s\,l;c'\,s'\,l'}^{J_{F}}$.  
The matrix element $S_{bB\,s\,l;nA\,s'\,l'}^{J_{F}}$ is given by Eq. (\ref{reactSmatrix1}). Substituting it into 
Eq. (\ref{extdwpriorres2})  gives
\begin{widetext}
\begin{align}
&M_{ext}^{DW(prior)}(P,\,{\rm {\bf k}}_{dA}) =  \frac{2\,\pi}{k_{bB}}\,\sqrt {\frac{v_{bB}}{v_{nA}}}\,         \sum\limits_{J_{F}\,M_{F}\,s'\,l\,l'\,m_{s'}\,m_{l}\,m_{l'}\,M_{n}}\,i^{l}\,< l\,m_{l}\,\,s\,{m_s}|J_{F}\,M_{F} >\, < l'\,m_{l'}\,\,s'\,m_{s'}|J_{F}\,M_{F} >                                                   \nonumber\\
&\times<J_{n}\,M_{n}\,\,J_{A}\,M_{A}|s'\,m_{s'}><J_{n}\,M_{n}\,\,J_{p}\,M_{p}|J_{d}\,M_{d}> Y_{l\,m_{l}}^{*}(-{\rm {\bf {\hat k}}}_{bB})\,     \nonumber\\
&\times e^{-\,i\,\delta_{bB\,l}^{hs}}\,e^{-\,i\,\delta_{nA\,l'}^{hs}}\,\sum\limits_{\nu,\tau = 1}^{N}\,[\Gamma_{\nu\,bB\,s\,l\,J_{F}}(E_{bB})]^{1/2}\,[{\bf A}^{ - 1} ]_{\nu \tau}\,\,[\Gamma_{\tau\,nA\,s'\,l'\,J_{F}}(E_{nA})]^{1/2}\,\frac{O_{l'}(k_{nA},\,R_{nA})}{R_{nA}}         \nonumber\\
&\times <\chi_{pF}^{(-)}\,\frac{O_{l'}^{*}(k_{nA},\,r_{nA})}{r_{nA}}\,
\frac{R_{nA}}{O_{l'}^{*}(k_{nA},\,R_{nA})}\,
Y_{l'\,m_{l'}}^{*}({\rm {\bf {\hat r}}}_{nA})|\,\Delta\,{\overline V}_{dA} |\varphi_{d}\,\chi_{dA}^{(+)}>\Big |_{r_{nA} > R_{nA}}.
\label{extdwpriorresS2}
\end{align}
\end{widetext}

Now we take into account that 
\begin{align}
&O_{\tilde l}(k_{\tilde c},\,R_{\tilde c})= \sqrt{F_{\tilde l}^{2}(k_{\tilde c},\,R_{\tilde c})+ G_{\tilde l}^{2}(k_{\tilde c},\,R_{\tilde c})} \nonumber\\
&\times e^{-\,i\,\omega_{{\tilde c}\,{\tilde l}}}\,e^{i\,\arctan{\frac{F_{\tilde l}(k_{\tilde c},\,R_{\tilde c})}{G_{\tilde l}(k_{\tilde c},\,R_{\tilde c})} }}   \nonumber\\
&= \sqrt{ F_{\tilde l}^{2}(k_{\tilde c},\,R_{\tilde c})+ G_{\tilde l}^{2}(k_{\tilde c},\,R_{\tilde c})}
\,e^{i\,\delta_{{\tilde c}\,{\tilde l}}^{hs}},  
\label{OFGphase1}
\end{align}
which for the channel ${\tilde c}= c' = n+A$ and ${\tilde l}=l'$ takes the form
\begin{align}
&O_{l'}(k_{nA},\,R_{nA})= \sqrt{F_{l'}^{2}(k_{nA},\,R_{nA})+ G_{l'}^{2}(k_{nA},\,R_{nA})} \nonumber\\
&\times \,e^{i\,\arctan{\frac{F_{l'}(k_{nA},\,R_{nA})}{G_{l'}(k_{nA},\,R_{nA})} }}   \nonumber\\
&= \sqrt{ F_{l'}^{2}(k_{nA},\,R_{nA})+ G_{l'}^{2}(k_{nA},\,R_{nA})}\,
e^{i\,\delta_{nA\,l'}^{hs}}, 
\label{OFGphasenA1}
\end{align}
where in the absence of the Coulomb interaction $F_{l}(\rho) = (\pi\,\rho/2)^{1/2}\,J_{l+1/2}(\rho)$ and 
$G_{l}(\rho) = (-1)^{l}\,(\pi\,\rho/2)^{1/2}\,J_{-(l+1/2)}(\rho)$, $\,\,J_{\pm(l+1/2)}(\rho)$ are Bessel functions.

Then using Eqs. (\ref{Gammaredwidth1}) and (\ref{OFGphasenA1}) we get 
\begin{widetext}
\begin{align}
&M_{ext}^{DW(prior)}(P,\,{\rm {\bf k}}_{dA}) =  2\,\pi\,\sqrt{\frac{2\,\mu_{nA}}{\mu_{bB}\,k_{bB}\,R_{nA}}}\,\sum\limits_{J_{F}\,M_{F}\,s'\,l\,l'\,m_{s'}\,m_{l}\,m_{l'}\,M_{n}}\,i^{l}\,< l\,m_{l}\,\,s\,{m_s}|J_{F}\,M_{F}>                                                   \nonumber\\
&\times  < l'\,m_{l'}\,\,s'\,m_{s'}|J_{F}\,M_{F} >\,<J_{n}\,M_{n}\,\,J_{A}\,M_{A}|s'\,m_{s'}><J_{n}\,M_{n}\,\,J_{p}\,M_{p}|J_{d}\,M_{d}> Y_{l\,m_{l}}^{*}(-{\rm {\bf {\hat k}}}_{bB})    \nonumber\\
&\times \,e^{-\,i\,\delta_{bB\,l}^{hs}} \sum\limits_{\nu,\tau = 1}^{N}\,[\Gamma_{\nu\,bB\,s\,l\,J_{F}}(E_{bB})]^{1/2}\,[{\bf A}^{ - 1} ]_{\nu \tau}\,\gamma_{\tau\,nA\,s'\,l'\,J}                             \nonumber\\
&\times <\chi_{pF}^{(-)}\,\frac{O_{l'}^{*}(k_{nA},\,r_{nA})}{r_{nA}}\,
\frac{R_{nA}}{O_{l'}^{*}(k_{nA},\,R_{nA})}\,
Y_{l'\,m_{l'}}^{*}({\rm {\bf {\hat r}}}_{nA})|\,\Delta\,{\overline V}_{dA} |\varphi_{d}\,\chi_{dA}^{(+)}>\Big |_{r_{nA} > R_{nA}}.
\label{extdwpriorresfinal1}
\end{align}
\end{widetext}
(ii) If $c=c'$, that is $b=n$ and $B=A$. Here two cases are possible: non-diagonal transition for which $s \not= s'$ or/and $l \not= l'$ and diagonal transition with $l=l'$ and $s=s'$. 
The  amplitude for the nondiagonal transition can be obtained from 
(\ref{extdwpriorres2}). Here we present the expression for the diagonal transition (elastic scattering) amplitude, which can be obtained taking into account Eq. (\ref{Upsilonelextc1}):
\begin{widetext}
\begin{align}
&M_{ext}^{DW(prior)}(P,\,{\rm {\bf k}}_{dA}) =   i\,\frac{2\,\pi}{k_{nA}\,R_{nA}}\,\sum\limits_{J_{F}\,M_{F}\,l\,m_{s'}\,m_{l}\,m_{l'}\,M_{n}}\,i^{l}\,< l\,m_{l}\,\,s\,{m_s}|J_{F}\,M_{F} >\,< l\,m_{l'}\,\,s\,m_{s'}|J_{F}\,M_{F} >  \nonumber\\
& \times \,<J_{n}\,M_{n}\,\,J_{p}\,M_{p}|J_{d}\,M_{d}>\,<J_{n}\,M_{n}\,\,J_{A}\,M_{A}|s\,m_{s'}>\, 
Y_{l\,m_{l}}^{*}(-{\rm {\bf {\hat k}}}_{nA})\,\Big[1 
- S_{(nA)\,s\,l;(nA)\,s\,l}^{J_{F}}\Big]\,O_{l}(k_{nA},\,R_{nA})       \nonumber\\
&\times <\chi_{pF}^{(-)}\,\frac{O_{l}^{*}(k_{nA},\,r_{nA})}{r_{nA}}\,\frac{R_{nA}}{O_{l}^{*}(k_{nA},\,R_{nA})}\,Y_{l\,m_{l'}}^{*}({\rm {\bf {\hat r}}}_{nA})|\,\Delta\,{\overline V}_{dA} |\varphi_{d}\,\chi_{dA}^{(+)}>\Big |_{r_{nA} > R_{nA}}.
\label{extdwpriorres3}
\end{align}
\end{widetext}
Substituting the expression for the elastic scattering $S$-matrix element $S_{(nA)\,s\,l;(nA)\,s\,l}^{J_{F}}$ given by Eq. (\ref{elscatSmatrix2}) we obtain
\begin{widetext}
\begin{align}
&M_{ext}^{DW(prior)}(P,\,{\rm {\bf k}}_{dA}) =    i\,\frac{2\,\pi}{k_{nA}\,R_{nA}}\,\sum\limits_{J_{F}\,M_{F}\,l\,m_{s'}\,m_{l}\,m_{l'}\,M_{n}}\,i^{l}\,< l\,m_{l}\,\,s\,{m_s}|J_{F}\,M_{F} >\,< l\,m_{l'}\,\,s\,m_{s'}|J_{F}\,M_{F}>  \nonumber\\
& \times \,<J_{n}\,M_{n}\,\,J_{p}\,M_{p}|J_{d}\,M_{d}>\,<J_{n}\,M_{n}\,\,J_{A}\,M_{A}|s\,m_{s'}>\, 
Y_{l\,m_{l}}^{*}(-{\rm {\bf {\hat k}}}_{nA})             \nonumber\\
&\times \,\Big[1 - e^{-2\,i\,\delta_{nA\,l}^{hs}}\,(1+ i\,\sum\limits_{\nu,\tau = 1}^N [\Gamma_{\nu\,nA\,s\,l\,J_{F}}(E_{nA})]^{1/2}\,[{\bf A}^{ - 1} ]_{\nu \tau}\,[\Gamma_{\tau\,nA\,s\,l\,J_{F}}(E_{nA})]^{1/2}) \Big]\,O_{l}(k_{nA},\,R_{nA})       \nonumber\\
&\times <\chi_{pF}^{(-)}\,\frac{O_{l}^{*}(k_{nA},\,r_{nA})}{r_{nA}}\,\frac{R_{nA}}{O_{l}^{*}(k_{nA},\,R_{nA})}Y_{l\,m_{l'}}^{*}({\rm {\bf {\hat r}}}_{nA})|\,\Delta\,{\overline V}_{dA} |\varphi_{d}\,\chi_{dA}^{(+)}>\Big |_{r_{nA} > R_{nA}}.
\label{extdwpriorreselast1}
\end{align}
\end{widetext} 
One-level, one channel case is the simplest one for which $M_{ext}^{DW(prior)}(P,\,{\rm {\bf k}}_{dA})$ boils down to
\begin{widetext}
\begin{align}
&M_{ext}^{DW(prior)}(P,\,{\rm {\bf k}}_{dA}) =   i\,\frac{2\,\pi}{k_{nA}\,R_{nA}}\,\sum\limits_{J_{F}\,M_{F}\,l\,m_{s'}\,m_{l}\,m_{l'}\,M_{n}}\,i^{l}\,< l\,m_{l}\,\,s\,{m_s}|J_{F}\,M_{F} >\,< l\,m_{l'}\,\,s\,m_{s'}|J_{F}\,M_{F} >  \nonumber\\
& \times \,<J_{n}\,M_{n}\,\,J_{p}\,M_{p}|J_{d}\,M_{d}>\,<J_{n}\,M_{n}\,\,J_{A}\,M_{A}|s\,m_{s'}>\, 
Y_{l\,m_{l}}^{*}(-{\rm {\bf {\hat k}}}_{nA})\,\Big[ 1 - e^{-2\,i\,\delta_{nA\,s\,l\,J_{F}}^{hs}}\,e^{2\,i\,\delta_{nA\,s\,l\,J_{F}}} \Big]\,O_{l}(k_{nA},\,R_{nA})       \nonumber\\
&\times <\chi_{pF}^{(-)}\,\frac{O_{l}^{*}(k_{nA},\,r_{nA})}{r_{nA}}\,\frac{R_{nA}}{O_{l}^{*}(k_{nA},\,R_{nA})}\,Y_{l'\,m_{l'}}^{*}({\rm {\bf {\hat r}}}_{nA})|\,\Delta\,{\overline V}_{dA} |\varphi_{d}\,\chi_{dA}^{(+)}>\Big |_{r_{nA} > R_{nA}},
\label{dwpronelevelonechres1}
\end{align} 
\end{widetext} 
where 
\begin{align}
&\delta_{nA\,s\,l\,J_{F}}= \arctan{\frac{\Gamma_{nA\,s\,l\,J_{F}}(E_{nA})}{2(E_{nA(0)\,s\,l\,J_{F}}- E_{nA})}}, \nonumber\\
& E_{nA(0)\,s\,l\,J_{F}} > E_{nA},
\label{CNphshift1}
\end{align}
is the resonant phase shift, $E_{nA(0)\,s\,l\,J_{F}}$ is the real part of the complex resonance energy of the resonance with the quantum numbers $s\,l\,J_{F}$ in the channel $n+A$. 
Now we derive the equation for $M_{S}^{DW}$ by transforming it into the surface integrals over variable ${\rm {\bf r}}_{nA}$.
We can repeat the discussion in  Section II A. The integration in Eq. (\ref{SDWnAres1}) over ${\rm {\bf r}}_{nA}$ is taken over the external volume restricted by two spherical surfaces: the inner surface with the radius $R_{nA}$ and the external surface with the radius $R_{nA}^{'} \to \infty$. As it has been shown in Appendix \ref{surfaceintegral1}
after regularization  the integral over the infinitely large sphere vanishes (see Eq. (\ref{intsurfint11})) and 
\begin{widetext}
\begin{align}
&M_{S}^{DW}(P,\,{\rm {\bf k}}_{dA}) = -M_{S_{R_{nA}}}^{DW}(P,\,{\rm {\bf k}}_{dA})
=R_{nA}^{2}\,\frac{1}{2\,\mu_{nA}}\,\int\,{\rm d}\,{\rm {\bf r}}_{pF}\,\int\,{\rm d}\,\Omega_{{\rm {\bf r}}_{nA}}\,
[\varphi_{d}(r_{pn})\,\chi_{ {\rm {\bf k}}_{dA} }^{(+)}({\rm {\bf r}}_{dA})\,\chi_{- {\rm {\bf k}}_{pF} }^{(+)}({\rm {\bf r}}_{pF})\,\frac{ \partial\,[\Upsilon^{(ext)(-)}_{nA}({\rm {\bf r}}_{nA})]^{*}}{ \partial\,r_{nA} }  \nonumber\\ 
&-\chi _{{-{\rm {\bf k}}_{pF}}}^{(+)}({\rm {\bf r}}_{pF})\,[\Upsilon_{nA}^{(ext)(-)}({\rm {\bf r}}_{nA})]^{*}\,\frac{ 
\partial\,\varphi_{d}(r_{pn})\,\chi_{{\rm {\bf k}}_{dA}}^{(+)})({\rm {\bf r}}_{dA})}{\partial \,r_{nA}}]\Big |_{r_{nA}= R_{nA}}. 
\label{SSDWnAres11}
\end{align}
\end{widetext} 
Here, $-M_{S_{R_{nA}}}^{DW}(P,\,{\rm {\bf k}}_{dA})$ is the surface integral encircling the inner surface of the external volume at $r_{nA} = R_{nA}$. A negative sign appears because the normal vector to the surface is directed to the center of the volume, i.e. opposite to the normal vector to the external surface (at infinitely large radius).  
For simplicity, we dropped the quantum numbers in Eq. (\ref{SSDWnAres11}) but they will be recovered below.  
Note that Eq. (\ref{SSDWnAres11}) can be obtained from Eq. (\ref{SdwdtnA12}) by substituting $\Upsilon_{nA}^{(ext)(-)}({\rm {\bf r}}_{nA})\,\,$ for the overlap function $I_{A\,\,J_{F}\,M_{F}\,\,J_{A}\,M_{A}\,\,m_{nA}}^{F}({\rm {\bf r}}_{nA})$.

For the exit channel $c=b+B$ in the resonant sub-process $n + A \to b + B$ different from channel $c'=n+ A$ using  Eq. (\ref{Upsiloncprext1}) we get
\begin{widetext}
\begin{align}
&M_{S}^{DW}({\rm {\bf k}}_{pF},\,{\rm {\bf k}}_{dA}) = -M_{S_{R_{nA}}}^{DW}(P,\,{\rm {\bf k}}_{dA})    
= -\,i\,\frac{2\,\pi}{k_{bB}}\,\sqrt {\frac{v_{bB}}{v_{nA}}}\,R_{nA}^{2}\,\frac{1}{2\,\mu_{nA}}\,\sum\limits_{J_{F}\,M_{F}\,l\,l'\,m_{l}\,m_{l'}\,s'\,M_{n}}\,i^{l}\, < l\,m_{l}\,\,s\,{m_s}|J_{F}\,M_{F} >     \nonumber\\
& \times \,\,< l'\,m_{l'}\,\,s'\,m_{s'}|JM >\,<J_{n}\,M_{n}\,\,J_{A}\,M_{A}|s'\,m_{s'}>
\,<J_{n}\,M_{n}\,\,J_{p}\,M_{p}|J_{d}\,M_{d}>\,Y_{l\,m_{l}}^{*}(-{\rm {\bf {\hat k}}}_{bB})\,S_{nA\,s'\,l';bB\,s\,l}^{J_{F}}\
\,\int{\rm d}\,{\rm {\bf r}}_{pF}\,\chi_{- {\rm {\bf k}}_{pF} }^{(+)}({\rm {\bf r}}_{pF})  \nonumber\\
&\times\int\,{\rm d}\,\Omega_{{\rm {\bf r}}_{nA}}\,Y_{l'\,m_{l'} }({\rm {\bf {\hat r}}}_{nA})\,
\Big[\varphi_{d}(r_{pn})\,\chi_{ {\rm {\bf k}}_{dA} }^{(+)}({\rm {\bf r}}_{dA})\, \frac{ \partial\,\frac{O_{l'}(k_{nA},\,r_{nA})}{r_{nA}}}{ \partial\,r_{nA} } 
-\chi_{{-{\rm {\bf k}}_{pF}}}^{(+)}({\rm {\bf r}}_{pF})\,\frac{O_{l'}(k_{nA},\,r_{nA})}{r_{nA}}\,\frac{ \partial\,\varphi_{d}(r_{pn})\,\chi_{{\rm {\bf k}}_{dA}}^{(+)})({\rm {\bf r}}_{dA})}{\partial \,r_{nA}} \Big]\Big |_{r_{nA}= R_{nA}} \nonumber\\
&=  -\,i\,\frac{2\,\pi}{k_{bB}}\,\sqrt {\frac{v_{bB}}{v_{nA}}}\,\frac{1}{2\,\mu_{nA}}\, \sum\limits_{J_{F}\,M\,l\,l'\,m_{l}\,m_{l'}\,s'\,M_{n}}\,i^{l}\, < l\,m_{l}\,\,s\,{m_s}|J_{F}M >\,< l'\,m_{l'}\,\,s'\,m_{s'}|J_{F}\,M_{F} >\,<J_{n}\,M_{n}\,\,J_{A}\,M_{A}|s'\,m_{s'}>           \nonumber\\
& \times <J_{n}\,M_{n}\,\,J_{p}\,M_{p}|J_{d}\,M_{d}>\,Y_{l\,m_{l}}^{*}(-{\rm {\bf {\hat k}}}_{bB})\,
S_{bB\,s\,l;nA\,s'\,l'}^{J_{F}}\, \,O_{l'}(k_{nA},\,R_{nA})\,
\,\int{\rm d}\,{\rm {\bf r}}_{pF}\,\chi_{- {\rm {\bf k}}_{pF} }^{(+)}({\rm {\bf r}}_{pF})\,\int\,{\rm d}\,\Omega_{{\rm {\bf r}}_{nA}}\,Y_{l'\,m_{l'} }({\rm {\bf {\hat r}}}_{nA})                  \nonumber\\
&\times \,\Big[\varphi_{d}(r_{pn})\,\chi_{ {\rm {\bf k}}_{dA} }^{(+)}({\rm {\bf r}}_{dA})\,(B_{nA} -1) - R_{nA}\,\frac{ \partial\,\varphi_{d}(r_{pn})\,\chi_{{\rm {\bf k}}_{dA}}^{(+)})({\rm {\bf r}}_{dA})}{\partial \,r_{nA}}\Big] \Big|_{r_{nA}= R_{nA}}.
\label{SDWnAres21}
\end{align}
\end{widetext}
Here,
\begin{align}
B_{nA} = {R_{nA}}\frac{{\frac{{\partial {\mkern 1mu} {O_{l'}}({k_{nA}},{\mkern 1mu} {r_{nA}})}}{{\partial {\mkern 1mu} {r_{nA}}}}{\Big|_{{r_{nA}} = {R_{nA}}}}}}{{{O_{l'}}({k_{nA}},{\mkern 1mu} {R_{nA}})}}
\label{BnAres1}
\end{align} 
is the boundary condition. Sum over $M_{n}$ is a formal because $M_{d}$ and $M_{p}$ are fixed. The coefficient 
$\,<J_{n}\,M_{n}\,\,J_{p}\,M_{p}|J_{d}\,M_{d}>\,$ appears from the vertex $d \to p+ n$ and the product 
$\,< l'\,m_{l'}\,\,s'\,m_{s'}|J_{F}\,M_{F} >\,<J_{n}\,M_{n}\,\,J_{A}\,M_{A}|s'\,m_{s'}>\,$ from the vertex $n + A \to F$. The matrix element $S_{bB\,s\,l;nA\,s'\,l'}^{J_{F}}$ is given by Eq. (\ref{reactSmatrix1}). Substituting it into 
Eq. (\ref{SDWnAres21})  gives
\begin{widetext}
\begin{align}
&M_{S}^{DW}({\rm {\bf k}}_{pF},\,{\rm {\bf k}}_{dA}) = -M_{S_{R_{nA}}}^{DW}(P,\,{\rm {\bf k}}_{dA}) 
=  \,\frac{\pi}{k_{bB}}\,\sqrt {\frac{v_{bB}}{v_{nA}}}\,\frac{1}{\mu_{nA}}\, \sum\limits_{J_{F}\,M_{F}\,l\,l'\,s'\,m_{l}\,m_{l'}\,m_{s'}\,M_{n}}\,i^{l}\, < l\,m_{l}\,\,s\,{m_s}|J_{F}\,M_{F} >        \nonumber\\
& \times < l'\,m_{l'}\,\,s'\,m_{s'}|J_{F}\,M_{F} >\,<J_{n}\,M_{n}\,\,J_{A}\,M_{A}|s'\,m_{s'}> \,<J_{n}\,M_{n}\,\,J_{p}\,M_{p}|J_{d}\,M_{d}>\,Y_{l\,m_{l}}^{*}(-{\rm {\bf {\hat k}}}_{bB})\,e^{-\,i\delta_{bB\,l}^{hs}}\,e^{-\,i\,\delta_{c'\,l'}^{hs}}\,     \nonumber\\
&\times \sum\limits_{\nu,\tau = 1}^{N}\,[\Gamma_{\nu\,bB\,s\,l\,J_{F}}(E_{bB})]^{1/2}\,[{\bf A}^{ - 1} ]_{\nu \tau}\,\,[\Gamma_{\tau\,nA\,s'\,l'\,J_{F}}(E_{nA})]^{1/2}\,O_{l'}(k_{nA},\,R_{nA})\,\int{\rm d}\,{\rm {\bf r}}_{pF}\,\chi_{- {\rm {\bf k}}_{pF} }^{(+)}({\rm {\bf r}}_{pF})\,\int\,{\rm d}\,\Omega_{{\rm {\bf r}}_{nA}}\,Y_{l'\,m_{l'} }({\rm {\bf {\hat r}}}_{nA})      \nonumber\\
&\times \Big[ \varphi_{d}(r_{pn})\,\chi_{ {\rm {\bf k}}_{dA} }^{(+)}({\rm {\bf r}}_{dA})\,(B_{nA} -1) - R_{nA}\,\frac{ \partial\,\varphi_{d}(r_{pn})\,\chi_{{\rm {\bf k}}_{dA}}^{(+)})({\rm {\bf r}}_{dA})}{\partial \,r_{nA}} \Big]\Big |_{r_{nA}= R_{nA}}.
\label{SDWnAresSm11}
\end{align}
\end{widetext}
Taking into account Eq. (\ref{Gammaredwidth1}) and  Eq. (\ref{OFGphase1}) we arrive at the final form for $M_{S}^{DW}({\rm {\bf k}}_{pF},\,{\rm {\bf k}}_{dA})$:
\begin{widetext}
\begin{align}
&M_{S}^{DW}({\rm {\bf k}}_{pF},\,{\rm {\bf k}}_{dA}) = -M_{S_{R_{nA}}}^{DW}(P,\,{\rm {\bf k}}_{dA})  
= \,\pi\,\sqrt{\frac{2\,R_{nA}}{\mu_{bB}\,\mu_{nA}\,k_{bB}}}\,\sum\limits_{J_{F}\,M_{F}\,l\,l'\,s'\,m_{l}\,m_{l'}\,m_{s'}\,M_{n}}\,i^{l}\, < l\,m_{l}\,\,s\,{m_s}|J_{F}\,M_{F} >         \nonumber\\
& \times \,< l'\,m_{l'}\,\,s'\,m_{s'}|J_{F}\,M_{F} >\,<J_{n}\,M_{n}\,\,J_{A}\,M_{A}|s'\,m_{s'}> \,<J_{n}\,M_{n}\,\,J_{p}\,M_{p}|J_{d}\,M_{d}>\,
Y_{l\,m_{l}}^{*}(-{\rm {\bf {\hat k}}}_{bB})\,e^{-\,i\,\delta_{bB\,l}^{hs}}     \nonumber\\
&\times \sum\limits_{\nu,\tau = 1}^{N}\,[\Gamma_{\nu\,bB\,s\,l\,J_{F}}(E_{bB})]^{1/2}\,[{\bf A}^{ - 1} ]_{\nu \tau}\,\gamma_{\tau\,nA\,s'\,l'\,J}\,\int{\rm d}\,{\rm {\bf r}}_{pF}\,\chi_{- {\rm {\bf k}}_{pF} }^{(+)}({\rm {\bf r}}_{pF})\,\int\,{\rm d}\,\Omega_{{\rm {\bf r}}_{nA}}\,Y_{l'\,m_{l'} }({\rm {\bf {\hat r}}}_{nA})      \nonumber\\
&\times \Big[ \varphi_{d}(r_{pn})\,\chi_{ {\rm {\bf k}}_{dA} }^{(+)}({\rm {\bf r}}_{dA})\,(B_{nA} -1) - R_{nA}\,\frac{ \partial\,\varphi_{d}(r_{pn})\,\chi_{{\rm {\bf k}}_{dA}}^{(+)})({\rm {\bf r}}_{dA})}{\partial \,r_{nA}} \Big]\Big |_{r_{nA}= R_{nA}}.
\label{SDWnAresSm12}
\end{align}
\end{widetext}

Now let us consider the diagonal transition $c\,s\,l \to c\,s\,l$, where $c=c'=n+ A$.  To get $M_{S}^{DW}$ once again we start from Eq. (\ref{SSDWnAres11}). Now in this equation $\Upsilon_{nA}^{(ext)(-)}$ should be replaced by $\Upsilon_{c\,s\,l\,m_{s};c\,s\,l\,m_{s''}}^{J(ext)(0)} + \Upsilon_{c\,s\,l\,m_{s};c\,s\,l\,m_{s''}}^{J(ext)(-)}$ given by Eqs. (\ref{Upsilonextdif01}) and (\ref{Upsilonelextc1}). 
Then the equation for surface matrix element for the diagonal transition takes the form
\begin{widetext}
\begin{align}
&M_{S}^{DW}({\rm {\bf k}}_{pF},\,{\rm {\bf k}}_{dA}) = i\,\frac{\pi}{\,\mu_{nA}\,k_{nA}}\,\sum\limits_{J_{F}\,M_{F}\,l\,m_{l}\,m_{l''}\,m_{s''}\,M_{n}}\,i^{l}\, < l\,m_{l}\,\,s\,{m_s}|J_{F}\,M_{F} >\,< l\,m_{l''}\,\,s\,m_{s''}|J_{F}\,M_{F} >         \nonumber\\
& \times <J_{n}\,M_{n}\,\,J_{A}\,M_{A}|s\,m_{s''}> \,<J_{n}\,M_{n}\,\,J_{p}\,M_{p}|J_{d}\,M_{d}>\,
Y_{l\,m_{l}}^{*}(-{\rm {\bf {\hat k}}}_{nA})\,   \nonumber\\
&\times \Big[1 -  e^{-i\,2\,\delta_{nA\,l}^{hs}} \Big(1 + i\,\sum\limits_{\nu,\tau = 1}^{N}\,[\Gamma_{\nu\,nA\,s\,l\,J_{F}}(E_{nA})]^{1/2}\,[{\bf A}^{ - 1} ]_{\nu \tau}\,\Gamma_{\tau\,nA\,s\,l\,J_{F}}(E_{nA})]^{1/2} \Big) \Big]\,O_{l}(k_{nA},\,R_{nA})      \nonumber\\
&\times \,\int{\rm d}\,{\rm {\bf r}}_{pF}\,\chi_{- {\rm {\bf k}}_{pF} }^{(+)}({\rm {\bf r}}_{pF})\,\int\,{\rm d}\,\Omega_{{\rm {\bf r}}_{nA}}\,Y_{l\,m_{l''} }({\rm {\bf {\hat r}}}_{nA})\,\Big[ \varphi_{d}(r_{pn})\,\chi_{ {\rm {\bf k}}_{dA} }^{(+)}({\rm {\bf r}}_{dA})\,(B_{nA} -1) - R_{nA}\,\frac{ \partial\,\varphi_{d}(r_{pn})\,\chi_{{\rm {\bf k}}_{dA}}^{(+)})({\rm {\bf r}}_{dA})}{\partial \,r_{nA}} \Big]\Big |_{r_{nA}= R_{nA}}.
\label{SDWnAreselastic12}
\end{align}
\end{widetext}
Summing up all three amplitudes $M_{int}^{DW(post)}(P,\,{\rm {\bf k}}_{dA})$, $\,M_{ext}^{DW(prior)}(P,\,{\rm {\bf k}}_{dA})$ and  $M_{S}^{DW}({\rm {\bf k}}_{pF},\,{\rm {\bf k}}_{dA})= -M_{S_{R_{nA}}}^{DW}({\rm {\bf k}}_{pF},\,{\rm {\bf k}}_{dA}) $ we get the total post DWBA for the $(d,p)$ stripping. \\
(i) Resonant reaction $n + A \to b + B$, that is $c= b + B \not= c'=n + A$.
 Then the total post form of the DWBA deuteron stripping amplitude is
\begin{widetext}
\begin{align}
&M^{DW(post)}(P,\,{\rm {\bf k}}_{dA})
= 2\,\pi\,\sqrt{\frac{1}{\mu_{bB}\,k_{bB}}} \sum\limits_{J_{F}\,M_{F}\,s'\,l\,l'\,m_{s'}\,m_{l}\,m_{l'}\,M_{n}}\,i^{l}\,\,<s\,m_{s}\,\,l\,m_{l}|J_{F}\,M_{F}>\,< s'\,m_{s'}\,l'\,m_{l'}\,|J_{F}\,M_{F} >\,  \nonumber\\
& \times <J_{n}\,M_{n}\,\,J_{A}\,M_{A}|s'\,m_{s'}>\,<J_{n}\,M_{n}\,\,J_{p}\,M_{p}|J_{d}\,M_{d}>\,    
e^{-i\,\delta_{bB\,l}^{hs}}\,Y_{l\,m_{l}}^{*}(-{\rm {\bf {\hat k}}}_{bB}) 
 \sum\limits_{\nu,\tau = 1}^N [\Gamma_{\nu\,bB\,s\,l\,J_{F}}(E_{bB})]^{1/2}\,[{\bf A}^{ - 1} ]_{\nu \tau} \nonumber\\
&\times \Bigg\{ \,<\chi_{pF}^{(-)}\,I_{A\,\,s'\,l'\, J_{F}}^{F}(r_{nA})\,|\Delta\,{\overline V}_{pF}| \varphi_{d}\,\chi_{dA}^{(+)}>\Big |_{r_{nA} \leq R_{nA}} + \,\sqrt{\frac{2\,\mu_{nA}}{R_{nA}}}\,\,\,\gamma_{\tau\,nA\,s'\,l'\,J}               \nonumber\\
&\times <\chi_{pF}^{(-)}\,\frac{O_{l'}^{*}(k_{nA},\,r_{nA})}{r_{nA}}\,
\frac{R_{nA}}{O_{l'}^{*}(k_{nA},\,R_{nA})}\,
Y_{l'\,m_{l'}}^{*}({\rm {\bf {\hat r}}}_{nA})|\,\Delta\,{\overline V}_{pF} |\varphi_{d}\,\chi_{dA}^{(+)}>\Big |_{r_{nA} > R_{nA}}\nonumber\\
&+ \,\sqrt{\frac{R_{nA}}{2\,\mu_{nA}}}\,\,\,\gamma_{\tau\,nA\,s'\,l'\,J}\,\int{\rm d}\,{\rm {\bf r}}_{pF}\,\chi_{- {\rm {\bf k}}_{pF} }^{(+)}({\rm {\bf r}}_{pF})\,\int\,{\rm d}\,\Omega_{{\rm {\bf r}}_{nA}}\,Y_{l'\,m_{l'} }({\rm {\bf {\hat r}}}_{nA})      \nonumber\\
&\times \bigg[\varphi_{d}(r_{pn})\,\chi_{ {\rm {\bf k}}_{dA} }^{(+)}({\rm {\bf r}}_{dA})\,(B_{nA} -1) - R_{nA}\,\frac{ \partial\,\varphi_{d}(r_{pn})\,\chi_{{\rm {\bf k}}_{dA}}^{(+)})({\rm {\bf r}}_{dA})}{\partial \,r_{nA}}\bigg]\Big |_{r_{nA}= R_{nA}} \Bigg \}.
\label{DWpostresfinal11}
\end{align}
\end{widetext}
Assuming in this equation $b=n$ and $B=A$, that is $c=c'$ but $l \not= l'$ and/or $s \not= s'$ we get the expression for the
DWBA deuteron stripping for the non-diagonal transition in the resonant subprocess $(n + A)_{l\,s} \to F \to (n+ A)_{l'\,s'}$.

Equation (\ref{DWpostresfinal11}) is very instructive for understanding the difference between the stripping to resonance states and on-shell binary resonant reactions. As we can see, the transfer reaction amplitude contains the resonance factors determining the resonant subprocess $n + A \to b + B$, the partial width amplitude 
$[\Gamma_{\nu\,bB\,s\,l\,J_{F}}(E_{bB})]^{1/2}$ of the level $\nu$ for the decay to the exit channel $b + B$, the matrix elements of the inverse $R$-matrix level matrix $[{\bf A}^{ - 1} ]_{\nu \tau}$ and the reduced width amplitude $\gamma_{\tau\,nA\,s'\,l'\,J}$ of the level $\tau$ for the entry channel $n + A$  rather than the corresponding partial width amplitude which would present if we consider the corresponding on-shell binary resonant reaction $n + A \to b +B$. The 
difference is crucial because the partial width amplitude $[\Gamma_{\nu\,bB\,s\,l\,J_{F}}(E_{bB})]^{1/2}$ contains the penetrability factor,
see Eq. (\ref{Gammaredwidth1}), which is missing in the reduced width amplitude and, hence, in Eq. (\ref{DWpostresfinal11}). The lower is the energy of the resonance, the stronger is its suppression due to the barrier penetrability in the entrance channel in the on-shell binary resonant reaction $n + A \to b + B$. Besides, if a few resonances do contribute with the different $l'$, then the higher is $l'$, the stronger is its suppression. However, it is not the case if one tries to populate low-energy resonances with different $l'$ using transfer reaction. Missing penetrability factor in the entry channel of the subresonance reaction $n + A \to b + B$ in the transfer amplitude makes it possible to populate low-lying resonances.
Moreover for the same reason, the resonances with higher $l'$ are not suppressed in the stripping. Hence, when a few resonances are populated in the transfer reaction, the measured experimental spectrum of the fragments $b$ and $B$ can be quite different from the one measured using the on-shell binary resonant reaction. The missing penetrability factor in the entry channel $n + A$ of the resonant subreaction $n + A \to b +B$ in the transfer reaction explains the power of the Trojan Horse method as indirect technique in nuclear astrophysics (see \cite{spitaleri,lacognata} and references therein).

(ii) Diagonal transition in the resonant subprocess  $(n + A)_{l\,s} \to F \to (n+ A)_{l\,s}$, that is, $c=c',\,l=l',\,s=s'$. The total post form of the deuteron stripping DWBA amplitude is 
\begin{widetext}
\begin{align}
&M^{DW(post)}(P,\,{\rm {\bf k}}_{dA})
= 2\,\pi\,\sum\limits_{J_{F}\,M_{F}\,l\,m_{s''}\,m_{l}\,m_{l''}\,M_{n}}\,i^{l}\,\,<s\,m_{s}\,\,l\,m_{l}|J_{F}\,M_{F}>\,< s\,m_{s''}\,l\,m_{l''}\,|J_{F}\,M_{F} >\,  \nonumber\\
& \times <J_{n}\,M_{n}\,\,J_{A}\,M_{A}|s\,m_{s''}>\,<J_{n}\,M_{n}\,\,J_{p}\,M_{p}|J_{d}\,M_{d}>\,    
e^{-i\,\delta_{nA\,l}^{hs}}\,Y_{l\,m_{l}}^{*}(-{\rm {\bf {\hat k}}}_{nA}) \nonumber\\
&\times \Bigg\{ \,\sqrt{\frac{1}{\mu_{nA}\,k_{nA}}} \,\sum\limits_{\nu,\tau = 1}^N [\Gamma_{\nu\,nA\,s\,l\,J_{F}}(E_{nA})]^{1/2}\,[{\bf A}^{ - 1} ]_{\nu \tau}\,<\chi_{pF}^{(-)}\,I_{A\,\,s\,l\, J_{F}}^{F}(r_{nA})\,|\Delta\,{\overline V}_{pF}| \varphi_{d}\,\chi_{dA}^{(+)}>\Big |_{r_{nA} \leq R_{nA}}              \nonumber\\
& + \,i\,\Bigg[1 -  e^{-i\,2\,\delta_{nA\,l}^{hs}} \Big(1 + i\,\sum\limits_{\nu,\tau = 1}^{N}\,[\Gamma_{\nu\,nA\,s\,l\,J_{F}}(E_{nA})]^{1/2}\,[{\bf A}^{ - 1} ]_{\nu \tau}\,\Gamma_{\tau\,nA\,s\,l\,J_{F}}(E_{nA})]^{1/2} \Big) \Bigg]                              \nonumber\\
&\times \Bigg(\,\frac{1}{k_{nA}\,R_{nA}}\,O_{l}(k_{nA},\,R_{nA})\,<\chi_{pF}^{(-)}\,\frac{O_{l}^{*}(k_{nA},\,r_{nA})}{r_{nA}}\,
\frac{R_{nA}}{O_{l}^{*}(k_{nA},\,R_{nA})}\,
Y_{l\,m_{l''}}^{*}({\rm {\bf {\hat r}}}_{nA})|\,\Delta\,{\overline V}_{pF} |\varphi_{d}\,\chi_{dA}^{(+)}>\Big |_{r_{nA} > R_{nA}}\nonumber\\
& + \,\frac{1}{2\,\mu_{nA}\,k_{nA}}\,\int{\rm d}\,{\rm {\bf r}}_{pF}\,\chi_{- {\rm {\bf k}}_{pF} }^{(+)}({\rm {\bf r}}_{pF})\,\int\,{\rm d}\,\Omega_{{\rm {\bf r}}_{nA}}\,Y_{l\,m_{l''} }({\rm {\bf {\hat r}}}_{nA})      \nonumber\\
&\times \Big[\varphi_{d}(r_{pn})\,\chi_{ {\rm {\bf k}}_{dA} }^{(+)}({\rm {\bf r}}_{dA})\,(B_{nA} -1) - R_{nA}\,\frac{ \partial\,\varphi_{d}(r_{pn})\,\chi_{{\rm {\bf k}}_{dA}}^{(+)})({\rm {\bf r}}_{dA})}{\partial \,r_{nA}} \Big]\Big |_{r_{nA}= R_{nA}} \Bigg) \Bigg\}.
\label{DWpostresfinalel11}
\end{align}
\end{widetext}

\subsection{Stripping to resonance states. Prior form of DWBA.}
\label{priorDWBAres} 
Here we show that starting from the prior form we are able to obtain the generalized DWBA $R$-matrix amplitude for the deuteron stripping to resonance states, Eq. (\ref{finaldwpostres1}), much easier than from the post form.  
The prior of the DWBA amplitude for deuteron stripping to resonance states is
\begin{align}
&M^{DW(prior)}(P,\,{\rm {\bf k}}_{dA})  \nonumber\\
&= <\chi_{pF}^{(-)}\,\Psi _{bB}^{(-)}|
 \,|\,\Delta\,{\overline V}_{dA}|\varphi_{d}\,\varphi_{A}\,\chi_{dA}^{(+)}>,
\label{dwpriorres1}
\end{align}
where $\Delta\,{\overline V}_{dA}$ is defined by Eq. (\ref{DeltaVdA1}).
As usually, we split the amplitude into internal and external parts
\begin{align}
M^{DW(prior)}(P,\,{\rm {\bf k}}_{dA}) 
= M_{int}^{DW(prior)}(P,\,{\rm {\bf k}}_{dA})    \nonumber\\ 
+ M_{ext}^{DW(prior)}(P,\,{\rm {\bf k}}_{dA}). 
\label{dwpriorintextres1}
\end{align}
with 
\begin{align}
&M_{int}^{DW(prior)}(P,\,{\rm {\bf k}}_{dA})           \nonumber\\                        
&=  <\chi_{pF}^{(-)}\,\Psi _{bB}^{(int)(-)}\,|\,\Delta\,{\overline V}_{dA}|\varphi_{d}\,\chi_{dA}^{(+)}>\Big |_{r_{nA} \leq R_{nA}}   
\label{dwpriorintres1}
\end{align}
and
\begin{align}
&M_{ext}^{DW(prior)}({\rm {\bf k}}_{pF},\,{\rm {\bf k}}_{dA})         \nonumber\\   
&= <\chi_{pF}^{(-)}\,\Psi _{bB}^{(ext)(-)}\,|\,\Delta\,{\overline V}_{dA}|\varphi_{d}\,\chi_{dA}^{(+)}>\Big |_{r_{nA} > R_{nA}}.
\label{dwextpriorres1}
\end{align}
The splitting of the amplitude into the internal and external parts in the subspace over the coordinate ${\rm {\bf r}}_{nA}$ is necessary to rewrite the prior DWBA amplitude in the generalized $R$-matrix approach for stripping to resonance states. As we have discussed in subsections \ref{postDWBA} and \ref{postDWBAres},  the external matrix element $M_{ext}^{DW(prior)}$ in the prior form is small and in some cases with reasonable choice of the channel radius $R_{nA}$ even can be neglected. It is important for analysis of the stripping to resonance states because the external part in the post form doesn't converge. In this sense the usage of the prior form in the external part has clear benefit. The main contribution to the prior form amplitude $M^{DW(prior)}$ comes from the internal part $M_{int}^{DW(prior)}$. Since the internal part is given by the volume integral, its calculation requires the knowledge of $\Psi _{bB}^{(int)(-)}$ in the internal region. The model dependence of this function in the nuclear interior ($r_{nA} \leq R_{nA}$), where different coupled channels do contribute, brings one of the main problems and main uncertainty in the calculation of the internal matrix element. 
Using the surface integral we can rewrite the volume integral of the internal matrix element in terms of the volume integral in the post form and dominant surface integral taken over the sphere at $r_{nA}=R_{nA}$. With reasonable choice of the channel radius $R_{nA}$ the contribution from the internal volume integral in the post form can be minimized to make it significantly smaller than the surface matrix element. The latter can be expressed in terms of the $R$-matrix parameters - the observable reduced width amplitude (ANC), boundary condition and channel radius.
Repeating the steps outlined in subsection \ref{priorDWBA} we get 
\begin{align}
&M_{int}^{DW(prior)}(P,\,{\rm {\bf k}}_{dA})   \nonumber\\
& =  M_{int}^{DW(post)}(P,\,{\rm {\bf k}}_{dA})  + M_{S}^{DW}(P,\,{\rm {\bf k}}_{dA}).   
\label{dwpriorintrespostS1}
\end{align}
Here, $M_{int}^{DW(post)}$ has been previously considered and is given by Eqs. (\ref{intdwpostres11}) and (\ref{intdwpostres12}) while $M_{S}^{DW}$ takes the form
\begin{align}
&M_{S}^{DW}(P,\,{\rm {\bf k}}_{dA})                 \nonumber\\
&  = -<\chi_{pF}^{(-)}\,\Upsilon_{nA}^{(ext)(-)}|\,{\overleftarrow T} - {\overrightarrow T}\,|                  \varphi_{d}\,\chi_{dA}^{(+)}>\Big |_{r_{nA} \leq R_{nA}},
\label{SintDWnAres22}
\end{align}
where $\Upsilon^{(ext)(-)}= <\Psi _{bB}^{(ext)(-)}|\varphi_{A}>$. 
The fact that the volume integral in this equation is the internal one makes transformation of this volume matrix element to the surface one much easier than for the post form. The transition operator $T= T_{pF} + T_{nA}$.
Since $r_{nA} \leq R_{nA}$ at $r_{pF} \to \infty$ the integrand in Eq. (\ref{SDWnAres1}) vanishes exponentially due to the presence of $\varphi_{d}$. Hence, the operator $T_{pF}$ is Hermitian, that is, applying the integration by parts over $r_{pF}$ twice we get  
\begin{align}
&<\chi_{pF}^{(-)}\,\Upsilon_{nA}^{(ext)(-)}|\,{\overleftarrow T} - {\overrightarrow T}\,|\,\varphi_{d}\,\chi_{dA}^{(+)}>\Big |_{r_{nA} \leq R_{nA}}                     \nonumber\\
& = <\chi_{pF}^{(-)}\,,<\Psi _{bB}^{(ext)(-)}|\varphi_{A}>|\,{\overrightarrow T} - {\overrightarrow T}\,|\,\varphi_{d}\,\chi_{dA}^{(+)}>\Big |_{r_{nA} \leq R_{nA}}                                     \nonumber\\
&=0.
\label{TpfHermitian1}
\end{align}
Thus $M_{S}^{DW}(P,\,{\rm {\bf k}}_{dA})$ reduces to
\begin{align}
&M_{S}^{DW}(P,\,{\rm {\bf k}}_{dA})                 \nonumber\\
& = <\chi_{pF}^{(-)}\,\Upsilon_{nA}^{(ext)(-)}|\,{\overleftarrow T}_{nA} - {\overrightarrow T}_{nA}\,|                                           \varphi_{d}\,\chi_{dA}^{(+)}>\Big |_{r_{nA} \leq R_{nA}}.
\label{SDWnAresTnA1}
\end{align}
Using the Green's theorem we can transform this volume integral into the surface one. Note that the volume integral over ${\rm {\bf r}}_{nA}$ is constrained by the sphere with the radius $r_{nA}= R_{nA}$. Hence, only one surface integral appears with $r_{nA}= R_{nA}$. Here we see an important advantage of using the prior form versus the post one. In the post form 
transformation of the external volume integral to the surface one led to two surface integrals at $r_{nA}= R_{nA}$ and $r_{nA}= R_{nA}^{'} \to \infty$. It required an elaborate proof, which included regularization, to demonstrate that the surface integral at $r_{nA}= R_{nA}^{'} \to \infty$ vanishes. After transformation to the surface integral we get 
\begin{align}
&M_{S}^{DW}(P,\,{\rm {\bf k}}_{dA})=- M_{S_{R_{nA}}}^{DW}(P,\,{\rm {\bf k}}_{dA}),       \nonumber\\
\label{SRnAWnAres1}
\end{align}
Eqs ({\ref{SSDWnAres11}), (\ref{SDWnAres21}) and (\ref{SDWnAresSm12}) determine this surface integral.

\subsection{Stripping to resonance states. Post CDCC formalism.}
\label{postCDCCres}
The CDCC approach for stripping to resonance states, which takes into account the deuteron breakup in the initial channel, definitely has advantage compared to a standard DWBA. The application of the surface formulation of the reaction theory for the DWBA has been done mainly for demonstration, but our main goal is the CDCC. 

Here we present the derivation of the post form CDCC amplitude using the surface integral. This amplitude is 
\begin{align}
&M^{CDCC(post)}({\rm {\bf k}}_{pF},\,{\rm {\bf k}}_{dA}) \nonumber\\
&= <\chi_{pF}^{(-)}\,\Psi _{bB}^{(ext)(-)}\,|\,\Delta\,{\overline V}_{pF}^{P_{pn}}|\,\Psi_{i}^{CDCC(+)}>.    
\label{postPCDCCres1}	
\end{align}
This equation is an extension of the post CDCC amplitude for stripping to bound states, see Eq. (\ref{postPCDCC1}).
$\Delta\,{\overline V}_{pF}^{P_{pn}}$ is defined by Eq. (\ref{deltavpfint1}).  
Now, as usually, we split $M^{CDCC(post)}$ into the internal and external parts in the subspace ${\rm {\bf r}}_{nA}$: 
\begin{align}
M^{CDCC(post)}(P,\,{\rm {\bf k}}_{dA}) = M_{int}^{CDCC(post)}(P,\,{\rm {\bf k}}_{dA}) \nonumber\\
+ \, M_{ext}^{CDCC(post)}(P,\,{\rm {\bf k}}_{dA}). 
\label{postCDCCresintext1}
\end{align}
The internal amplitude $M_{int}^{CDCC(post)}$ is given by
\begin{align}
&M_{int}^{CDCC(post)}(P,\,{\rm {\bf k}}_{dA})    \nonumber\\
&= <\chi_{pF}^{(-)}\,\Psi _{bB}^{(ext)(-)}|\,\Delta\,{\overline V}_{pF}^{P_{pn}}|  \,\Psi_{i}^{CDCC(+)}>\Big |_{r_{nA} \leq R_{nA}}. 
\label{intCDCCpostres1}
\end{align}
Correspondingly, the external amplitude is 
\begin{align}
&M_{ext}^{CDCC(post)}(P,\,{\rm {\bf k}}_{dA})     \nonumber\\
&= <\chi_{pF}^{(-)}\,\Psi _{bB}^{(ext)(-)}|\,\Delta\,{\overline V}_{pF}^{P_{pn}}| \Psi_{i}^{CDCC(+)}>\Big |_{r_{nA} > R_{nA}}.
\label{extCDCCpostres1}
\end{align}
Now we repeat the steps outlined in subsection \ref{postCDCC}.  Taking into account Eqs (\ref{extdltvcdcc1}),  (\ref{shreqcdccpsi1}) and (\ref{shreqphif11}) we arrive at
\begin{align}
M_{ext}^{CDCC(post)}(P,\,{\rm {\bf k}}_{dA}) \equiv
M_{S}^{CDCC(post)}(P,\,{\rm {\bf k}}_{dA})              \nonumber\\
= <\chi_{pF}^{(-)}\,\Psi _{bB}^{(ext)(-)}|{\overleftarrow T} - {\overrightarrow T}|\,\Psi_{i}^{CDCC(+)}>\Big |_{r_{nA} > R_{nA}},
\label{extCDCCpostres2}
\end{align}
where $T = T_{pF} + T_{nA}$. 
It is shown in Appendix \ref{CDCCres} that $M_{S}^{CDCC(post)}$ can be reduced to
\begin{align}
M_{S}^{CDCC(post)}(P,\,{\rm {\bf k}}_{dA}) = - M_{S_{R_{nA}}}^{CDCC(post)}(P,\,{\rm {\bf k}}_{dA})                         \nonumber\\
= <\chi_{pF}^{(-)}\,\Psi _{bB}^{(ext)(-)}
|{\overleftarrow T}_{nA} - {\overrightarrow T}_{nA}|\,\Psi_{i}^{CDCC(+)}>\Big |_{r_{nA} \leq R_{nA}}. 
\label{CDCCpostresTnA1}
\end{align}
This integral can be directly transformed into the surface integral with $r_{nA}= R_{nA}$ encircling the internal volume, 
while the integral over ${\rm {\bf r}}_{pF}$ is taken over all the coordinate space. 
Thus we have shown that the post CDCC amplitude for stripping to resonance states is given by the difference of two terms, internal post CDCC amplitude and the surface integral:
\begin{align}
M^{CDCC(post)}(P,\,{\rm {\bf k}}_{dA}) = M_{int}^{CDCC(post)}(P,\,{\rm {\bf k}}_{dA}) \nonumber\\
- M_{S_{R_{nA}}}^{CDCC(post)}({\rm {\bf k}}_{pF},\,{\rm {\bf k}}_{dA}). 
\label{postCDCCresintSRnA1}
\end{align}
The internal amplitude $M_{int}^{CDCC(post)}$ can be minimized by a proper choice of $U_{pF}$ and the channel radius $R_{nA}$, while the surface integral is dominant. If the channel radius is larger than the $n-A$ nuclear interaction radius the second term is parametrized in terms of the reduced width amplitude and the boundary condition at $r_{nA}= R_{nA}$. 
Thus we succeeded to parametrize the post form of the CDCC amplitude in terms of the $R$-matrix parameters. 
It is one of the main results of this paper.   
Eq. (\ref{postCDCCresintSRnA1}) is the most important result of this paper. Due to the absence of the external term, which is present in the DWBA and which causes the convergence issue, the convergence problem in the post CDCC approach is resolved: the integration in the surface matrix element is performed over the full coordinate space only over one coordinate ${\rm {\bf r}}_{pF}$  rather than over two coordinates, ${\rm {\bf r}}_{pF}$ and ${\rm {\bf r}}_{nA}$.  

Expression for $M_{int}^{CDCC(post)}$ for different cases can be obtained from 
Eq. (\ref{intdwpostres12}) by replacing the initial channel wave function $\varphi_{d}(r_{pn})\,\chi_{{\rm {\bf k}}_{dA}}^{(+)})({\rm {\bf r}}_{dA})$ by $\Psi_{i}^{CDCC(+)}({\rm {\bf r}}_{pF},\,{\rm {\bf r}}_{nA}) $:
\begin{widetext}
\begin{align}
&M_{int}^{CDCC(post)}(P,\,{\rm {\bf k}}_{dA}) = \frac{2\,\pi}{k_{bB}}\,\sqrt{\frac{k_{bB}}{\mu_{bB}}} \,
\sum\limits_{J_{F}\,M_{F}\,s'\,l\,l'\,m_{s'}\,m_{l}\,m_{l'}\,M_{n}}\,i^{l}\,\,<s\,m_{s}\,\,l\,m_{l}|J_{F}\,M_{F}>\,  \nonumber\\
& \times\,< s'\,m_{s'}\,l'\,m_{l'}\,|J_{F}\,M_{F} >\,<J_{n}\,M_{n}\,\,J_{A}\,M_{A}|s'\,m_{s'}>\,e^{-i\,\delta_{bB\,l}^{hs}}\,Y_{l\,m_{l}}^{*}(-{\rm {\bf {\hat k}}}_{bB}) 
 \sum\limits_{\nu,\tau = 1}^N [\Gamma_{\nu\,bB\,s\,l\,J_{F}}(E_{bB})]^{1/2}\,[{\bf A}^{ - 1} ]_{\nu \tau} \nonumber\\
&\times <\chi_{pF}^{(-)}\,I_{A\,\,s'\,l'\, J_{F}}^{F}(r_{nA})\,|\Delta\,{\overline V}_{pF}| \Psi_{i}^{CDCC(+)}({\rm {\bf r}}_{pF},\,{\rm {\bf r}}_{nA}) 
>\Big |_{r_{nA} \leq R_{nA}}.
\label{intpostCDCCres12}
\end{align}
\end{widetext}
Note that the CDCC wave function itself also depends on quantum numbers of $p-n$ and $d-A$ subsystems, which we don't specify here. It will be done in the following up paper where concrete calculations will be presented.
Natural Jacobian variables for $\Psi_{i}^{CDCC(+)}$ are ${\rm {\bf r}}_{dA}$ and ${\rm {\bf r}}_{pn}$, but we use here another
set of Jacobian variables, ${\rm {\bf r}}_{pF}$ and ${\rm {\bf r}}_{nA}$. 

To write down explicitly $M_{S_{R_{nA}}}^{CDCC(post)}({\rm {\bf k}}_{pF},\,{\rm {\bf k}}_{dA})$ in terms of the surface integral we can use Eq. (\ref{SSDWnAres11}) replacing the initial channel wave function by the CDCC one:
\begin{widetext}
\begin{align}
&M_{S}^{CDCC(post)}({\rm {\bf k}}_{pF},\,{\rm {\bf k}}_{dA})= - M_{S_{R_{nA}}}^{CDCC(post)}({\rm {\bf k}}_{pF},\,{\rm {\bf k}}_{dA}) \qquad\qquad\qquad 
\nonumber\\
&= \,\frac{R_{nA}^{2}}{2\,\mu_{nA}}\,\int\,{\rm d}\,{\rm {\bf r}}_{pF}\,\chi_{-{\rm {\bf k}}_{pF}}^{(+)}({\rm {\bf r}}_{pF})\,\int{\rm d}\,\Omega_{{\rm {\bf r}}_{nA}}\,{\hat r}_{nA}\, \left [\Upsilon_{nA}^{(ext)(-)*}\big({\overleftarrow {\rm {\bf \nabla }}}_{{\rm {\bf r}}_{nA}} - {\overrightarrow {\rm {\bf \nabla }}}_{{\rm {\bf r}}_{nA}}\big )\Psi_{i}^{CDCC(+)}({\rm {\bf r}}_{pF},\,{\rm {\bf r}}_{nA})  \,\right]\,\Big |_{r_{nA} < R_{nA}}  \nonumber\\
&= \frac{R_{nA}^{2}}{2\,\mu_{nA}}\,\int\,{\rm d}\,{\rm {\bf r}}_{pF}\,\chi_{-{\rm {\bf k}}_{pF}}^{(+)}({\rm {\bf r}}_{pF})\,\int{\rm d}\,\Omega_{{\rm {\bf r}}_{nA}}\,
\left [\Psi_{i}^{CDCC(+)}({\rm {\bf r}}_{pF},\,{\rm {\bf r}}_{nA})\,\frac{\partial\,\Upsilon_{nA}^{(ext)(-)*}}{{\partial {r_{nA}}}}\, - \,\Upsilon_{nA}^{(ext)(-)*}\,\frac{\partial\,\Psi_{i}^{CDCC(+)}({\rm {\bf r}}_{pF},\,{\rm {\bf r}}_{nA})}{{\partial {r_{nA}}}}\right]\,\Big |_{r_{nA} = R_{nA}}.
\label{SCDCCpostresTnA1}
\end{align}
\end{widetext} 
We can extend corresponding equations from subsection \ref{postDWBAres} by replacing the initial channel wave function by the CDCC one. In particular, for the nodiagonal transition in the resonant subreaction $c'\,s'\,l' \to c\,s\,l$, where $c=b + B$ and $c'=n + A$, we get from Eq (\ref{SDWnAresSm12})
\begin{widetext}
\begin{align}
&M_{S}^{CDCC(post)}({\rm {\bf k}}_{pF},\,{\rm {\bf k}}_{dA}) = -M_{S_{R_{nA}}}^{CDCC(post)}(P,\,{\rm {\bf k}}_{dA})  
= \,\pi\,\sqrt{\frac{2\,R_{nA}}{\mu_{bB}\,\mu_{nA}\,k_{bB}}}\,\sum\limits_{J_{F}\,M_{F}\,l\,l'\,s'\,m_{l}\,m_{l'}\,m_{s'}\,M_{n}}\,i^{l}        \nonumber\\
& \times < l\,m_{l}\,\,s\,{m_s}|J_{F}\,M_{F} >\,< l'\,m_{l'}\,\,s'\,m_{s'}|J_{F}\,M_{F} > \,<J_{n}\,M_{n}\,\,J_{A}\,M_{A}|s'\,m_{s'}>           
 \nonumber\\
&\times Y_{l\,m_{l}}^{*}(-{\rm {\bf {\hat k}}}_{bB})\,e^{-\,i\,\delta_{bB\,l}^{hs}}\,\sum\limits_{\nu,\tau = 1}^{N}\,[\Gamma_{\nu\,bB\,s\,l\,J_{F}}(E_{bB})]^{1/2}\,[{\bf A}^{ - 1} ]_{\nu \tau}\,\gamma_{\tau\,nA\,s'\,l'\,J}\,\int{\rm d}\,{\rm {\bf r}}_{pF}\,\chi_{- {\rm {\bf k}}_{pF} }^{(+)}({\rm {\bf r}}_{pF})\,\int\,{\rm d}\,\Omega_{{\rm {\bf r}}_{nA}}\,Y_{l'\,m_{l'} }({\rm {\bf {\hat r}}}_{nA})      \nonumber\\
&\times \Bigg[\Psi_{i}^{CDCC(+)}({\rm {\bf r}}_{pF},\,{\rm {\bf r}}_{nA})\,(B_{nA} -1) - R_{nA}\,\frac{ \partial\,\Psi_{i}^{CDCC(+)}({\rm {\bf r}}_{pF},\,{\rm {\bf r}}_{nA})}{\partial \,r_{nA}} \Bigg]\Big |_{r_{nA}= R_{nA}}.
\label{SDWnAreactrespostCDCC1}
\end{align}
\end{widetext}
Correspondingly, the surface integral for the diagonal transition $c\,s\,l \to c\,s\,l$ can be obtained from 
Eq. (\ref{SDWnAreselastic12}):
\begin{widetext}
\begin{align}
&M_{S}^{CDCC(post)}({\rm {\bf k}}_{pF},\,{\rm {\bf k}}_{dA}) = i\,\frac{\pi}{\,\mu_{nA}\,k_{nA}}\,\sum\limits_{J_{F}\,M_{F}\,l\,m_{l}\,m_{l''}\,m_{s''}\,M_{n}}\,i^{l}\, < l\,m_{l}\,\,s\,{m_s}|J_{F}\,M_{F} >\,< l\,m_{l''}\,\,s\,m_{s''}|J_{F}\,M_{F} >         \nonumber\\
& \times <J_{n}\,M_{n}\,\,J_{A}\,M_{A}|s\,m_{s''}>\, 
Y_{l\,m_{l}}^{*}(-{\rm {\bf {\hat k}}}_{nA})\,   \nonumber\\
&\times \Big[1 -  e^{-i\,2\,\delta_{nA\,l}^{hs}} \Big(1 + i\,\sum\limits_{\nu,\tau = 1}^{N}\,[\Gamma_{\nu\,nA\,s\,l\,J_{F}}(E_{nA})]^{1/2}\,[{\bf A}^{ - 1} ]_{\nu \tau}\,\Gamma_{\tau\,nA\,s\,l\,J_{F}}(E_{nA})]^{1/2} \Big) \Big]\,O_{l}(k_{nA},\,R_{nA})      \nonumber\\
&\times \,\int{\rm d}\,{\rm {\bf r}}_{pF}\,\chi_{- {\rm {\bf k}}_{pF} }^{(+)}({\rm {\bf r}}_{pF})\,\int\,{\rm d}\,\Omega_{{\rm {\bf r}}_{nA}}\,Y_{l\,m_{l''} }({\rm {\bf {\hat r}}}_{nA})\,\Big[ \Psi_{i}^{CDCC(+)}({\rm {\bf r}}_{pF},\,{\rm {\bf r}}_{nA})\,(B_{nA} -1) - R_{nA}\,\frac{ \partial\,\Psi_{i}^{CDCC(+)}({\rm {\bf r}}_{pF},\,{\rm {\bf r}}_{nA})}{\partial \,r_{nA}} \Big]\Big |_{r_{nA}= R_{nA}}.
\label{SnAreselastpostCDCC1}
\end{align}
\end{widetext}

Summing up two amplitudes $M_{int}^{CDCC(post)}(P,\,{\rm {\bf k}}_{dA})$ and  $M_{S}^{DW}({\rm {\bf k}}_{pF},\,{\rm {\bf k}}_{dA})= -M_{S_{R_{nA}}}^{DW}({\rm {\bf k}}_{pF},\,{\rm {\bf k}}_{dA}) $ we get the total post CDCC amplitude for the $(d,p)$ stripping. \\
(i) Resonant reaction $n + A \to b + B$, that is $c= b + B \not= c'=n + A$.
 The total post form of the CDCC deuteron stripping amplitude can be obtained from Eq. (\ref{DWpostresfinal11}):
\begin{widetext}
\begin{align}
&M^{CDCC(post)}(P,\,{\rm {\bf k}}_{dA})(P,\,{\rm {\bf k}}_{dA})
= 2\,\pi\,\sqrt{\frac{1}{\mu_{bB}\,k_{bB}}} \sum\limits_{J_{F}\,M_{F}\,s'\,l\,l'\,m_{s'}\,m_{l}\,m_{l'}\,M_{n}}\,i^{l}\,\,<s\,m_{s}\,\,l\,m_{l}|J_{F}\,M_{F}>\,< s'\,m_{s'}\,l'\,m_{l'}\,|J_{F}\,M_{F} >\,  \nonumber\\
& \times <J_{n}\,M_{n}\,\,J_{A}\,M_{A}|s'\,m_{s'}>\    
e^{-i\,\delta_{bB\,l}^{hs}}\,Y_{l\,m_{l}}^{*}(-{\rm {\bf {\hat k}}}_{bB}) 
 \sum\limits_{\nu,\tau = 1}^N [\Gamma_{\nu\,bB\,s\,l\,J_{F}}(E_{bB})]^{1/2}\,[{\bf A}^{ - 1} ]_{\nu \tau} \nonumber\\
&\times \Bigg\{ \,<\chi_{pF}^{(-)}\,I_{A\,\,s'\,l'\, J_{F}}^{F}(r_{nA})\,|\Delta\,{\overline V}_{pF}| \Psi_{i}^{CDCC(+)}({\rm {\bf r}}_{pF},\,{\rm {\bf r}}_{nA})>\Big |_{r_{nA} \leq R_{nA}}              \nonumber\\
& + \,\sqrt{\frac{R_{nA}}{2\,\mu_{nA}}}\,\,\,\gamma_{\tau\,nA\,s'\,l'\,J}\,\int{\rm d}\,{\rm {\bf r}}_{pF}\,\chi_{- {\rm {\bf k}}_{pF} }^{(+)}({\rm {\bf r}}_{pF})\,\int\,{\rm d}\,\Omega_{{\rm {\bf r}}_{nA}}\,Y_{l'\,m_{l'} }({\rm {\bf {\hat r}}}_{nA})      \nonumber\\
&\times \bigg[\Psi_{i}^{CDCC(+)}({\rm {\bf r}}_{pF},\,{\rm {\bf r}}_{nA})\,(B_{nA} -1) - R_{nA}\,\frac{ \partial\,\Psi_{i}^{CDCC(+)}({\rm {\bf r}}_{pF},\,{\rm {\bf r}}_{nA})}{\partial \,r_{nA}}\bigg]\Big |_{r_{nA}= R_{nA}} \Bigg \}.
\label{CDCCpostresfinal11}
\end{align}
\end{widetext}
Assuming in this equation $b=n$ and $B=A$, that is $c=c'$ but $l \not= l'$ and/or $s \not= s'$ we get the expression for the
post CDCC deuteron stripping for the non-diagonal transition in the resonant subprocess $(n + A)_{l\,s} \to F \to (n+ A)_{l'\,s'}$. 

(ii) Diagonal transition, $c=c',\,l=l',\,s=s'$. The total post form of the CDCC amplitude is 
\begin{widetext}
\begin{align}
&M^{CDCC(post)}(P,\,{\rm {\bf k}}_{dA})
= 2\,\pi\,\sum\limits_{J_{F}\,M_{F}\,l\,m_{s''}\,m_{l}\,m_{l''}\,M_{n}}\,i^{l}\,\,<s\,m_{s}\,\,l\,m_{l}|J_{F}\,M_{F}>\,< s\,m_{s''}\,l\,m_{l''}\,|J_{F}\,M_{F} >\,  \nonumber\\
& \times <J_{n}\,M_{n}\,\,J_{A}\,M_{A}|s\,m_{s''}>    
e^{-i\,\delta_{nA\,l}^{hs}}\,Y_{l\,m_{l}}^{*}(-{\rm {\bf {\hat k}}}_{nA}) \nonumber\\
&\times \Bigg\{ \,\sqrt{\frac{1}{\mu_{nA}\,k_{nA}}} \,\sum\limits_{\nu,\tau = 1}^N [\Gamma_{\nu\,nA\,s\,l\,J_{F}}(E_{nA})]^{1/2}\,[{\bf A}^{ - 1} ]_{\nu \tau}\,<\chi_{pF}^{(-)}\,I_{A\,\,s\,l\, J_{F}}^{F}(r_{nA})\,|\Delta\,{\overline V}_{pF}| \Psi_{i}^{CDCC(+)}({\rm {\bf r}}_{pF},\,{\rm {\bf r}}_{nA})>\Big |_{r_{nA} \leq R_{nA}}              \nonumber\\
& + \,i\,\Bigg[1 -  e^{-i\,2\,\delta_{nA\,l}^{hs}} \Big(1 + i\,\sum\limits_{\nu,\tau = 1}^{N}\,[\Gamma_{\nu\,nA\,s\,l\,J_{F}}(E_{nA})]^{1/2}\,[{\bf A}^{ - 1} ]_{\nu \tau}\,\Gamma_{\tau\,nA\,s\,l\,J_{F}}(E_{nA})]^{1/2} \Big) \Bigg]                              \nonumber\\
& \times\,\frac{1}{2\,\mu_{nA}\,k_{nA}}\,\int{\rm d}\,{\rm {\bf r}}_{pF}\,\chi_{- {\rm {\bf k}}_{pF} }^{(+)}({\rm {\bf r}}_{pF})\,\int\,{\rm d}\,\Omega_{{\rm {\bf r}}_{nA}}\,Y_{l\,m_{l''} }({\rm {\bf {\hat r}}}_{nA})      \nonumber\\
&\times \Big[\Psi_{i}^{CDCC(+)}({\rm {\bf r}}_{pF},\,{\rm {\bf r}}_{nA})\,(B_{nA} -1) - R_{nA}\,\frac{ \partial\,\Psi_{i}^{CDCC(+)}({\rm {\bf r}}_{pF},\,{\rm {\bf r}}_{nA})}{\partial \,r_{nA}} \Big]\Big |_{r_{nA}= R_{nA}} \Bigg\}.
\label{CDCCpostresfinalel11}
\end{align}
\end{widetext} 

Eqs (\ref{CDCCpostresfinal11}) and (\ref{CDCCpostresfinalel11}) are the final and main result of this paper.  Both matrix elements consist of only terms, the internal post CDCC and the surface term. The internal term contains the integration over ${\rm {\bf r}}_{nA}$ in the internal volume $r_{nA} \leq R_{nA}$. Hence, at $r_{pF} \to \infty$  variables $r_{dA} \sim r_{pF} \to \infty$  and $r_{pn} \sim r_{pF} \to \infty$. But then $\Psi_{i}^{CDCC(+)}({\rm {\bf r}}_{pF},\,{\rm {\bf r}}_{nA}) \sim r_{pF}^{-3}$ \cite{sawada} the integral over ${\rm {\bf r}}_{pF}$ does converge. The same conclusion is true for the surface integral in which $r_{nA}=R_{nA}$. Hence, in this matrix element also $\Psi_{i}^{CDCC(+)}({\rm {\bf r}}_{pF},\,{\rm {\bf r}}_{nA}) \sim r_{pF}^{-3}$ and integral over ${\rm {\bf r}}_{pF}$ converges. Both amplitudes are parametrized 
in terms of the parameters used in the conventional $R$-matrix approach and providea tool to analyze the stripping into resonance states using generalized $R$-matrix approach. Finally, both amplitudes, (\ref{CDCCpostresfinal11}) and (\ref{CDCCpostresfinalel11}), don't have penetration factor in the entry channel $n + A$ of the resonance formation in the resonant subreactions $n + A \to b + B$   and $n + A \to n + A$. That is why stripping to resonantstates provides a powerful tool to measure resonances in the subsystem $n+ A$ very close to the threshold, which can be suppressed in the on-shell binary resonant reaction but not in the stripping to resonance states.

\section{Summary}
The theory of the deuteron stripping populating bound and resonance states based on the surface integral formalism is presented. To demonstrate the theory I first develop it for the DWBA. Since the DWBA is outdated and, definitely, deficient compared to the CDCC,
the theory is extended to the CDCC formalism. The theory is applied for stripping to bound and resonance states. The eventual goal of this paper is to deliver the theory of the deuteron stripping to resonance states within the CDCC formalism using the surface integral formulation of the reaction theory \cite{kadyrov09}.  Transformation of the volume matrix element to the surface one (in the subspace over ${\rm {\bf r}}_{nA}$)
and $R$-matrix representation of the scattering wave function of the fragments formed by the resonance decay allows one to parametrize the reaction amplitude in terms of the $R$-matrix parameters used in the analysis of the binary resonant reactions. Since the reaction under consideration is the deuteron stripping, the presence of the deuteron in the initial state and exiting proton causes the distortions. That is why the reaction amplitude, in addition to the $R$-matrix parameters describing the binary subprocess, contains additional factors - CDCC wave function describing the $d-A$ scattering
in the initial channel (coupled to the deuteron breakup channel) and the proton distorted wave in the final state. Hence, the
approach can be called a generalized $R$ matrix for the stripping to resonance states. The advantage of the approach is that the reaction amplitude for stripping to resonance states in the post CDCC formalism doesn't have convergence problem
and is parametrized in terms of the same observables as binary resonant reactions. Hence, the formalism provides experimentalists a consistent tool to analyze binary resonant reactions and stripping reactions populating resonant states 
extracting the same observable parameters, namely, reduced widths (ANCs). The power of the method has been demonstrated in the analysis of the Trojan Horse reaction ${}^{19}{\rm F}(d,\,n\,\alpha){}^{16}{\rm O}$ \cite{lacognata}. The numerical application of the method will be demonstrated in the following up papers.

\acknowledgments
The work was supported by the US Department of Energy under Grant Nos. DE-FG02-93ER40773, DE-FG52-06NA26207 and DE-SC0004958
(topical collaboration TORUS), NSF under Grant No. PHY-0852653. The author express his thanks to A. S. Kadyrov and the members of the TORUS collaboration, I. Thompson, F. M. Nunes, J. Escher, C. Elster and G. Arbanas.

\appendix
\section{$\bold{b+ B}$ scattering wave function $\bold{\Psi_{bB}^{(+)}}$}
\label{psibB1}

In this Appendix we consider the representation of the scattering $\Psi_{bB}^{(+)}$ wave function used in the $R$-matrix approach for binary resonance processes \cite{lanethomas58,blattbiedenharn}

\subsection{Internal scattering wave function $\bold{\Psi_{bB}^{(+)}}$}

A general equation for the internal wave function contains the sum over total angular momentum $J_{F}$ and its projection $M_{F}$. Since we are interested in a wave function $\Psi_{bB}^{(+)}$ describing a resonance in the system $F=b+B$,
we consider only the internal wave function at given $J_{F}$, at which resonance occurs.   
In the internal region in the state with the total momentum $J_{F}$, channel spin $s$ (its projection $m_{s}$) in the initial channel $c=b+B$  the wave function $\Psi_{bB}^{(+)}$ can be written as 
\cite{lanethomas58} 
\begin{align}
&\Psi _{c\,s\,m_{s}}^{J_{F}(int)(+)}  = \,\frac{2\,\pi}{k_{c}}\,\sqrt{\frac{k_{c}}{\mu_{c}}} \,\sum\limits_{M\,l\,m_{l}}\,
e^{-i\,\delta_{c\,l}^{hs}}\,i^{l}\,<s\,m_{s}\,\,l\,m_{l}|J_{F}\,M_{F}>  \nonumber\\
&\times Y_{l\,m_{l}}^{*}({\rm {\bf {\hat k}}}_{c})\,\sum\limits_{\nu,\tau = 1}^N [\Gamma_{\nu\,c\,s\,l\,J_{F}}(E_{c})]^{1/2}\,[{\bf A}^{ - 1} ]_{\nu \tau}\,X_{\tau\,}^{J_{F}M_{F}}. 
\label{PsibBresint1}
\end{align}
Here, $\,E_{c}=E_{bB}$ and ${\rm {\bf k}}_{c}={\rm {\bf k}}_{bB}$ are the relative energy and momentum of particles $b$ and $B$, $\,\mu_{c}= \mu_{bB}$, $\, \Gamma_{\nu\,c}(E_{c})$ is the formal ($R$-matrix) partial resonance width of the level $\nu$ in the channel $\,c=b + B$, $\,{\bf A}$ is the $R$-matrix level matrix, $\,N$ is the number of the levels included, $\,\sigma_{c\,l}$ is the Coulomb scattering phase shift in the channel $c$ and the partial wave $l$,
 $\,\delta_{c\,l}^{hs}$ is the hard-sphere scattering phase shift in the channel $c$ given by 
\begin{equation}
\delta_{c\,l}^{hs}= -\omega_{c\,l} +  \arctan\frac{F_{l}(k_{c},\,R_{c})}{G_{l}(k_{c},\,R_{c})},
\label{deltacss1}
\end{equation}
where $F_{l}(k_{c},\,r_{c})$ and $G_{l}(k_{c},\,r_{c})$ are regular and singular Coulomb solutions of the radial Schr\"odinger 
equation, 
\begin{align}
\omega_{l}= \sigma_{c\,l} - \sigma_{c\,0}= \sum\limits_{n=1}^{l}\,\arctan{\frac{\eta_{c}}{n}},
\label{omega1}
\end{align}
$\sigma_{c\,l}$ is the Coulomb scattering phase shift in the $l$-partial wave, 
$\,\eta_{c}$ is the Coulomb 
parameter for the scattering of the fragments in the channel $c$.

We consider only two coupled channels $c=b+ B$ and $c'=n + A$. Also $\,X_{\tau}^{J_{F}\,M_{F}}$ is an eigenfunction of the Hamiltonian describing the compound system $F=n+ A= b+ B$ in the internal region excited to the discrete level $\tau$  with the total angular momentum $J_{F}$ and its projection $M_{F}$ \footnote[2]{It is shown in \cite{mahaux} how to calculate $X_{\tau}^{J_{F}\,M_{F}}$ in the shell-model approach.}. 
A separable form for $\Psi _{c\,s\,m_{s}}^{J_{F}(int)}$ reflects the fact that we consider the $b+ B$ interaction proceeding through resonance states.
The entry channel of this scattering is the channel $c=b + B$. The inverse level matrix contains contribution from all $N$ resonance levels. In a simple one level case it reduces to the well-known Breit-Wigner resonance propagator. All the open channels coupled to $c$ contribute to $X_{\tau}^{J_{F}\,M_{F}}$ and determine possible exit channel contributions into resonance scattering. Hence, in the internal region, where different open channels are coupled, $X_{\tau}^{J_{F}\,M_{F}}$ can be written as a nonorthogonal sum of these channels \cite{lanethomas58}:
\begin{align}
X_{\tau}^{J_{F}\,M_{F}} = \sum\limits_{{\tilde c}\,{\tilde s}\,{\tilde l}\,m_{\tilde s}\,j}\,\frac{1}{r_{\tilde c}}\,w_{\tau\,{\tilde c}\,j}\,{\hat A}\{\xi_{{\tilde c}}\,\phi_{{\tilde c}\,{\tilde s}\,{\tilde l}\,m_{\tilde s}}^{J_{F}\,M_{F}}\,u_{{\tilde c}\,{\tilde s}\,{\tilde l}\,J_{F}\,j}\}, 
\label{XtauJM1}
\end{align}
where $\xi_{\tilde c}$ is the product of the internal bound state wave functions of the fragments in the channel ${\tilde c}$, $\,{\tilde c}=c,\,c'$, $\,\,u_{{\tilde c}\,{\tilde s}\,{\tilde l}\,J_{F}\,j}(r_{{\tilde c}})$ is a set of the radial wave functions of the relative motion of the fragments in the channel ${\tilde c}$ with the channel spin ${\tilde s}$, orbital angular momentum ${\tilde l}$ and total angular momentum $J_{F}$ in some adopted potential, $\,\phi_{{\tilde c}\,{\tilde s}\,{\tilde l}\,m_{\tilde s}}^{J_{F}\,M_{F}}$, where $m_{\tilde s}$ is the projection of ${\tilde s}$, is the channel wave function (in $LS$-coupling). Also ${\hat A}$ is the antisymmetrization operator between the nucleons of the fragments in the channel ${\tilde c}$. We consider only two coupled channels, $c= b + B$ and $c'= n + A$. Thus the initial channel $c$ can propagate into two final channels $c$ and $c'$ via the intermediate resonances. 
Although Eq. (\ref{XtauJM1}) contains the sum over all channel spins ${\tilde s}$ and projections in each open channel, in what follows consider the contribution to  $X_{\tau}^{J_{F}\,M_{F}} $ only from the channel with fixed channel spin and its projection.

First, let us consider the contribution of the channel $c\,s''\,m_{s''}$ into $X_{\tau}^{J_{F}\,M_{F}}$. In this channel $\xi_{c}=\varphi_{b}\,\varphi_{B}$ and 
\begin{align}
&\phi_{c\,s''\,l''\,m_{s''}}^{J_{F}\,M_{F}}= \,\sum\limits_{m_{l''}}\,< s''\,m_{s''}\,\,l''\,
m_{l''}|\,J_{F}\,M_{F} >  \nonumber\\
&\times Y_{l''\,m_{l''}}({\rm {\bf \widehat r}}_{c})\,\phi_{c\,s''\,m_{s''}},
\label{chancwavef1}
\end{align}
\begin{align}
&\phi_{c\,s''\,m_{s''}} = \sum\limits_{M_{b}\,M_{B}}\,<J_{b}\,M_{b}\,\,J_{B}\,M_{B}|s''\,m_{s''}>                                \nonumber\\
& \times \psi_{J_{b}\, M_{b}}\,\psi_{J_{B}\, M_{B}}.
\label{spincwavefunct1}
\end{align}
Here, $\phi _{c\, s''\,m_{s''}}$ is the channel spin wave function in the channel $c\,s''\,m_{s''}$, $\,\,\psi_{J_{i}\, M_{i}}$ is the spin wave function of particle $i$, $\,l''\,$ ($m_{l''}$) is the relative orbital angular momentum (its projection) of the fragments in the channel $c$, $\,{\rm {\bf r}}_{c}= {\rm {\bf r}}_{bB}$ is the radius-vector connecting $b$ and the center-of-mass of $B$.  We adopt the channel radius $R_{c}$ large enough to neglect antisymmetrization between nucleons of $b$ and  $B$ at $r_{c}= R_{c}$, that is 
\begin{align}
{\hat A}\{w_{\tau\,c\,j}\,\xi_{c}\phi_{c\,s''\,l''\,m_{s''}}^{J_{F}\,M_{F}}\,u_{c\,s''\,l''\,J_{F}\,j}\}\Big |_{r_{c}= R_{c}}  \nonumber\\
\approx N_{c}\,\xi_{c}\,\phi_{c\,s''\,l''}^{J_{F}\,M_{F}}\,u_{c\,s''\,l''\,J_{F}\,j}\Big |_{r_{c} = R_{c}},
\label{periphapprox1}
\end{align}
where $N_{c}= {\left( {\frac{{(b + B)!}}{{b!}{B!}}} \right)^{-1/2}}$.   

Assuming that the overlap of the channel $c$ at the channel radius $R_{c}$ with the channel $c'$ is negligible
we get for the component of $X_{\tau\,c\,s''\,m_{s''}}^{J_{F}\,M_{F}}$ projected on $\xi_{c}=\varphi_{b}\,\varphi_{B}$  at $r_{c}= R_{c}$ \cite{lanethomas58}
\begin{align}
&\Xi_{\tau\,c\,s''\,m_{s''}}^{J_{F}\,M_{F}}(R_{c}\,{\rm {\bf {\hat r}}}_{c}) =  <\xi_{c}|X_{\tau\,s''\,m_{s''} }^{J_{F}\,M_{F}}>\Big |_{r_{c}= R_{c} }  \nonumber\\
&= \frac{1}{R_{c}}\,\sum\limits_{l''}\,\phi_{c\,s''\,l''\,m_{s''}}^{J_{F}\,M_{F}}\,u_{\tau\,c\,s'',\,l''\,J_{F}}(R_{c}),               \nonumber\\
\end{align} 
where
\begin{align}
u_{\tau\,c\,s'',\,l''\,J_{F}}(r_{c}) = N_{c}\,\sum\limits_{j}\,w_{\tau\,c\,j}\,u_{c\,s'',\,l''\,J_{F}\,j}(r_{c}).
\label{ucslJl1}
\end{align}
At $r_{c}= R_{c}$ by definition \cite{lanethomas58}
\begin{align}
u_{\tau\,c\,s''\,l''\,J_{F}}(R_{c}) = \sqrt{2\,\mu_{c}\,R_{c}}\,\gamma_{\tau\,c\,s''\,l''\,J_{F}},
\label{uclredwidthampl1}
\end{align}
where $\,\,\gamma_{\tau\,c\,s''\,l''\,J_{F}}$ is the reduced width amplitude of the level $\tau$ in the channel $c\,s''\,l''\,J_{F}$.  I remind that the system of units ${\hbar}=c=1$ is being used throughout the paper if not specified otherwise.  Then 
\begin{align}
&\Xi_{\tau\,c\,s''\,m_{s''}}^{J_{F}\,M_{F}}(R_{c}\,{\rm {\bf {\hat r}}}_{c})= \frac{1}{R_{c}}\,\sum\limits_{l''}\,
\sqrt{2\,\mu_{c}\,R_{c}}\,\gamma_{\tau\,c\,s''\,l''\,J_{F}}    \nonumber\\
&\times \phi_{c\,s''\,l''\,m_{s''}}^{J_{F}\,M_{F}}. 
\label{xtauredwidthaml1}
\end{align}

Thus we can express the component $\Xi_{\tau\,c\,s''\,m_{s''}}^{J_{F}\,M_{F}}({\rm {\bf r}}_{c})$ taken at the channel radius $r_{c}= R_{C}$ 
in terms of the sum of the reduced width amplitudes, where the sum is taken over all allowed in the channel $c$ partial waves $l''$ at given $J_{F}$ and $s''$.
Then the component of $\Psi _{c\,s\,m_{s}}^{J_{F}(int)}$ in the exit channel $c\,s''\,m_{s''}$ projected onto $\xi_{c}= \varphi_{b}\,\varphi_{B}$ at $r_{c}= R_{c}$ takes the form
\begin{widetext} 
\begin{align}
&  \Upsilon _{c\,s\,m_{s}; c\,s''\,m_{s''}}^{J_{F}(int)(+)}(R_{c}\,{\rm {\bf {\hat r}}}_{c}) =  <\xi_{c}|\Psi _{c\,s\,m_{s}}^{J_{F}(int)(+)}>  \nonumber\\  
&= \,\frac{2\,\pi}{k_{c}\,R_{c}}\,\sqrt{\frac{k_{c}}{\mu_{c}}}\,\sum\limits_{M,\,l\,m_{l}}\,\,e^{-i\,\delta_{c\,l}^{hs}}\,i^{l}\,<s\,m_{s}\,\,l\,m_{l}|J_{F}\,M_{F}>\,Y_{l\,m_{l}}^{*}({\rm {\bf {\hat k}}}_{c})\,\sum\limits_{\nu,\tau = 1}^N [\Gamma_{\nu\,c\,s\,l\,J_{F}}(E_{c})]^{1/2}\,[{\bf A}^{ - 1} ]_{\nu \tau}\,\Xi_{\tau\,c\,s''\,m_{s''}}^{J_{F}\,M_{F}}(R_{c}\,{\rm {\bf {\hat r}}}_{c}) \nonumber\\
&= \,\frac{2\,\pi}{k_{c}\,R_{c}}\,\sqrt{\frac{k_{c}}{\mu_{c}}}\,\sum\limits_{M,\,l\,l''\,m_{l}\,m_{l''}}\,e^{-i\,\delta_{c\,l}^{hs}}\,i^{l}\,<s\,m_{s}\,\,l\,m_{l}|J_{F}\,M_{F}>\,< s''\,m_{s''}\,\,l''\,m_{l''}|\,J_{F}\,M_{F} >\,Y_{l\,m_{l}}^{*}({\rm {\bf {\hat k}}}_{c})  \nonumber\\
& \times \sum\limits_{\nu,\tau = 1}^N [\Gamma_{\nu\,c\,s\,l\,J_{F}}(E_{c})]^{1/2}\,[{\bf A}^{ - 1} ]_{\nu \tau}\sqrt{2\,\mu_{c}\,R_{c}}\,\gamma_{\tau\,c\,s''\,l''\,J_{F}}\,Y_{l''\,m_{l''}}({\rm {\bf \widehat r}}_{c})\,\phi_{c\,s''\,m_{s''}}.
\label{projcPsibBresint1}
\end{align}
\end{widetext}
Here, $s''$ is any channel spin value in the channel $c=b + B$ allowed by the spin and angular momentum conservation law.
In particular, $s''$ may coincide with $s$, that is $s''=s$.  

The diagonal component, $l''=l$ and $s''=s$, which is needed to determine the elastic scattering amplitude (see below) is 
\begin{widetext} 
\begin{align}
& \Upsilon _{c\,s\,l\,m_{s}; c\,s\,l\,m_{s''}}^{J_{F}(int)(+)}(R_{c}\,{\rm {\bf {\hat r}}}_{c}) = \,\frac{2\,\pi}{k_{c}\,R_{c}}\,\sqrt{\frac{k_{c}}{\mu_{c}}}\,e^{-i\,\delta_{c\,l}^{hs}}\,i^{l}\,\sum\limits_{M\,m_{l}\,m_{l''}}\,<s\,m_{s}\,\,l\,m_{l}|J_{F}\,M_{F}>\,< s\,m_{s''}\,\,l\,m_{l''}|\,J_{F}\,M_{F} >\,Y_{l\,m_{l}}^{*}({\rm {\bf {\hat k}}}_{c}) \nonumber\\
&\times \sum\limits_{\nu,\tau = 1}^N [\Gamma_{\nu\,c\,s\,l\,J_{F}}(E_{c})]^{1/2}\,[{\bf A}^{ - 1} ]_{\nu \tau}\,\sqrt{2\,\mu_{c}\,R_{c}}\,\gamma_{\tau\,c\,s\,l\,J_{F}}\,Y_{l\,m_{l''}}({\rm {\bf \widehat r}}_{c})\,\phi_{c\,s\,m_{s''}}.
\label{projcdiagint1}
\end{align}
\end{widetext}

A similar consideration can be applied when we consider the contribution of the channel $c'\,s'\,m_{s'}$ into $X_{\tau}^{J_{F}\,M_{F}}$.
In this channel $\,\xi_{c'}= \varphi_{A}$ and 
\begin{align}
&\phi_{c'\,s'\,l'\,m_{s'}}^{J_{F}\,M_{F}}=\,\sum\limits_{m_{l'}}\,< s'\,m_{s'}\,\,l'\,
m_{l'}|\,J_{F}\,M_{F} >  \nonumber\\
&\times Y_{l'\,m_{l'}}({\rm {\bf \widehat r}}_{c'})\,\phi_{c'\,s'\,m_{s'}},
\label{chanwavef1}
\end{align}
\begin{align}
&\phi_{c'\,s'\,m_{s'}} = \sum\limits_{M_{n}\,M_{A}}\,<J_{n}\,M_{n}\,\,J_{A}\,M_{A}|s'\,m_{s'}>                                \nonumber\\
& \times \psi_{J_{n}\, M_{n}}\,\psi_{J_{A}\, M_{A}}.
\label{spinwavefunct1}
\end{align}
Here, $\phi _{c'\, s'\,m_{s'}}$ is the channel spin wave function in the channel $c'$ with the channel spin $s'$ and its projection $m_{s'}$, $\,\,l'\,$ ($m_{l'}$) is the relative orbital angular momentum (its projection) of the fragments in the channel $c'$, ${\rm {\bf r}}_{c'}= {\rm {\bf r}}_{nA}$ is the radius-vector connecting $n$ and the center-of-mass of $A$.
We adopt the channel radius $R_{c'}$ large enough to neglect antisymmetrization between $n$ and nucleons of $A$
at $r_{c'}= R_{c'}$, that is
\begin{align}
{\hat A}\{w_{\tau\,c'\,j}\,\xi_{c'}\phi_{c'\,s'\,l'\,m_{s'}}^{J_{F}\,M_{F}}\,u_{c'\,s'\,l'\,J_{F}\,j}\}\Big |_{r_{c'}= R_{c'}}  \nonumber\\
\approx N_{c'}\,\xi_{c'}\,\phi_{c'\,s'\,l'}^{J_{F}\,M_{F}}\,u_{\tau\,c'\,s'\,l'\,J_{F}\,j}\Big |_{r_{c'} = R_{c'}},
\label{periphapprox1}
\end{align}
where $N_{c'}= {\left( {\frac{{(A + 1)!}}{{A!}}} \right)^{-1/2}}= (A + 1)^{-1/2}$.   

Assuming that the overlap of the channel $c'$ at the channel radius $R_{c'}$ with the channel $c$ is negligible
we get for the component of $X_{\tau\,c'\,s'\,m_{s'}}^{J_{F}\,M_{F}}$  projected onto $\xi_{c'}=\varphi_{A}$  at $r_{c'}= R_{c'}$ 
\begin{align}
&\Xi_{\tau\,c'\,s'\,m_{s'}}^{J_{F}\,M_{F}}(R_{c'}\,{\rm {\bf {\hat r}}}_{c'}) =  <\varphi_{A}|X_{\tau}^{J_{F}\,M_{F}}>\Big |_{r_{c'}= R_{c'} }  \nonumber\\
&=  \frac{1}{R_{c'}}\sum\limits_{l'}\,\phi_{c'\,s'\,l'\,m_{s'}}^{J_{F}\,M_{F}}\,u_{\tau\,c'\,s'\,l'\,J_{F}}(R_{c'}),               \nonumber\\
\end{align} 
where
\begin{align}
u_{\tau\,c'\,s'\,l'\,J_{F}}(r_{c'}) = N_{c'}\,\sum\limits_{j}\,w_{\tau\,c'\,j}\,u_{c'\,s'\,l'\,J_{F}\,j}(r_{c'}).
\label{ucslJl1}
\end{align}
At
 $r_{c'}= R_{c'}$ 
\begin{align}
u_{\tau\,c'\,s'\,l'\,J_{F}}(R_{c'}) = \sqrt{2\,\mu_{c'}\,R_{c'}}\,\gamma_{\tau\,c'\,s'\,l'\,J_{F}},
\label{uclredwidthampl1}
\end{align}
where $\mu_{c'}= \mu_{nA}$, $\,\,\gamma_{\tau\,c'\,s'\,l'\,J_{F}}$ is the reduced width amplitude of the level $\tau$ in the channel $c'\,s'\,l'\,J_{F}$.  
Then 
\begin{align}
&\Xi_{\tau\,c'\,s'\,m_{s'}}^{J_{F}\,M_{F}}(R_{c'}\,{\rm {\bf {\hat r}}}_{c'})  
=  \frac{1}{R_{c'}}\,\sum\limits_{l'}\,\sqrt{2\,\mu_{c'}\,R_{c'}}\,\gamma_{\tau\,c'\,s'\,l'\,J_{F}} \nonumber\\
&\times\phi_{c'\,s'\,l'\,m_{s'}}^{J_{F}\,M_{F}}, 
\label{xtauredwidthaml1}
\end{align}
that is it can be expressed in terms of the sum of the reduced widths amplitudes in all allowed partial waves $l'$ in the channel $c'$ at given $J_{F}$ and $s'$. Then the component $\Psi _{c\,s\,m_{s};\,c'\,s'\,m_{s'}}^{J_{F}(int)(+)}$ projected on $\xi_{c'}=\varphi_{A}$ at $r_{c'}= R_{c'}$ takes the form
\begin{widetext} 
\begin{align}
&\Upsilon_{c\,s\,m_{s};\,c'\,s'\,m_{s'}}^{J_{F}(int)(+)}(R_{c'}\,{\rm {\bf {\hat r}}}_{c'})   \nonumber\\  
&= \,\frac{2\,\pi}{k_{c}}\,\sqrt{\frac{k_{c}}{\mu_{c}}}\,\sum\limits_{M,\,l\,m_{l}}\,e^{-i\,\delta_{c\,l}^{hs}}\,i^{l}\,<s\,m_{s}\,\,l\,m_{l}|J_{F}\,M_{F}>\,Y_{l\,m_{l}}^{*}({\rm {\bf {\hat k}}}_{c})\,\sum\limits_{\nu,\tau = 1}^N [\Gamma_{\nu\,c\,s\,l\,J_{F}}(E_{c})]^{1/2}\,[{\bf A}^{ - 1} ]_{\nu \tau}\,\Xi_{\tau c'\,s'\,m_{s'}}^{J_{F}\,M_{F}}(R_{c'}\,{\rm {\bf {\hat r}}}_{c'}) \nonumber\\
&= \,\frac{2\,\pi}{k_{c}\,R_{c'}}\,\sqrt{\frac{k_{c}}{\mu_{c}}}\,\sum\limits_{M\,l\,l'\,m_{l}\,m_{l'}}\,e^{-i\,\delta_{c\,l}^{hs}}\,i^{l}\,<s\,m_{s}\,\,l\,m_{l}|J_{F}\,M_{F}>\,< s'\,m_{s'}\,\,l'\,m_{l'}|\,J_{F}\,M_{F} >\,\nonumber\\
&\times \,Y_{l\,m_{l}}^{*}({\rm {\bf {\hat k}}}_{c})\sum\limits_{\nu,\tau = 1}^N [\Gamma_{\nu\,c\,s\,l\,J_{F}}(E_{c})]^{1/2}\,[{\bf A}^{ - 1} ]_{\nu \tau}\,\sqrt{2\,\mu_{c'}\,R_{c'}}\,\gamma_{\tau\,c'\,s'\,l'\,J_{F}}\,  
Y_{l'\,m_{l'}}({\rm {\bf \widehat r}}_{c'})\,\phi_{c'\,s'\,m_{s'}}.
\label{projcnAresint1}
\end{align}
\end{widetext}

The component $\Upsilon_{c\,s\,l\,m_{s};c'\,s'\,l'\,m_{s'}}^{J_{F}(int)(+)}(R_{c'}\,{\rm {\bf {\hat r}}}_{c'})$ is given by
\begin{widetext}
\begin{align}
&\Upsilon_{c\,s\,l\,m_{s};\,c'\,s'\,l'\,m_{s'}}^{J_{F}(int)(+)}(R_{c'}\,{\rm {\bf {\hat r}}}_{c'})= \,\frac{2\,\pi}{k_{c}\,R_{c'}}\,\sqrt{\frac{k_{c}}{\mu_{c}}}\,e^{-i\,\delta_{c\,l}^{hs}}\,i^{l}\,\sum\limits_{M\,m_{l}\,m_{l'}}\,<s\,m_{s}\,\,l\,m_{l}|J_{F}\,M_{F}>\,< s'\,m_{s'}\,\,l'\,m_{l'}|\,J_{F}\,M_{F} > \nonumber\\
&\times Y_{l\,m_{l}}^{*}({\rm {\bf {\hat k}}}_{c})\sum\limits_{\nu,\tau = 1}^N [\Gamma_{\nu\,c\,s\,l\,J_{F}}(E_{c})]^{1/2}\,[{\bf A}^{ - 1} ]_{\nu \tau}\,\sqrt{2\,\mu_{c'}\,R_{c'}}\,\gamma_{\tau\,c'\,s'\,l'\,J_{F}}\,Y_{l'\,m_{l'}}({\rm {\bf \widehat r}}_{c'})\,\phi_{c'\,s'\,m_{s'}}.
\label{projcnAresint2}
\end{align}
\end{widetext}

\subsection{External scattering wave function $\bold{\Psi_{bB}^{(+)}}$ }

Now we proceed to the expression for the $\Psi_{c}^{(+)}$ in the external region, where $r_{c} > R_{c}$ or $r_{c'} > R_{c'}$. 
In the external region the wave function $\Psi_{c\,s\,m_{s}}^{(ext)(+)}$ with fixed channel spin and its projection in the incident channel $c$ can be written as 
\begin{align}
\Psi_{c\,s\,m_{s}}^{(ext)(+)} = \Psi_{c\,s\,m_{s}}^{(0)} + \Psi_{c\,s\,m_{s};r}^{(ext)(+)},
\label{psicincreact1}
\end{align}
where the first term is the incident wave and the second term  is the sum of the outgoing waves in all the open channels. 
The incident term is
\begin{align}
&\Psi_{c\,s\,m_{s}}^{(ext)(0)} = \,4\,\pi\,\xi_{c}\,\sum\limits_{J_{F}\,M_{F}}\,\sum\limits_{l\,m_{l}\,m_{s''}}\,i^{l}\,<s\,m_{s}\,\,l\,m_{l}\,|J_{F}\,M_{F}>\,   \nonumber\\
&\times  <s\,m_{s''}\,\,l\,m_{l}|J_{F}\,M_{F}>\,Y_{l\,m_{l}}^{*}({\rm {\bf {\hat k}}}_{c})\,e^{i\,\omega_{c\,l}}\,\frac{ F_{l}(k_{c},\,r_{c}) }{k_{c}\,r_{c}}\,                                  \nonumber\\
&\times Y_{l\,m_{l}}({\hat {\rm {\bf r}}}_{c})\,\phi_{c\,s\,m_{s''}},
\label{psicext01}
\end{align}
where the subscript $c$ means that the incident wave is in the channel $c$.  
The sum over $m_{s''}$ is a formal because 
\begin{align}
&\sum\limits_{J_{F}\,M_{F}}<s\,m_{s}\,\,l\,m_{l}\,|J_{F}\,M_{F}>\,<s\,m_{s''}\,\,l\,m_{l}|J_{F}\,M_{F}>  \nonumber\\
& = \delta_{m_{s}\,m_{s''}}.
\label{clebschgordanorth1}
\end{align}
Note that here we use the incident wave with the unit amplitude rather than with the unit flux density.
The component $\Psi_{c\,s\,l\,m_{s};c\,s\,l\,m_{s''}}^{J_{F}(ext)(0)}$, which corresponds to the exit channel $c\,s\,l\,m_{s''}$ and fixed $J_{F}$, projected on $\xi_{c}$ reduces to
\begin{align}
&\Upsilon_{c\,s\,l\,m_{s};\,c\,s\,l\,m_{s''}}^{J_{F}(ext)(0)}({\rm {\bf r}}_{c}) = \,4\,\pi\,\sum\limits_{M\,m_{l}}\,i^{l} \nonumber\\ 
&\times <s\,m_{s}\,\,l\,m_{l}\,|J_{F}\,M_{F}>\, <s\,m_{s''}\,\,l\,m_{l}|J_{F}\,M_{F}>\,Y_{l\,m_{l}}^{*}({\rm {\bf {\hat k}}}_{c})   \nonumber\\
&\times e^{i\,\omega_{c\,l}}\,\frac{ F_{l}(k_{c},\,r_{c}) }{k_{c}\,r_{c}}\,Y_{l\,m_{l}}({\hat {\rm {\bf r}}}_{c})\,\phi_{c\,s\,m_{s''}}.
\label{Upsilonincext01}
\end{align}
Now we take into account that
\begin{align}
F_{l}(k_{c},\,r_{c})= \frac{e^{i\,\omega_{c\,l}}\,O_{l}(k_{c},\,r_{c}) - e^{-i\,\omega_{c\,l}}\,I_{l}(k_{c},\,r_{c}) }{2\,i}.
\label{flolil1}
\end{align}
Here, $O_{l}(k_{c},r_{c})$ and $I_{l}(k_{c},r_{c})$ are the Coulomb Jost singular solution of the Schr\"odinger equation 
with outgoing and ingoing asymptotic behavior (we follow the definitions used in  \cite{lanethomas58}):
\begin{align}
O_{l}(k_{c},r_{c})  
 \stackrel{r_{c} \to \infty}{\approx} e^{ i\,[k_{c}\,r_{c} - \eta_{c}\,\ln(2\,k_{c}\,r_{c}) 
- l\,\pi/2 + \sigma_{c\,0}]},
\label{singsolut1}
\end{align} 
and
\begin{align}
I_{l}(k_{c},r_{c})
\stackrel{r_{c} \to \infty}{\approx} e^{- i\,[k_{c}\,r_{c} - \eta_{c}\,\ln(2\,k_{c}\,r_{c}), 
- l\,\pi/2 + \sigma_{c\,0}]}.
\label{singsolut1}
\end{align} 
Then we can rewrite $\Upsilon_{c\,s\,l\,m_{s};c\,s\,l\,m_{s''}}^{J_{F}\,(ext)(0)}$ in the form
\begin{align}
&\Upsilon_{c\,s\,l\,m_{s};\,c\,s\,l\,m_{s''}}^{J_{F}(ext)(0)}({\rm {\bf r}}_{c}) = i\,\frac{2\,\pi}{k_{c}\,r_{c} }\,i^{l}
\nonumber\\
&\times \sum\limits_{M\,m_{l}}\, <s\,m_{s}\,\,l\,m_{l}\,|J_{F}\,M_{F}>\, <s\,m_{s''}\,\,l\,m_{l}|J_{F}\,M_{F}>   \nonumber\\
&\times \,Y_{l\,m_{l}}^{*}({\rm {\bf {\hat k}}}_{c})\,\Big[I_{l}(k_{c},\,r_{c}) - e^{i\,2
\,\omega_{c\,l}}O_{l}(k_{c},\,r_{c})\Big]\,Y_{l\,m_{l}}({\widehat {\rm {\bf r}}}_{c})\,\phi_{c\,s\,m_{s''}}.
\label{Upsilonextdif01}
\end{align}
Thus the incident wave is the pure Coulomb scattering wave function in the incident channel $c$. 
The second term in Eq. (\ref{psicincreact1}) is given by the sum of the outgoing waves in the open channels \cite{blattbiedenharn}:
\begin{align}
&\Psi _{c\,s\,m_{s};\,r}^{(ext)( + )} = \,i\,\frac{2\,\pi}{k_{c}}\,\sum\limits_{\tilde c}\,\sqrt {\frac{v_{c}}{v_{\tilde c}}}\,\frac{1}{r_{\tilde c}}\,\xi_{\tilde c}\,
\sum\limits_{J_{F}\,M_{F}\,l\,{\tilde l}\,m_{l}\,m_{\tilde l}}\,\,i^{l}\,   \nonumber\\
&\times <s\,{m_s}\,\,l\,m_{l}|J_{F}\,M_{F} >\, Y_{l\,m_{l}}^{*}({\rm {\bf {\hat k}}}_{c})\,
\Big[ e^{i\,2\,\omega_{c\,l} }\,\delta_{{\tilde c}\,c}\,\delta_{{\tilde s}\,s}\,\delta_{{\tilde l}\,l} \nonumber\\
&-  S_{ {\tilde c}\,{\tilde s}\,{\tilde l};c\,s\,l}^{J_{F}} \Big]\,O_{l}(k_{c},\,r_{c}) \,< {\tilde s}\,m_{\tilde s}\,\,{\tilde l}\,m_{\tilde l}|J_{F}\,M_{F} >\,
Y_{ {\tilde l}\,m_{\tilde l} }({\rm {\bf {\hat r}}}_{\tilde c})           \nonumber\\
&\times \phi _{{\tilde c}\,{\tilde s}\,m_{\tilde s}}.
\label{psirc11}
\end{align}
Here, $\xi_{\tilde c}$ is the product of the bound state wave functions in the channel ${\tilde c}=c,\,c'$,
$\,S_{ {\tilde c}\,{\tilde s}\,{\tilde l};c\,s\,l}^{J_{F}}$ is the $S$-matrix element for transition $c\,s\,l \to {\tilde c}\,{\tilde s}\,{\tilde l}$. Note that we consider the outgoing waves in the channel with given total angular momentum $J_{F}$, initial channel spin $s$ (its projection $m_{s}$) and final channel spin ${\tilde s}$ (its projection $m_{\tilde s}$). Since only two open channels are taken into account here, we will write explicitly the outgoing waves in both channels. First consider the elastic scattering, that is the outgoing channel ${\tilde c}=c= b + B$ and the channel spin and orbital angular momentum coincide with the incident channel values, that is ${\tilde s}=s$ and ${\tilde l}=l$. The component of the outgoing elastic scattered wave ($c\,s\,l \to c\,s\,l$) is  
\begin{align}
&\Psi _{c\,s\,l\,m_{s};c\,s\,l\,m^{"}_{s}}^{(ext)( + )} = \,i\,\frac{2\,\pi}{k_{c}\,r_{c}}\,\xi_{c}\,
\sum\limits_{J_{F}\,M_{F}\,m_{l}\,m_{l''}} \nonumber\\
&\times \,<s\,{m_s}\,\,l\,m_{l}|J_{F}\,M_{F}>\, < s\,m^{"}_{s}\,\,l\,m_{l''}|J_{F}\,M_{F}>  \nonumber\\
&\times i^{l}\,Y_{l\,m_{l}}^{*}({\rm {\bf {\hat k}}}_{c})\,[e^{i\,2\,\omega_{c\,l}}\, -  S_{c\,s\,l;c\,s\,l}^{J_{F}}]\,O_{l}(k_{c},\,r_{c})\,Y_{l\,m_{l''}}({\rm {\bf {\hat r}}}_{c})                                                 \nonumber\\
&\times \,\phi _{c\,s\,m^{"}_{s}}.
\label{psielc1}
\end{align}
Hence, the projection of $\Psi _{c\,s\,l\,m_{s};\,c\,s\,l\,m^{"}_{s}}^{(ext)( + )}$ on $\xi_{c}$ leads to
\begin{align}
&\Upsilon_{c\,s\,l\,m_{s};c\,s\,l\,m^{"}_{s}}^{(ext)( + )}({\rm {\bf r}}_{c}) = \,i\,\frac{2\,\pi}{k_{c}\,r_{c}}\,
\sum\limits_{J_{F}\,M_{F}\,m_{l}\,m_{l''}} \nonumber\\
&\times \,<s\,{m_s}\,\,l\,m_{l}|J_{F}\,M_{F}>\, < s\,m^{"}_{s}\,\,l\,m_{l''}|J_{F}\,M_{F}>  \nonumber\\
&\times i^{l}\,Y_{l\,m_{l}}^{*}({\rm {\bf {\hat k}}}_{c})\,[e^{i\,2\,\omega_{c\,l}}\, -  S_{c\,s\,l;c\,s\,l}^{J_{F}}]\,O_{l}(k_{c},\,r_{c})\,Y_{l\,m_{l''}}({\rm {\bf {\hat r}}}_{c})                                                 \nonumber\\
&\times \,\phi _{c\,s\,m^{"}_{s}}.
\label{Upsilonelextc1}
\end{align}
Correspondingly, for the inelastic scattering, ${\tilde c}=c$ but either ${\tilde s} \not= s$ or ${\tilde l} \not= l$ or both differ from the entry values, we get  
\begin{align}
&\Psi _{c\,s\,l\,m_{s};\,c\,s''\,l''\,m_{s''}}^{(ext)( + )} = -\,i\,\frac{2\,\pi}{k_{c}\,r_{c}}\,\xi_{c}\,
\sum\limits_{J_{F}\,M_{F}\,m_{l}\,m_{l''}}  \nonumber\\
&\times \,< l\,m_{l}\,\,s\,{m_s}|J_{F}\,M_{F} >\,< l''\,m_{l''}\,\,s''\,m_{s''}|J_{F}\,M_{F} >
  \nonumber\\
&\times \,i^{l}\, Y_{l\,_{l}}^{*}({\rm {\bf {\hat k}}}_{c})\,S_{c\,s''\,l'';c\,s\,l}^{J_{F}}\,O_{l''}(k_{c},\,r_{c})\,Y_{l''\,m_{l''} }({\rm {\bf {\hat r}}}_{c})\,\phi _{c\,s''\,m_{s''}}.
\label{psiinelextc1}
\end{align}
Then the projection of $\Psi _{c\,s\,l\,m_{s};\,c\,s''\,l''\,m_{s''}}^{(ext)( + )}$ on $\xi_{c}$ is
\begin{align}
&\Upsilon_{c\,s\,l\,m_{s};c\,s''\,l''\,m_{s''}}^{(ext)( + )}({\rm {\bf r}}_{c}) = -\,i\,\frac{2\,\pi}{k_{c}\,r_{c}}\,
\sum\limits_{J_{F}\,M_{F}\,m_{l}\,m_{l''}}  \nonumber\\
&\times \,< l\,m_{l}\,\,s\,{m_s}|J_{F}\,M_{F} >\,< l''\,m_{l''}\,\,s''\,m_{s''}|J_{F}\,M_{F} >
  \nonumber\\
&\times \,i^{l}\, Y_{l\,_{l}}^{*}({\rm {\bf {\hat k}}}_{c})\,S_{c\,s''\,l'';c\,s\,l}^{J_{F}}\,O_{l''}(k_{c},\,r_{c})\,Y_{l''\,m_{l''} }({\rm {\bf {\hat r}}}_{c})\,\phi _{c\,s''\,m_{s''}}.
\label{Upsiloninelextc1}
\end{align}
Finally, for the outgoing scattered wave in the reaction channel ${\tilde c}= c'=n+ A$ we have
\begin{align}
&\,\,\,\Psi _{c\,s\,l\,m_{s};c'\,s'\,l'\,m_{s'}}^{(ext)( + )} = -\,i\,\frac{2\,\pi}{k_{c}\,r_{c'}}\,\sqrt {\frac{v_{c}}{v_{c'}}}\,\xi_{c'}\,i^{l}\,\sum\limits_{J_{F}\,M_{F}\,m_{l}\,m_{l'}}                                                        \nonumber\\
&\times \,< l\,m_{l}\,\,s\,{m_s}|J_{F}\,M_{F} >\,< l'\,m_{l'}\,\,s'\,m_{s'}|J_{F}\,M_{F} >\,   \nonumber\\
& \times \,Y_{l\,m_{l}}^{*}({\rm {\bf {\hat k}}}_{c})\,S_{c'\,s'\,l';c\,s\,l}^{J_{F}}\,O_{l'}(k_{c'},\,r_{c'})\,Y_{l'\,m_{l'} }({\rm {\bf {\hat r}}}_{c'})\,\phi_{c'\,s'\,m_{s'}}.
\label{psiccprext1}
\end{align}
It leads to its projection on $\xi_{c'}$:
\begin{align}
&\Upsilon_{c\,s\,l\,m_{s};\,c'\,s'\,l'\,m_{s'}}^{(ext)( + )}({\rm {\bf r}}_{c'}) = -\,i\,\frac{2\,\pi}{k_{c}\,r_{c'}}\,\sqrt {\frac{v_{c}}{v_{c'}}}\,i^{l}\,\sum\limits_{J_{F}\,M_{F}\,m_{l}\,m_{l'}}                                                        \nonumber\\
&\times \,< l\,m_{l}\,\,s\,{m_s}|J_{F}\,M_{F} >\,< l'\,m_{l'}\,\,s'\,m_{s'}|J_{F}\,M_{F} >\,   \nonumber\\
& \times \,Y_{l\,m_{l}}^{*}({\rm {\bf {\hat k}}}_{c})\,S_{c'\,s'\,l';c\,s\,l}^{J_{F}}\,O_{l'}(k_{c'},\,r_{c'})\,Y_{l'\,m_{l'} }({\rm {\bf {\hat r}}}_{c'})\,\phi_{c'\,s'\,m_{s'}}.
\label{Upsiloncprext1}
\end{align}

Now we can derive the expression for the matrix elements of the $S$ matrix. 
Since the wave function $\Psi_{c}^{(+)}$ is continuous using Eqs. (\ref{projcdiagint1}), (\ref{Upsilonextdif01}) and (\ref{Upsilonelextc1}) we get the equality
\begin{align}
&\Upsilon _{c\,s\,l\,\,m_{s};\,c\,s\,l\,m_{s''}}^{J_{F}(int)}(R_{c}\,{\rm {\bf {\hat r}}}_{c})= \Upsilon_{c\,s\,l\,m_{s};\,c\,s\,l\,m_{s''}}^{J_{F}(ext)(0)}(R_{c}\,{\rm {\bf {\hat r}}}_{c})
\nonumber\\
&+ \Upsilon_{c\,s\,l\,m_{s};c\,s\,l\,m_{s''}}^{(ext)( + )}(R_{c}\,{\rm {\bf {\hat r}}}_{c}),
\label{Upsilonintextelscat1}
\end{align}
which boils down to
\begin{align}
&e^{-i\,\delta^{hs}_{c\,l}}\,\sum\limits_{\nu,\tau = 1}^N [\Gamma_{\nu\,c\,s\,l\,J_{F}}(E_{c})]^{1/2}\,[{\bf A}^{ - 1} ]_{\nu \tau}\,\sqrt{2\,k_{c}\,R_{c}}\,\gamma_{\tau\,c\,s\,l\,J_{F}} \nonumber\\
&= i\,\big[ I_{l}(k_{c},\,R_{c}) - S_{c\,s\,l;c\,s\,l}^{J_{F}}\,O_{l}(k_{c},\,R_{c}) \big].
\label{elscatSmatrix1}
\end{align}
Taking into account that \cite{lanethomas58}
\begin{align}
\frac{I_{l}(k_{c},\,R_{c})}{O_{l}(k_{c},\,R_{c})}= \frac{G_{l}(k_{c},\,R_{c}) - i\,F_{l}(k_{c},\,R_{c})}{G_{l}(k_{c},\,R_{c}) + i\,F_{l}(k_{c},\,R_{c})}\,e^{i\,2\,\omega_{c\,l}}  =e^{-2\,i\,\delta_{c\,l}^{hs}}
\label{IOdeltass1}
\end{align}
and
\begin{align}
\Gamma_{\tau\,c\,s\,l\,J_{F}}(E_{c})= 2\,P_{c\,l}(E_{c},\,R_{c})\,\gamma_{\tau\,c\,s\,l\,J_{F}}^{2},
\label{Gammaredwidth1}
\end{align}
where 
\begin{align}
P_{c\,l}(E_{c},\,R_{c}) = \frac{k_{c}\,R_{c}}{F_{l}^{2}(k_{c},\,R_{c})
+ G_{l}^{2}(k_{c},\,R_{c})}
\label{penetrab1}
\end{align}
is the Coulomb-centrifugal barrier penetrability,
we get the elastic scattering matrix $S$-matrix element:
\begin{align}
&S_{c\,s\,l;\,c\,s\,l}^{J_{F}}= e^{-2\,i\,\delta_{c\,l}^{hs}}\,(1+ i\,\sum\limits_{\nu,\tau = 1}^N [\Gamma_{\nu\,c\,s\,l\,J_{F}}(E_{c})]^{1/2}\,[{\bf A}^{ - 1} ]_{\nu \tau}                   \nonumber\\
& \times [\Gamma_{\tau\,c\,s\,l\,J_{F}}(E_{c})]^{1/2}).
\label{elscatSmatrix2}
\end{align}
From equality of Eqs. (\ref{projcnAresint2}) and  (\ref{Upsiloncprext1}) at $r_{c'}= R_{c'}$ 
\begin{align}
\Upsilon_{c\,s\,m_{s};c'\,s'\,m_{s'}}^{J_{F}(int)}(R_{c'}\,{\rm {\bf {\hat r}}}_{c'}) = \Upsilon_{c\,s\,l\,m_{s};c'\,s'\,l'\,m_{s'}}^{(ext)( + )}(R_{c'}\,{\rm {\bf {\hat r}}}_{c'})
\label{upsilintupsilext1}
\end{align} 
we obtain the reaction matrix element:
\begin{align}
&S_{c\,s\,l;\,c'\,s'\,l'}^{J_{F}}= i\,e^{-\,i\,\delta_{c\,l}^{hs}}\,e^{-\,i\,\delta_{c'\,l'}^{hs}}\,\sum\limits_{\nu,\tau = 1}^N [\Gamma_{\nu\,c\,s\,l\,J_{F}}(E_{c})]^{1/2}                  \nonumber\\
& \times [{\bf A}^{ - 1} ]_{\nu \tau}\,[\Gamma_{\tau\,c'\,s'\,l'\,J_{F}}(E_{c'})]^{1/2}.
\label{reactSmatrix1}
\end{align}
Both obtained matrix elements coincide with the corresponding matrix elements from \cite{blattbiedenharn}. The only difference 
is in the definition of the solid scattering phase shifts. The obtained matrix elements of the $S$ matrix confirm that the relative normalization of the internal and external wave parts of $\Psi_{bB}^{(+)}$ are correct and we can use them
to calculate the reaction amplitude of the deuteron stripping proceeding through resonance states.

\section{Matrix element $M_{S}^{DW}$}
\label{surfaceintegral1}
Let us consider the DWBA surface (in the subspace over ${\rm {\bf r}}_{nA}$) matrix element 
\begin{align}
&M_{S}^{DW}(P,\,{\rm {\bf k}}_{dA}) = <\chi_{pF}^{(-)}\,\Upsilon_{nA}^{(ext)(-)}|\,{\overleftarrow T} - {\overrightarrow T}\,|   \nonumber\\
&\times \varphi_{d}\,\chi_{dA}^{(+)}>\Big |_{r_{nA} > R_{nA}}       \nonumber\\
&= M_{S(pF)}^{DW}(P,\,{\rm {\bf k}}_{dA})  + M_{S(nA)}^{DW}({\rm {\bf k}}_{pF},\,{\rm {\bf k}}_{dA}),
\label{mspfmsna1}
\end{align}
where $\Upsilon_{nA}^{(ext)(+)}= <\varphi_{A}|\Psi_{bB}^{(ext)(+)}>$,
\begin{align}
&M_{S(pF)}^{DW}(P,\,{\rm {\bf k}}_{dA})                              
= \int\limits_{r_{nA} > R_{nA}}\,{\rm d}{\rm {\bf r}}_{nA}\,\int {\rm d}{\rm {\bf r}}_{pF} \chi_{pF}^{(-)*}\,\Upsilon_{nA}^{(ext)(-)*}({\rm {\bf r}}_{nA})
\nonumber\\
&\times [{\overleftarrow T}_{pF} - {\overrightarrow T}_{pF}]\varphi_{d}\,\chi_{dA}^{(+)}>
\label{msrespf1}
\end{align}
and
\begin{align}
&M_{S(nA)}^{DW}(P,\,{\rm {\bf k}}_{dA}) = \int {\rm d}{\rm {\bf r}}_{pF}\,\int\limits_{r_{nA} > R_{nA}}\,{\rm d}{\rm {\bf r}}_{nA}\,\chi_{pF}^{(-)*}\,\Upsilon_{nA}^{(ext)(-)*}({\rm {\bf r}}_{nA})
\nonumber\\
&\times [{\overleftarrow T}_{nA} - {\overrightarrow T}_{nA}]\varphi_{d}\,\chi_{dA}^{(+)}>.
\label{msresna1}
\end{align}
 $M_{S(pF)}^{DW}$ can be written as 
\begin{widetext}
\begin{align}
&M_{S(pF)}^{DW}(P,\,{\rm {\bf k}}_{dA})                              
= \int {\rm d}{\rm {\bf r}}_{nA}\,\int {\rm d}{\rm {\bf r}}_{pF} \chi_{pF}^{(-)*}\,\Upsilon_{nA}^{(-)*}({\rm {\bf r}}_{nA})[{\overleftarrow T}_{pF} - {\overrightarrow T}_{pF}]\varphi_{d}\,\chi_{dA}^{(+)}   \nonumber\\
&- \int\limits_{r_{nA} \leq R_{nA}}\,{\rm d}{\rm {\bf r}}_{nA}\,\int {\rm d}{\rm {\bf r}}_{pF} \chi_{pF}^{(-)*}\,\Upsilon_{nA}^{(int)(-)*}({\rm {\bf r}}_{nA})\,[{\overleftarrow T}_{pF} - {\overrightarrow T}_{pF}]\varphi_{d}\,\chi_{dA}^{(+)}>  \nonumber\\
&= \int {\rm d}{\rm {\bf r}}_{nA}\,\int {\rm d}{\rm {\bf r}}_{pF} \chi_{pF}^{(-)*}\,\Upsilon_{nA}^{(ext)(-)*}({\rm {\bf r}}_{nA})
[{\overleftarrow T}_{pF} - {\overrightarrow T}_{pF}]\varphi_{d}\,\chi_{dA}^{(+)}. 
\label{msrespf2}
\end{align}
\end{widetext}
We took into account that for any finite volume $r_{nA} \leq R_{nA}$ the matrix element containing ${\overleftarrow T}_{pF} - {\overrightarrow T}_{pF}$ vanishes as it has been discussed in Section II A for deuteron stripping to bound states. To estimate 
$M_{S(pF)}^{DW}$  we need equations connecting different variables: 
\begin{align}
{\rm {\bf r}}_{dA}= 1/2\,{\rm {\bf r}}_{pn}  + {\rm {\bf r}}_{nA},  
\label{rdArpnrnA1} \\
{\rm {\bf r}}_{pF}= A/(A+1)\,{\rm {\bf r}}_{nA} + {\rm {\bf r}}_{pn}. 
\label{rpFrnArpn1}
\end{align}
Now in the matrix element (\ref{msrespf2}) we replace the variable ${\rm {\bf r}}_{nA}$ by ${\rm {\bf r}}_{pn}$. Then we get
\begin{align}
&M_{S(pF)}^{DW}(P,\,{\rm {\bf k}}_{dA}) = -(\frac{A+1}{A})^{3}\,\int {\rm d}{\rm {\bf r}}_{pn}\,\int {\rm d}{\rm {\bf r}}_{pF}\,\chi_{pF}^{(-)*}({\rm {\bf r}}_{pF})                     \nonumber\\
&\times\Upsilon_{nA}^{(ext)(-)*}(\frac{A+1}{A}[{\rm {\bf r}}_{pF} - {\rm {\bf r}}_{pn}])\,[{\overleftarrow T}_{pF} - {\overrightarrow T}_{pF}]               \nonumber\\
& \times\,\varphi_{d}({\rm {\bf r}}_{pn})\,\chi_{dA}^{(+)}(\frac{A + 1}{A}\,{\rm {\bf r}}_{pF} - \frac{A + 2}{2A}\,{\rm {\bf r}}_{pn}). 
\label{msrespf11}
\end{align}  
This matrix element can be rewritten in the form, in which the integral over ${\rm {\bf r}}_{pF}$ is transformed to the surface integral:
\begin{widetext}
\begin{align}
&M_{S(pF)}^{DW}(P,\,{\rm {\bf k}}_{dA}) = (\frac{A+1}{A})^{3}\,\frac{A}{A+1}\,\lim\limits_{R_{pF} \to \infty }\, R_{pF}^{2}\,\frac{1}{2\,\mu_{pF}}
\nonumber\\
&\times \int{\rm d}{\rm {\bf r}}_{pn}\,\varphi_{d}(r_{pn})\,\int{\rm d}{\Omega}_{{\rm {\bf r}}_{pF}}\Big[\,\chi_{dA}^{(+)}(\frac{A + 1}{A}\,{\rm {\bf r}}_{pF} - \frac{A + 2}{2A}\,{\rm {\bf r}}_{pn})\,\frac{  \partial\,\chi_{pF}^{(-)*}({\rm {\bf r}}_{pF})\,\Upsilon_{nA}^{(ext)(-)*}(\frac{A+1}{A}[{\rm {\bf r}}_{pF} - {\rm {\bf r}}_{pn}])}{\partial\,r_{pF} }             \nonumber\\
&- \chi_{pF}^{(-)*}({\rm {\bf r}}_{pF})\,\Upsilon_{nA}^{(ext)(-)*}(\frac{A+1}{A}[{\rm {\bf r}}_{pF} - {\rm {\bf r}}_{pn}])\,
\frac{ \partial\,\chi_{dA}^{(+)}(\frac{A + 1}{A}\,{\rm {\bf r}}_{pF} - \frac{A + 2}{2A}\,{\rm {\bf r}}_{pn}) }{ \partial\,r_{pF} } \Big]\Big |_{r_{pF}=R_{pF} \to \infty}.
\label{msrespf11}
\end{align} 
\end{widetext}
Due to the presence of the deuteron bound state wave function the integration over $r_{pn}$ is limited.  At $r_{pF} \to \infty$ and $r_{pn} < \infty$ we can replace the distorted waves in the initial and final channels by their leading asymptotic terms: 
\begin{align}
\chi_{dA}^{(+)}({\rm{\bf r}}_{dA}) \stackrel{r_{dA} \to \infty}{\sim}\,e^{i\,{\rm {\bf k}}_{dA} \cdot {\rm{\bf r}}_{dA} 
+ i\,\eta_{dA}\,\ln(k_{dA}\,r_{dA} - {\rm {\bf k}}_{dA} \cdot {\rm {\bf r}}_{dA})}.
\label{asymchidA1}
\end{align} 
and 
\begin{align}
\chi_{pF}^{(-)*}\,\, \stackrel{r_{pF} \to \infty}{\to} e^{-i\,{\rm {\bf k}}_{pF} \cdot {\rm {\bf r}}_{pF} + i\,\eta_{pF}\,\ln(k_{pF}\,r_{pF}
+ {\rm {\bf k}}_{pF} \cdot {\rm {\bf r}}_{pF})}.
\label{asympchipF1}
\end{align}  
Here, $\eta_{ij}$ is the Coulomb parameter of particles $i$ and $j$ in the continuum. 
Note that ${\rm{\bf r}}_{dA}= \frac{A + 1}{A}\,{\rm {\bf r}}_{pF} - \frac{A + 2}{2A}\,{\rm {\bf r}}_{pn}$, and 
at $r_{pF} \to \infty$ and $r_{pn} < \infty$  $\,\,r_{dA} \to \infty$.
Then
\begin{widetext}
\begin{align}
&\frac{  \partial\,e^{i\,{\rm {\bf k}}_{dA} \cdot {\rm {\bf r}}_{dA} + i\,\eta_{dA}\,\ln(k_{dA}\,r_{dA}
- {\rm {\bf k}}_{dA} \cdot {\rm {\bf r}}_{dA})}}{\partial\,r_{pF} }                  \nonumber\\
&\stackrel{r_{pF} \to \infty}{\to}
i\,\frac{A + 1}{A}\,{\rm {\bf k}}_{dA}\cdot {\rm {\bf {\hat r}}}_{pF}\,e^{i\,{\rm {\bf k}}_{dA}\cdot (\frac{A + 1}{A}\, \cdot {\rm {\bf r}}_{pF} - \frac{A + 2}{2A}\,{\rm {\bf r}}_{pn}) + i\,\eta_{dA}\,\ln(k_{dA}\,r_{dA} - {\rm {\bf k}}_{dA} \cdot {\rm {\bf r}}_{dA})}
\label{derchidAF2}
\end{align}
\end{widetext}
and
\begin{align}
&\frac{  \partial\,e^{-i\,{\rm {\bf k}}_{pF} \cdot {\rm {\bf r}}_{pF} + i\,\eta_{pF}\,\ln(k_{pF}\,r_{pF}
+ {\rm {\bf k}}_{pF} \cdot {\rm {\bf r}}_{pF})}}{\partial\,r_{pF} }                  \nonumber\\
&\stackrel{r_{pF} \to \infty}{\approx}
-i\,{\rm {\bf k}}_{pF}\cdot {\rm {\bf {\hat r}}}_{pF}\,e^{-i\,{\rm {\bf k}}_{pF} \cdot {\rm {\bf r}}_{pF} + i\,\eta_{pF}\,\ln(k_{pF}\,r_{pF} + {\rm {\bf k}}_{pF} \cdot {\rm {\bf r}}_{pF})}.
\label{derchipF1}
\end{align}
For $\Upsilon_{nA}^{(ext)(-)*}(\frac{A+1}{A}[{\rm {\bf r}}_{pF} - {\rm {\bf r}}_{pn}])$ we can take only the external part, which 
contains the resonant $S$ matrix element, see Eq.  (\ref{Upsiloncprext1}). Neglecting all the spin-dependent and angular parts
and leaving only its radial part, which is $O_{nA}(r_{nA})/r_{nA}$, we get for its leading asymptotic term:
\begin{align}
&\frac{O_{nA}(k_{nA},\,r_{nA})}{r_{nA}}  
\stackrel{r_{pF} \to \infty}{\to} \frac{A}{A+1}\,\frac{1}{r_{pF}}\,e^{ i\,\frac{A+1}{A}\,(k_{nA}\,r_{pF} - k_{nA}\,{\rm {\bf {\hat r}}}_{pF}\cdot {\rm {\bf r}}_{pn})}       \nonumber\\
& \times\, e^{-i[\,\eta_{nA}\,\ln(2\,k_{nA}\,r_{nA}) + l_{nA}\,\pi/2 - \sigma_{nA\,0}] } .
\label{Onaas1}
\end{align} 
The leading term of its derivative  at $r_{pF} \to \infty$  is 
\begin{align}
&\frac{\partial{O_{nA}(k_{nA},\,r_{nA})/r_{nA}}}{\partial\,r_{pF} } 
\stackrel{r_{pF} \to \infty}{\to}  \,i\,k_{nA}\,\frac{1}{r_{pF}}   \nonumber\\
&\times e^{ i\,\frac{A+1}{A}\,(k_{nA}\,r_{pF} - k_{nA}\,{\rm {\bf {\hat r}}}_{pF}\cdot {\rm {\bf r}}_{pn})}
\nonumber\\
&\times e^{-i\,[\,\eta_{nA}\,\ln(2\,\frac{A+1}{A}\,k_{nA}\,r_{pF}) + \,l_{nA}\,\pi/2 - \,\sigma_{nA\,0}] } .
\label{derOnaas1}
\end{align} 

Then $M_{S(pF)}^{DW}$ reduces to 
\begin{widetext}
\begin{align} 
&M_{S(pF)}^{DW}(P,\,{\rm {\bf k}}_{dA}) \sim \lim\limits_{R_{pF} \to \infty }\, R_{pF}\,\int{\rm d}{\rm {\bf r}}_{pn}\,\varphi_{d}(r_{pn})\,e^{-i\,\frac{A + 2}{2A}\,{\rm {\bf k}}_{dA}\cdot {\rm {\bf r}}_{pn}}\,\int\,{\rm d}{\Omega}_{{\rm {\bf r}}_{pF}}\,
[(\frac{A + 1}{A}\,{\rm {\bf k}}_{dA}  + {\rm {\bf k}}_{pF})\cdot {\rm {\bf {\hat r}}}_{pF} - \frac{A+1}{A}\,k_{nA}]
\nonumber\\
& \times e^{i\,(\frac{A + 1}{A}\,{\rm {\bf k}}_{dA} - {\rm {\bf k}}_{pF}) \cdot {\rm {\bf {\hat  r}}}_{pF}\,R_{pF}}\,
e^{ i\,\frac{A+1}{A}\,k_{nA}\,R_{pF}}\,
e^{ i\,\eta_{dA}\,\ln \big(k_{dA}\,r_{dA}(R_{pF}) - {\rm {\bf k}}_{dA} \cdot {\rm {\bf r}}_{dA}(R_{pF})\big)
 + i\,\eta_{pF}\,\ln(k_{pF}\,R_{pF} + {\rm {\bf k}}_{pF} \cdot {\rm {\bf {\hat r}}}_{pF}\,R_{pF})}      \nonumber\\
& \times e^{ -i[\,\frac{A+1}{A}\,k_{nA}\,{\rm {\bf {\hat r}}}_{pF}\cdot {\rm {\bf r}}_{pn}  + \eta_{nA}\,\ln(2\,\frac{A+1}{A}\,k_{nA}\,R_{pF})] }.
\label{integrandrpF1}
\end{align}
\end{widetext}
Taking into account the asymptotic behavior of the plane wave 
\begin{align}
&e^{i\,{\rm {\bf q}}\cdot {\rm {\bf r}}_{pF}} \stackrel{r_{pF} \to \infty}{\to} \frac{2\,\pi}{i\,q\,r_{pF}}\,[e^{i\,q\,r_{pF}}\,\delta({\rm {\bf {\hat q}}} - {\rm {\bf {\hat r}}}_{pF})        \nonumber\\
& - e^{-i\,q\,r_{pF}}\,\delta({\rm {\bf {\hat q}}} + {\rm {\bf {\hat r}}}_{pF})],
\label{asymplwave1}
\end{align} 
where ${\rm {\bf q}} =  \frac{A + 1}{A}\,{\rm {\bf k}}_{dA} - {\rm {\bf k}}_{pF}$ 
we obtain that the matrix element 
\begin{align}
&M_{S(pF)}^{DW}(P,\,{\rm {\bf k}}_{dA}) \sim  \lim\limits_{R_{pF} \to \infty } f_{1}(R_{pF})\,e^{i\,q\,R_{pF}} \nonumber\\
&+ f_{2}(R_{pF})\,e^{-i\,q\,R_{pF}}].
\label{mspFdw1}
\end{align}
Thus $M_{S(pF)}^{DW}$ has no limit at $R_{pF} \to \infty$ but regularization of this matrix element by integrating the matrix element over an infinitesimal bin in the momentum plane leads to disappearance of $M_{S(pF)}^{DW}$:
\begin{align}
&\frac{1}{2\,\epsilon}\,\int\limits_{q- \epsilon}^{q+ \epsilon}\,{\rm d}q\,M_{S(pF)}^{DW}(P,\,{\rm {\bf k}}_{dA}) \to \lim\limits_{R_{pF} \to \infty } \,\frac{sin\,(\epsilon\,R_{pF})}{\epsilon\,R_{pF}}  \nonumber\\
& \times \big[e^{i\,q\,R_{pF}}\,f_{1}(R_{pF}) - e^{-i\,q\,R_{pF}}\,f_{2}(R_{pF}) \big] =0,
\label{limMSpFpostres1}
\end{align}
where $\epsilon << q$. 

A similar prove can be applied to estimate $M_{S(nA)}^{DW}$ given by Eq. (\ref{msresna1}).
Since the integral over $r_{nA}$ is taken over external volume with $r_{nA} > R_{nA}$ the transformation of the volume integral into the surface one leads to two surface integrals:
\begin{align}
&M_{S(nA)}^{DW}(P,\,{\rm {\bf k}}_{dA}) = - M_{S_{R_{nA}}}^{DW}(P,\,{\rm {\bf k}}_{dA}) 
+ M_{S_{\infty}}^{DW}(P,\,{\rm {\bf k}}_{dA}).
\label{twosurfint1}
\end{align}
The first term is the surface integral encircling the inner surface of the external volume at $r_{nA}= R_{nA}$, while the second term is the surface integral taken at $r_{nA}= R_{nA}^{'} \to \infty$. A negative sign in front of the first term  appears because the normal to the surface is directed inward to the center of the volume, that is opposite to the normal to the external surface (at infinitely large radius). 
The surface integral over the infinitely large sphere in the subspace over ${\rm {\bf r}}_{nA}$ is 
\begin{widetext}
\begin{align}
&M_{S_{\infty}}^{DW}(P,\,{\rm {\bf k}}_{dA}) = -\lim\limits_{R_{nA}^{'} \to \infty }\, {R_{nA}^{'}}^{2}\,\frac{1}{2\,\mu_{nA}}\,\int{\rm d}{\rm {\bf r}}_{pn}\,\varphi_{d}(r_{pn})
 \,[\,\chi_{dA}^{(+)}((1/2)\,{\rm {\bf r}}_{pn}  + {\rm {\bf r}}_{nA})\,\frac{  \partial\,\chi_{pF}^{(-)*}({ A/(A+1)\,{\rm {\bf r}}_{nA} + {\rm {\bf r}}_{pn}})\,\Upsilon_{nA}^{(ext)(-)*}({\rm {\bf r}}_{nA})}{\partial\,r_{nA} }             \nonumber\\
&- \chi_{pF}^{(-)*}({ A/(A+1)\,{\rm {\bf r}}_{nA} + {\rm {\bf r}}_{pn}})\,\Upsilon_{nA}^{(ext)(-)*}({\rm {\bf r}}_{nA})\,
\frac{ \partial\,\chi_{dA}^{(+)}((1/2)\,{\rm {\bf r}}_{pn}  + {\rm {\bf r}}_{nA}) }{ \partial\,r_{nA} } ]\Big |_{r_{nA}=R_{nA}^{'} \to \infty}. 
\label{MSDWinfty11}
\end{align}
\end{widetext} 
Here, the Jacobian variable ${\rm {\bf r}}_{pF}$ is replaced by 
${\rm {\bf r}}_{pn}$ by ${\rm {\bf r}}_{pn}= {\rm {\bf r}}_{pF} - A/(A+1)\,{\rm {\bf r}}_{nA}$.
The disappearance of the matrix element (\ref{MSDWinfty11}) can be proved similarly to the proof for $M_{S(pF)}^{DW}$. 
Due to the presence of the bound state $\varphi_{d}(r_{pn})$ the integration over $r_{pn}$ is limited by finite distances. Hence,  $r_{pF} \to \infty$ and $r_{dA} \to \infty$ at $r_{nA} \to \infty$.    
Replacing the distorted waves by their leading asymptotic terms (\ref{asymchidA1}) and (\ref{asympchipF1}),
singling out the plane wave containing ${\rm {\bf r}}_{nA}$ and using the asymptotic representation of this plane wave,
see Eq. (\ref{asymplwave1}), integrating over $\Omega_{{\rm {\bf r}}_{nA}}$ we eventually arrive at
\begin{align}
&M_{S_{\infty}}^{DW}(P,\,{\rm {\bf k}}_{dA}) \sim\, \lim\limits_{R_{nA}^{'} \to \infty }[ e^{i\,q'\,R_{nA}^{'}}\,g_{1}(R_{nA}^{'})                                               \nonumber\\
&+  e^{-i\,q'\,R_{nA}^{'}}\,g_{2}(R_{nA}^{'})].
\label{MSnADWlim1}
\end{align}
Regularization of this matrix element by integrating it over an infinitesimal bin in the momentum plane $q'$ leads to disappearance of $M_{S_{\infty}}^{DW}$, that is
\begin{align}
&M_{S(nA)}^{DW}(P,\,{\rm {\bf k}}_{dA}) = - M_{S_{R_{nA}}}^{DW}(P,\,{\rm {\bf k}}_{dA}).
\label{intsurfint11}
\end{align}

 \section{Matrix element $M_{S}^{CDCC(post)}(P,\,{\rm {\bf k}}_{dA})$} 
\label{CDCCres}
Here we show how to transform $M_{S}^{CDCC(post)}$ into the surface integral over the coordinate ${\rm {\bf r}}_{nA}$.  $\,M_{S}^{CDCC(post)}$ can be written as 
\begin{widetext}
\begin{align}
&M_{S}^{CDCC}(P,\,{\rm {\bf k}}_{dA}) = \int\limits_{r_{nA} > R_{nA}} {\rm d}{\rm {\bf r}}_{nA}\,\int {\rm d}{\rm {\bf r}}_{pF} \chi_{-{\rm {\bf k}}_{pF}}^{(+)}({\rm {\bf r}}_{pF}) \,\Upsilon_{nA}^{(ext)(-)*}({\rm {\bf r}}_{nA})
[{\overleftarrow T} - {\overrightarrow T}]\Psi_{i}^{CDCC(+)}({\rm {\bf r}}_{pF},\,{\rm {\bf r}}_{nA}) \nonumber\\ 
& = M_{Stot}^{CDCC}(P,\,{\rm {\bf k}}_{dA})  - M_{Sint}^{CDCC}(P,\,{\rm {\bf k}}_{dA}),  
\label{postCDCCresApS1}
\end{align}
\end{widetext}
where 
\begin{align}
&M_{Stot}^{CDCC}(P,\,{\rm {\bf k}}_{dA}) = \int {\rm d}{\rm {\bf r}}_{nA}\,\int {\rm d}{\rm {\bf r}}_{pF} \chi_{-{\rm {\bf k}}_{pF}}^{(+)}({\rm {\bf r}}_{pF})   \nonumber\\
&\times\,\Upsilon_{nA}^{(-)*}({\rm {\bf r}}_{nA})[{\overleftarrow T} - {\overrightarrow T}]\Psi_{i}^{CDCC(+)}({\rm {\bf r}}_{pF},\,{\rm {\bf r}}_{nA})
\label{MStotCDCCres1}
\end{align}
and
\begin{align}
&M_{Sint}^{CDCC}(P,\,{\rm {\bf k}}_{dA}) =  \int\limits_{r_{nA} \leq R_{nA}} {\rm d}{\rm {\bf r}}_{nA}\,\int {\rm d}{\rm {\bf r}}_{pF} \chi_{-{\rm {\bf k}}_{pF}}^{(+)}({\rm {\bf r}}_{pF})          \nonumber\\
&\times \Upsilon_{nA}^{(int)(-)*}({\rm {\bf r}}_{nA})\,[{\overleftarrow T} - {\overrightarrow T}]\,\Psi_{i}^{CDCC(+)}({\rm {\bf r}}_{pF},\,{\rm {\bf r}}_{nA}). 
\label{MSintCDCCpostres1}
\end{align}
Note that in the matrix element $M_{Stot}^{DW}$ the integration is carried over ${\rm {\bf r}}_{pF}$ and ${\rm {\bf r}}_{nA}$ in all the coordinate space while in $M_{Sint}^{DW}$ the external region 
in the subspace over ${\rm {\bf r}}_{nA}$ is excluded.   
Let us first consider $M_{Stot}^{DW}$.  The CDCC wave function is given by Eq. (\ref{cdcc1}). If we substitute the first term,$\,n=0$, which contains the deuteron bound state wave function,  the transformation leads to the surface integrals with $r_{pF}= R_{pF} \to \infty$ and $r_{nA} = R_{nA} \to \infty$. Both surface integrals vanish and the proof is similar to the one presented in the previous section. For the rest of the CDCC wave function corresponding to the sum with $n>0$, which we call $\Psi_{i\,c}^{CDCC(+)}$, transformation to the surface integrals gives 
\begin{widetext}
\begin{align}
&M_{Stot}^{DW(c)}(P,\,{\rm {\bf k}}_{dA}) = \int {\rm d}{\rm {\bf r}}_{nA}\,\int {\rm d}{\rm {\bf r}}_{pF} \chi_{-{\rm {\bf k}}_{pF}}^{(+)}({\rm {\bf r}}_{pF})\,\Upsilon_{nA}^{(-)*}({\rm {\bf r}}_{nA})[{\overleftarrow T}_{pF} - {\overrightarrow T}_{pF}]\Psi_{i\,c}^{CDCC(+)}({\rm {\bf r}}_{pF},\,{\rm {\bf r}}_{nA})    \nonumber\\
&+ \int {\rm d}{\rm {\bf r}}_{nA}\,\int {\rm d}{\rm {\bf r}}_{pF}\,\chi_{-{\rm {\bf k}}_{pF}}^{(+)}({\rm {\bf r}}_{pF}) \,\Upsilon_{nA}^{(-)*}({\rm {\bf r}}_{nA})[{\overleftarrow T}_{nA} - {\overrightarrow T}_{nA}]\Psi_{i\,c}^{CDCC(+)}({\rm {\bf r}}_{pF},\,{\rm {\bf r}}_{nA}) \nonumber\\
&= \lim\limits_{R_{pF} \to \infty}\,\frac{R_{pF}^{2}}{2\,\mu_{pF}}\,\int {\rm d}{\Omega_{{\rm {\bf r}}_{pF}}}\,
\int {\rm d}{\rm {\bf r}}_{nA}\,\Bigg[\Upsilon_{nA}^{(-)*}({\rm {\bf r}}_{nA})\,\Psi_{i\,c}^{CDCC(+)}({\rm {\bf r}}_{pF},\,{\rm {\bf r}}_{nA})\,
\frac{\partial\,\chi_{-{\rm {\bf k}}_{pF}}^{(+)}({\rm {\bf r}}_{pF})}{\partial\,r_{pF} }             \nonumber\\
&- \chi_{-{\rm {\bf k}}_{pF}}^{(+)}({\rm {\bf r}}_{pF})\,\Upsilon_{nA}^{(-)*}({\rm {\bf r}}_{nA})\,\,\frac{\partial\,\Psi_{i\,c}^{CDCC(+)}({\rm {\bf r}}_{pF},\,{\rm {\bf r}}_{nA})}{\partial\,r_{pF} } \Bigg]       \nonumber\\
&+ \lim\limits_{R_{nA} \to \infty}\,\frac{R_{nA}^{2}}{2\,\mu_{nA}}\,\int {\rm d}{\Omega}_{{\rm {\bf r}}_{nA}}\,
\int {\rm d}{\rm {\bf r}}_{pF}\,\Bigg[\chi_{-{\rm {\bf k}}_{pF}}^{(+)}({\rm {\bf r}}_{pF})\,\Psi_{i\,c}^{CDCC(+)}({\rm {\bf r}}_{pF},\,{\rm {\bf r}}_{nA})\,\frac{\partial\,\Upsilon_{nA}^{(-)*}({\rm {\bf r}}_{nA})}{\partial\,r_{nA} }               \nonumber\\   
&- \chi_{-{\rm {\bf k}}_{pF}}^{(+)}({\rm {\bf r}}_{pF})\,\Upsilon_{nA}^{(-)*}({\rm {\bf r}}_{nA})\,\,\frac{\partial\,\Psi_{i\,c}^{CDCC(+)}({\rm {\bf r}}_{pF},\,{\rm {\bf r}}_{nA})}{\partial\,r_{nA} }  \Bigg]. 
\label{MScCDCCres1}
\end{align} 
\end{widetext}
Let us, first, consider the first term, in which $R_{pF} \to \infty$. Let us divide the integration region over ${\rm {\bf r}}_{nA}$ into the region $r_{nA}/R_{pF} \to 0$ and the region where $r_{nA} \gtrsim R_{pF} \to \infty$. In the first region 
we get that $r_{dA} \sim R_{pF} \to \infty$ and $r_{pn} \sim R_{pF} \to \infty$. Taking into account the asymptotic behavior of
$\Psi_{i\,c}^{CDCC(+)}({\rm {\bf r}}_{pF},\,{\rm {\bf r}}_{nA}) \sim r_{pF}^{-3}$ and Eq. (\ref{asymplwave1}) we get that the first term goes to zero as $R_{pF}^{-2} \to 0$. In the remained region $r_{nA} \sim R_{pF} \to \infty$ and we consider it later. The second term of Eq. (\ref{MScCDCCres1}), in which $R_{nA} \to \infty$, we also separate into two regions: 
$r_{pF}/R_{nA} \to 0$ and $r_{pF} \gtrsim R_{nA} \to \infty$.  In the first region $r_{pn} \sim R_{nA} \to \infty$ and $r_{dA} \sim R_{nA} \to \infty$ and $\Psi_{i\,c}^{CDCC(+)}({\rm {\bf r}}_{pF},\,{\rm {\bf r}}_{nA}) \sim r_{nA}^{-3}$. Hence 
the matrix element goes to zero as $R_{nA}^{-2} \to 0$. To consider the behavior of the first and second terms of Eq. (\ref{MScCDCCres1}) in the second regions, where $r_{nA},\,r_{pF} \to \infty$, it is more convenient to introduce the hyper-spherical coordinates in the six-dimensional hyper-space:
\begin{align}
&\rho= \sqrt{\frac{\mu_{nA}}{m}\,r_{nA}^{2} +  \frac{\mu_{pF}}{m}\,r_{pF}^{2}}, \nonumber\\
&r_{nA}= \rho\,\sqrt{\frac{m}{\mu_{nA}}}\,\sin\,\alpha,   \qquad 
r_{pF}= \rho\,\sqrt{\frac{m}{\mu_{pF}}}\,\cos\,\alpha,              \nonumber\\
& 0 \leq \alpha \leq \pi/2.     
\label{hypspcoord1}   
\end{align}
Here, $m$ is the scaling mass parameter, for example, the nucleon mass.
Then $M_{Stot}^{DW}$ in the region, where $r_{nA},\,r_{pF} \to \infty$, can be written as the integral over the hypersphere encircling the volume integral with the radius of the hyper-sphere $\rho \to \infty$ \cite{kadyrov09}: 
\begin{widetext}
\begin{align}
&M_{Stot}^{DW}(P,\,{\rm {\bf k}}_{dA}) = \frac{1}{2}\,\frac{m^{2}}{(\mu _{n\,A 
}\,\mu_{pF})^{3/2}}\,\lim_{\rho \rightarrow \infty }\,\rho^{5}\,\int 
\mathrm{d}{\hat{\rm {\bf r}}_{pF}}\,\int\,\mathrm{d}{\hat{\mathrm{\mathbf{r}}}}_{nA } 
\int\limits_{0}^{\pi /2}\mathrm{d}{\alpha }\sin ^{2}{\alpha 
}\,\cos^{2}{\alpha }    \nonumber \\
& \Big[\chi_{pF}^{(-)*}({\rm {\bf r}}_{pF})\,\Upsilon_{nA}^{(-)*}({\rm {\bf r}}_{nA})\,\frac{\partial }{\partial \rho}\Psi_{i}^{CDCC(+)}({\rm {\bf r}}_{pF},\,{\rm {\bf r}}_{nA})-\Psi_{i}^{CDCC(+)}({\rm {\bf r}}_{pF},\,{\rm {\bf r}}_{nA})\,\frac{\partial }{\partial \rho}\,\chi_{pF}^{(-)*}
({\rm {\bf r}}_{pF})\,\Upsilon_{nA}^{(-)*}({\rm {\bf r}}_{nA}) \Big].
\label{MStotCDCCres2}
\end{align}
\end{widetext}
Here the hyper-radius $\rho$ is the parameter going to infinity. The integrand is contains highly oscillating (actually infinitely 
oscillating) functions. The behavior of the integral at $\rho \to \infty$ depends on the asymptotic behavior of the integrand. 
The integration over $\mathrm{d}{\hat{\rm {\bf r}}_{pF}}$ can be performed directly using the asymptotic form of  
$\chi_{pF}^{(-)*}({\rm {\bf r}}_{pF})$. It is given by the Coulomb distorted plane wave, but for simplicity, what does not affect the final result we neglect, as in the previous section, the Coulomb effects. Then the asymptotic form of the plane wave 
is given by Eq. (\ref{asymplwave1}) and, hence, integration over ${\hat{\rm {\bf r}}_{pF}}$ using $\delta$-functions is trivial leading to ${\rm {\bf {\hat r}}}_{pF}= \pm {\rm {\bf {\hat k}}}_{pF}$. After performing the integration over $\mathrm{d}{\hat{\rm {\bf r}}_{pF}}$ only two integrals are left. From Eqs (\ref{rdArpnrnA1}), (\ref{rpFrnArpn1}) and 
(\ref{hypspcoord1}) we get for 
\begin{widetext}
\begin{align}
&r_{pn}= \sqrt{r_{pF}^{2} - \,2\,\frac{A}{A+1}\,{\rm {\bf r}}_{pF}\,{\rm {\bf r}}_{nA}  + \frac{A^{2}}{(A+1)^{2}}\,r_{nA}^{2}} \nonumber\\
&= \rho\,\sqrt{\frac{m}{\mu_{pF}}\,\cos^{2}\alpha  \mp\,\frac{A}{A+1}\,\sqrt{\frac{m}{\mu_{pF}}}\,\sqrt{\frac{m}{\mu_{nA}}}\,z\,\sin2\,\alpha   + \frac{A^{2}}{(A+1)^{2}}\,\frac{m}{\mu_{nA}}\,\sin^{2}\alpha}
\label{rpnrpFrnA1}
\end{align}
\end{widetext}
and 
\begin{widetext}
\begin{align}
&r_{dA}= \sqrt{\frac{1}{4}\,r_{pF}^{2} + \,\frac{A+2}{2\,(A+1)}\,{\rm {\bf r}}_{pF}\,{\rm {\bf r}}_{nA}  + \frac{(A +2)^{2}}{4(A+1)^{2}}\,r_{nA}^{2}} \nonumber\\
&= \rho\,\sqrt{\frac{1}{4}\,\frac{m}{\mu_{pF}}\,\cos^{2}\alpha  \mp\,\frac{A+2}{4(A+1)}\,\sqrt{\frac{m}{\mu_{pF}}}\,\sqrt{\frac{m}{\mu_{nA}}}\,z\,\sin2\,\alpha   + \frac{(A +2)^{2}}{4(A+1)^{2}}\,\frac{m}{\mu_{nA}}\,\sin^{2}\alpha} \,.
\label{rdArpFrnA1}
\end{align}
\end{widetext}
Here, $z= {\rm {\bf {\hat r}}}_{nA} \cdot {\rm {\bf {\hat k}}}_{pF}$. 
We recall also that in Eq. (\ref{cdcc1}) at $n>0$ $\,\,\psi_{pn}^{(n)}({\rm {\bf r}}_{pn})$ at $r_{pn} \to \infty$ contains the asymptotic terms $\frac{e^{\pm i\,k_{pn}\,r_{pn}}}{r_{pn}^{2}}$, while $\chi_{i}^{(n)(+)}({\rm {\bf r}}_{dA}) \sim \frac{e^{i\,k_{dA}\,r_{dA} }}{r_{dA}}$, where we, for simplicity, neglected the Coulomb distortion. 
Then after integration over $\mathrm{d}{\hat{\rm {\bf r}}_{pF}}$ the leading asymptotic form of the integrand with omitted Coulomb effects is a product of highly oscillating at $\rho \to \infty$ exponents:
\begin{align} 
&\frac{e^{\pm\,i\,k_{pF}\,r_{pF}}}{r_{pF}}\,\frac{e^{i\,k_{nA}\,r_{nA}}}{r_{nA}}\,\frac{e^{\pm\,i\,k_{pn}\,r_{pn}}}{r_{pn}^{2}}\frac{e^{i\,k_{dA}\,r_{dA}}}{r_{dA}}                                             \nonumber\\
&\stackrel{\rho \to \infty}{\sim}\,\frac{1}{\rho^{5}}\,e^{i\,\rho\,g(\alpha,\,z)}. 
\label{integr1}
\end{align}
Thus we need to estimate a highly oscillatory integral:
\begin{align}
&J_{1}  \sim \,\lim_{\rho \rightarrow \infty }\,\int\limits_{-1}^{1}\,{\rm d}z 
\int\limits_{0}^{\pi /2}\mathrm{d}{\alpha }\sin^{2}{\alpha}\,\cos^{2}{\alpha }\,\,e^{i\,\rho\,g(z,\,\alpha)}.
\label{inthosc1}
\end{align}
Evidently that this integral and, hence, $M_{Stot}^{DW}(P,\,{\rm {\bf k}}_{dA})$ vanishes at $\rho \to \infty$, whether a stationary phase point does exist or not, because the integration brings $\rho$ to the denominator.  

Now we proceed to $M_{Sint}^{DW}(P,\,{\rm {\bf k}}_{dA})$. We rewrite is as 
\begin{align}
&M_{Sint}^{DW}(P,\,{\rm {\bf k}}_{dA}) =  \int\limits_{r_{nA} \leq R_{nA}} {\rm d}{\rm {\bf r}}_{nA}\,\int {\rm d}{\rm {\bf r}}_{pF} \chi_{pF}^{(-)*}({\rm {\bf r}}_{pF})         \nonumber\\
&\times \Upsilon_{nA}^{(int)(-)*}({\rm {\bf r}}_{nA})\,\big[{\overleftarrow T}_{pF} - {\overrightarrow T}_{pF} \big]\,\Psi_{i}^{CDCC(+)}({\rm {\bf r}}_{pF},\,{\rm {\bf r}}_{nA})                       \nonumber\\
&+ \int\limits_{r_{nA} \leq R_{nA}} {\rm d}{\rm {\bf r}}_{nA}\,\int {\rm d}{\rm {\bf r}}_{pF} \chi_{pF}^{(-)*}({\rm {\bf r}}_{pF})         \nonumber\\
&\times \Upsilon_{nA}^{(int)(-)*}({\rm {\bf r}}_{nA})\,\big[{\overleftarrow T}_{nA} - {\overrightarrow T}_{nA} \big]\,\Psi_{i}^{CDCC(+)}({\rm {\bf r}}_{pF},\,{\rm {\bf r}}_{nA}). 
\label{MSintCDCCpostres11}
\end{align}
Let first consider the first matrix element  containing $T_{pF}$.  It is easy to show that this matrix element vanishes. 
After transforming it into the surface integral over ${\rm {\bf r}}_{pF}$ we get
\begin{align}
&\int\limits_{r_{nA} \leq R_{nA}} {\rm d}{\rm {\bf r}}_{nA}\,\int {\rm d}{\rm {\bf r}}_{pF} \chi_{pF}^{(-)*}({\rm {\bf r}}_{pF}) \,\Upsilon_{nA}^{(int)(-)*}({\rm {\bf r}}_{nA})           \nonumber\\      
&\times \Big[{\overleftarrow T}_{pF} - {\overrightarrow T}_{pF} \Big]\,\Psi_{i}^{CDCC(+)}({\rm {\bf r}}_{pF},\,{\rm {\bf r}}_{nA})  \nonumber\\
&= -\frac{1}{2\,\mu_{pF}}\,\lim_{R_{pF} \rightarrow \infty}\,R_{pF}^{2}\,\int\,{\rm d}{\Omega}_{{\rm{\bf r}}_{pF}}\,\int\limits_{r_{nA} \leq R_{nA}}{\rm d}\,{\rm {\bf r}}_{nA}\,\Upsilon_{nA}^{(-)*}({\rm {\bf r}}_{nA})                              \nonumber\\
&\times \Big[\Psi_{i}^{CDCC(+)}({\rm {\bf r}}_{pF},\,{\rm {\bf r}}_{nA})\,\frac{\partial \chi_{pF}^{(-)*}({\rm {\bf r}}_{pF})  }{\partial r_{pF}}                                             \nonumber\\       
&- \chi_{pF}^{(-)*}
({\rm {\bf r}}_{pF})\,\frac{\partial \Psi_{i}^{CDCC(+)}({\rm {\bf r}}_{pF},\,{\rm {\bf r}}_{nA})}{\partial r_{pF}}\,\Big]\Big|_{r_{pF}=R_{pF}}
\label{MintCDCCresaux1}
\end{align}
The matrix element containing $n=0$ term of the CDCC wave function vanishes because in the subspace $r_{nA} \leq R_{nA}$ at $r_{pF}\to \infty$ the deuteron bound state wave function exponentially fades away. The terms of the CDCC wave function with $n \geq 1$
also produce vanishing matrix element because the CDCC wave function corresponding to these terms in the subspace $r_{nA} \leq R_{nA}$ at $r_{pF}\to \infty$ decays as $1/r_{pF}^{3}$, that the whole matrix element vanishes as $\lim\limits_{R_{pF} \to \infty} R_{pF}^{2}/R_{pF}^{3} \to 0$. 
Thus we arrive at
\begin{widetext}
\begin{align}
&M_{S}^{DW}(P,\,{\rm {\bf k}}_{dA})= -M_{Sint}^{DW}(P,\,{\rm {\bf k}}_{dA}) = -\int\limits_{r_{nA} \leq R_{nA}} {\rm d}{\rm {\bf r}}_{nA}\,\int {\rm d}{\rm {\bf r}}_{pF}\, \chi_{pF}^{(-)*}({\rm {\bf r}}_{pF})\,\Upsilon_{nA}^{(int)(-)*}({\rm {\bf r}}_{nA})\,[{\overleftarrow T}_{nA} - {\overrightarrow T}_{nA}]\,\Psi_{i}^{CDCC(+)}({\rm {\bf r}}_{pF},\,{\rm {\bf r}}_{nA})    \nonumber\\
&= \frac{1}{2\,\mu_{nA}}\,R_{nA}^{2}\,\int\,{\rm d}\,{\rm {\bf r}}_{pF}\, \chi_{pF}^{(-)*}({\rm {\bf r}}_{pF})\,\Big[\Psi_{i}^{CDCC(+)}({\rm {\bf r}}_{pF},\,{\rm {\bf r}}_{nA})\,\frac{\partial }{\partial r_{nA}}\Upsilon_{nA}^{(-)*}({\rm {\bf r}}_{nA}) - \Upsilon_{nA}^{(-)*}({\rm {\bf r}}_{nA}) \,\frac{\partial }{\partial r_{nA}}\,\Psi_{i}^{CDCC(+)}({\rm {\bf r}}_{pF},\,{\rm {\bf r}}_{nA})\Big]\Big|_{r_{nA}=R_{nA}}.
\label{MSintCDCCresnA1}
\end{align}
\end{widetext}

\end{document}